\documentclass[11pt]{article}
\usepackage{amsfonts,amsmath,amssymb}
\usepackage{enumerate}
\usepackage[hidelinks]{hyperref}
\usepackage{bbm}
\usepackage{nicefrac}
\usepackage[all]{xy}
\usepackage{graphicx}
\usepackage{subcaption}
\usepackage{bm}
\usepackage{makecell}
\usepackage[table]{xcolor}

\usepackage{upgreek}
\usepackage{slashed}

\usepackage{float}			
\usepackage{lscape}                 
\usepackage{pdflscape}               

\usepackage{booktabs}
\usepackage{longtable,array}
\usepackage{multicol}

\newcommand{\be}{\begin{equation}}
\newcommand{\ee}{\end{equation}}

\newcommand{\half}{\tfrac{1}{2}}  

\newcommand{\bea}{\begin{eqnarray}}
\newcommand{\eea}{\end{eqnarray}}

\newcommand{\bes}{\begin{subequations}}
\newcommand{\ees}{\end{subequations}}

\newcommand{\cG}{{\cal G}}
\newcommand{\cH}{{\cal H}}

\newcommand{\cL}{{\cal L}}
\newcommand{\cM}{{\cal M}}
\newcommand{\cN}{{\cal N}}

\newcommand{\cP}{{\cal P}}
\newcommand{\cQ}{{\cal Q}}
\newcommand{\cR}{{\cal R}}
\newcommand{\cS}{{\cal S}}

\newcommand{\cY}{{\cal Y}}

\def\sst#1{{\scriptscriptstyle #1}}

\def\0{{\sst{(0)}}}
\def\1{{\sst{(1)}}}
\def\2{{\sst{(2)}}}
\def\3{{\sst{(3)}}}
\def\4{{\sst{(4)}}}
\def\5{{\sst{(5)}}}
\def\6{{\sst{(6)}}}
\def\7{{\sst{(7)}}}
\def\8{{\sst{(8)}}}

\def\cH{{{\cal H}}}

\def\cM{{{\cal M}}}
\def\cY{{{\cal Y}}}

\allowdisplaybreaks  

\usepackage{multirow}
\usepackage{rotating}

\newcommand{\ba}{\begin{align}}
\newcommand{\ea}{\end{align}}

\newcommand{\bse}{\begin{subequations}}
\newcommand{\ese}{\end{subequations}}

\allowdisplaybreaks

\usepackage{lipsum}

\newlength\Colsep
\newcommand{\uM}{\underline{M}}
\newcommand{\uN}{\underline{N}}

\newcommand{\LL}{L\bar L}
\newcommand{\LAOne}{L\bar A_1}
\newcommand{\AOneL}{A_1\bar L}
\newcommand{\LATwo}{L\bar A_2}
\newcommand{\ATwoL}{A_2\bar L}
\newcommand{\LBone}{L\bar B_1}
\newcommand{\BOneL}{B_1\bar L}
\newcommand{\AOneAOne}{A_1\bar A_1}
\newcommand{\ATwoATwo}{A_2\bar A_2}

\begin{document}

\addtocontents{toc}{\setcounter{tocdepth}{2}}

\makeatletter
\renewcommand{\theequation}{\thesection.\arabic{equation}}
\@addtoreset{equation}{section}
\makeatother

\hypersetup{pageanchor=false}
\begin{titlepage}

\begin{flushright}
%
%
\end{flushright}

\vspace{20pt}

   \begin{center}
   \baselineskip=16pt

   \begin{Large}\textbf{
  %
%
%
%
A universal sector of $\mathcal{N}=1$ class $\mathcal{S}$ spectra
}
   \end{Large}

\vspace{25pt}

{\large Ritabrata Bhattacharya$^{1,2}$, Abhay Katyal$^{2}$ and Oscar Varela$^{2,3}$ }

\vspace{30pt}

	\begin{small}

	{\it $^1$ Department of Physics, Indian Institute of Science Education and Research Bhopal \\
	Bhopal, 462066, India} 



	\vspace{15pt}
          
   {\it $^2$ Department of Physics, Utah State University, Logan, UT 84322, USA} 

	\vspace{15pt}
          
	{\it $^3$  Instituto de F\'\i sica Te\'orica UAM/CSIC, 28049 Madrid, Spain} 
		
	\end{small}

\vskip 70pt

\end{center}

\begin{center}
\textbf{Abstract}
\end{center}

\begin{quote}

We determine a universal sector of the spectrum of light operators in four-dimensional $\mathcal{N}=1$ class $\mathcal{S}$ superconformal field theories arising from M5-branes wrapped on a compact hyperbolic Riemann surface $\Sigma_2$. We work at large $N$, analysing Kaluza--Klein fluctuations about the dual Bah--Beem--Bobev--Wecht anti-de Sitter backgrounds for general values of the twist parameter $z$. To obtain these spectra, we specialise recently derived mass operators to modes that are constant on $\Sigma_2$ by imposing a suitable $z$-dependent generalised $G$-structure. We then organise the resulting eigenstates into long and protected \mbox{$\mathcal{N}=1$} supermultiplets with definite flavour quantum numbers. As $z$ varies, rich patterns emerge in the spectrum, including both isolated multiplets and infinite towers of multiplets, which we characterise analytically.

\end{quote}

\vfill

\end{titlepage}
\hypersetup{pageanchor=true}

\tableofcontents


\section{Introduction}
\label{sec:Introduction}

Four-dimensional $\mathcal{N}=1$ superconformal field theories (SCFTs) of class $\mathcal{S}$ provide a large family of interacting fixed points obtained by twisted compactification of the six-dimensional $(2,0)$ theory on Riemann surfaces \cite{Benini:2009mz,Bah:2011vv,Bah:2012dg}. These theories extend the familiar $\mathcal{N}=2$ class $\mathcal{S}$ construction \cite{Gaiotto:2009we,Gaiotto:2009gz} and, like the latter, inherit a geometric description from their six-dimensional ancestor. In this paper we focus on the type A$_{N-1}$ theories constructed by Bah, Beem, Bobev and Wecht (BBBW) \cite{Bah:2011vv,Bah:2012dg}, arising from $N$ M5-branes wrapped on a smooth, unpunctured, compact hyperbolic surface $\Sigma_2=H^2/\Gamma$ of genus $g\geq2$. Here $\Gamma$ is a discrete group of isometries acting freely on $H^2$. The surface $\Sigma_2$ sits inside a Calabi--Yau threefold, and its normal bundle splits as $\mathcal L_p\oplus\mathcal L_q$, where the two line bundles have degrees $p$ and $q$ and satisfy $\mathcal L_p\otimes\mathcal L_q=K_{\Sigma_2}$, with $p+q=2g-2$ and $K_{\Sigma_2}$ the canonical line bundle of $\Sigma_2$.

For generic $(p,q)$, the topological twist preserves four supercharges. The rational 
\begin{equation}
\label{eq:IntroTwistingData}
z \equiv \frac{p-q}{p+q}
\end{equation}
parametrises the topological twisting data and labels the resulting families of SCFTs at fixed genus $g$ \cite{Bah:2011vv,Bah:2012dg}. The corresponding $\cN=1$ R-symmetry, as well as a generic flavour symmetry, arise from this geometric construction as follows. The generators $T_p$ and $T_q$ of phase rotations of the fibres of $\mathcal L_p$ and $\mathcal L_q$ span the Cartan subgroup $\textrm{U}(1)_+\times \textrm{U}(1)_-$ of the SO$(5)$ R-symmetry of the parent $(2,0)$ theory. More precisely, the (anti)diagonal combinations $J_\pm$,
\begin{equation}
\label{eq:R0FfromJpm}
R_0 \equiv J_- \equiv T_p+T_q\, ,\qquad
F \equiv J_+ \equiv T_p-T_q\; ,
\end{equation}
generate $\textrm{U}(1)_\pm$. The generator $F \equiv J_+$ commutes with the preserved $\cN=1$ supercharge and therefore singles out $\textrm{U}(1)_+$ as flavour. In contrast, $R_0  \equiv J_-$ rotates the supercharges and hence defines a (reference) R-symmetry. The actual R-charge $r$ is determined by $a$-maximisation \cite{Intriligator:2003jj} which, at leading order in large $N$, gives the $z$-dependent mixing \cite{Bah:2011vv,Bah:2012dg}
\begin{equation}
\label{eq:SpectrumExactRCharge}
r=R_0+\epsilon(z)F\, ,\qquad \textrm{with} \quad
\epsilon(z)=\frac{1-\sqrt{1+3z^2}}{3z}\; ,
\end{equation}
and $\epsilon(0)=0$ defined by continuity.

Being related to the M5-brane, these $\cN=1$ SCFTs have holographic duals in M-theory. At large $N$, these are provided by the family, again labelled by the parameter $z$ in (\ref{eq:IntroTwistingData}), of AdS$_5$ backgrounds of eleven-dimensional supergravity also constructed by BBBW \cite{Bah:2011vv,Bah:2012dg}. In these solutions, the internal six-dimensional geometry is a squashed $S^4$ bundle over the hyperbolic, unpunctured Riemann surface $\Sigma_2$. Two angular coordinates on $S^4$, dual to $\textrm{U}(1)_+\times \textrm{U}(1)_-$, encode the $z$-dependent fibration. The family contains $\mathcal{N}=1$ BBBW AdS$_5$ backgrounds in two ranges of $z$, which we will refer to as
\begin{equation} \label{eq:zRegions}
\textrm{ the interior region: $0<|z|<1$}, \qquad \textrm{and} \qquad \textrm{the exterior region: $|z|>1$}.
\end{equation}
At distinguished values of $z$594
, the family reproduces the Maldacena--N\'u\~nez (MN) AdS$_5$ solutions \cite{Maldacena:2000mw}. Setting $z=0$ yields MN1, which retains $\mathcal{N}=1$ supersymmetry but has flavour symmetry enhanced to $\mathrm{SU}(2)_+$. At $|z|=1$, the family yields MN2, where supersymmetry is enhanced to $\mathcal{N}=2$ and no continuous flavour symmetry remains.

A substantial amount of information about these $\cN=1$ class $\cS$ SCFTs is available from protected observables: their anomalies, central charges \cite{Bah:2012dg,Baggio:2014hua,Bah:2020uev} and superconformal index \cite{Beem:2012yn} have all been determined. However, an explicit enumeration of the single-trace operators in these theories remains unavailable. In this paper, we take a significant step towards such an enumeration by characterising large families of light operators whose conformal dimensions remain of order one as $N$ becomes large, within a sector accessible to classical supergravity. Our analysis uses the standard holographic dictionary, which identifies these light single-trace operators in the large-$N$ limit with Kaluza--Klein (KK) states of the dual AdS$_5$ supergravity solutions \cite{Maldacena:1997re,Gubser:1998bc,Witten:1998qj}. The results we obtain apply universally to any hyperbolic Riemann surface $\Sigma_2$ wrapped by the M5-branes and, in particular, are independent of both its genus $g$ and the Fuchsian group $\Gamma$ defining it. The spectrum nevertheless depends on the specific BBBW background through the value of the twist parameter $z$ in (\ref{eq:IntroTwistingData}).

In \cite{Bhattacharya:2024tjw,Bhattacharya:2026ypg}, we carried out this spectral analysis for the MN2 and MN1 backgrounds. Here, we extend those results to the full BBBW family at general values of $z$. We perform our KK analysis using the spectral methods developed in \cite{Malek:2019eaz,Malek:2020yue,Varela:2020wty,Cesaro:2020soq,Pico:2026kch}, which build on the exceptional generalised geometry (ExGG) \cite{Coimbra:2011ky,Coimbra:2012af} and exceptional field theory (ExFT) \cite{Hohm:2013pua,Hohm:2013vpa} reformulations of $D=11$ supergravity. Applying these methods at face value requires a maximally supersymmetric consistent truncation whose linearised field content can be identified with KK level $k=0$. For the BBBW backgrounds, we constructed such a truncation in \cite{Bhattacharya:2024tjw,Bhattacharya:2026ypg}. As discussed in detail in those references, the resulting $D=5$ $\cN=8$ supergravity involves gaugings of the local trombone scaling symmetry in the sense of \cite{LeDiffon:2008sh,Varela:2025xeb}. The corresponding BBBW spectra are generally defined only locally and, following the terminology of \cite{Pico:2026kch,Bhattacharya:2026ypg}, we refer to them as putative spectra. Such local spectra are known to exhibit pathologies for related AdS$_4$ wrapped M5-brane solutions \cite{Pico:2025cmc,Pico:2026rji,Pico:2026kch}. The putative BBBW spectra likewise contain pathological modes: as we show in the appendices, some fall below the unitarity bound.

As argued in \cite{Bhattacharya:2026ypg} in the case at hand, physical modes can be extracted  from the putative spectrum by restricting attention to modes that are constant on $\Sigma_2$. This restriction selects modes invariant under the globally defined generalised $\mathrm{U}(1)_z$-structure \cite{Bhattacharya:2026ypg} defined on the BBBW solutions \cite{Cassani:2019vcl,Cassani:2020cod}, see also \cite{Josse:2025uro}. Here, in contrast to \cite{Bhattacharya:2026ypg}, we impose $\mathrm{U}(1)_z$-invariance from the outset, before diagonalising the mass matrices. This approach is computationally much more efficient than first computing the putative spectra and subsequently selecting the physical modes, as done in \cite{Bhattacharya:2026ypg} for the MN1 case and in \cite{Pico:2026kch} for closely related AdS$_4$ cases. We therefore begin, in section \ref{sec:U1zMassMatrices}, by constructing $\Sigma_2$-constant, $\mathrm{U}(1)_z$-invariant mass matrices whose spectra are guaranteed to contain only physical modes, after removal of Goldstone directions. In section \ref{sec:UniversalSpectrum}, we diagonalise these mass matrices and organise the resulting eigenstates into multiplets of the $\cN=1$ superconformal algebra with definite flavour. The efficiency of this approach allows us to study the spectral problem up to very high KK levels $k$ and thereby obtain closed-form expressions for the supermultiplet content at fixed $z$ for all $k$. A rich pattern of long and protected multiplets emerges, which we characterise analytically. Examples  at representative values of $z$ are discussed in section \ref{sec:RepresentativeSpectra}, before concluding in section \ref{sec:Discussion}.


\section{\texorpdfstring{$\mathrm{U}(1)_z$-invariant mass matrices}{U(1)z-invariant mass matrices}}
\label{sec:U1zMassMatrices}

A basic ingredient of the ExFT-based KK spectral programme of \cite{Malek:2019eaz,Malek:2020yue,Varela:2020wty,Cesaro:2020soq,Pico:2026kch} is the existence of a consistent truncation to a maximal lower-dimensional gauged supergravity. In our case, this is a $D=5$, $\cN=8$ gauged supergravity constructed in \cite{Varela:2025xeb}. The linearised theory provides the lowest-lying modes in the KK expansion, while the towers above them are constructed by expanding the ExFT fields in graviton mass eigenstates. This approach has proved highly effective, yielding complete spectral results in examples such as those studied in \cite{Malek:2019eaz,Malek:2020yue,Varela:2020wty,Cesaro:2020soq,Guarino:2020flh,Cesaro:2021haf,Giambrone:2021zvp,Cesaro:2021tna,Cesaro:2022mbu}. In these cases, the maximal consistent truncations are performed on spheres or products thereof, as in {\it e.g.}~\cite{deWit:1986iy,Lee:2014mla,Guarino:2015jca,Guarino:2015vca,Inverso:2016eet}, and the resulting lower-dimensional theories do not involve trombone gaugings of the non-compact scaling symmetry, as in {\it e.g.}~\cite{deWit:1982ig,DallAgata:2011aa,Guarino:2015qaa}. The internal space of the BBBW solutions, however, is not spherical: it contains a four-sphere $S^4$ fibred over the Riemann surface $\Sigma_2$. A maximal consistent truncation to $D=5$ gauged supergravity nevertheless exists \cite{Bhattacharya:2024tjw,Bhattacharya:2026ypg}, but the construction is defined only locally on the total space of this fibration. The local character of this construction is reflected in the presence of trombone gaugings in the lower-dimensional supergravity \cite{Bhattacharya:2024tjw,Varela:2025xeb}. It may also manifest itself through the appearance of pathological KK modes, as discussed in \cite{Pico:2025cmc,Pico:2026rji,Pico:2026kch} for related AdS$_4$ M5-brane examples. Any such modes must be removed from the physical spectrum.

In this context, the term \emph{putative} was coined in \cite{Pico:2026kch,Bhattacharya:2026ypg} to describe spectra obtained at face value, before their constituent modes are screened to identify those that belong to the actual physical spectrum. For the M5-brane solutions considered here, one criterion for selecting physical modes is to require that the fluctuations be constant over the hyperbolic Riemann surface $\Sigma_2$. This condition admits a geometric formulation within ExGG. Viewed in this framework, the BBBW AdS$_5$ solutions \cite{Bah:2011vv,Bah:2012dg} of $D=11$ supergravity admit a $z$-dependent generalised $\mathrm{U}(1)_z$-structure \cite{Cassani:2019vcl,Cassani:2020cod}. The associated structure group $\mathrm{U}(1)_z$ is the subgroup of $\mathrm{USp}(8) \subset \mathrm{E}_{6(6)}$ generated by \cite{Cassani:2020cod}
\begin{equation} \label{eq:Q_zCharge}
Q_z \equiv R_0 + z F + J_0 \,,
\end{equation}
where $R_0$ and $F$ are the generators of $\mathrm{U}(1)_\pm$ defined in (\ref{eq:R0FfromJpm}), while $J_0$ is the compact generator of the $\mathrm{SL}(2,\mathbb{R})$ that commutes with $\mathrm{SL}(6,\mathbb{R})$ inside $\mathrm{E}_{6(6)}$. As discussed at length in \cite{Bhattacharya:2026ypg} following \cite{Cassani:2019vcl,Cassani:2020cod}, constancy on $\Sigma_2$ is equivalent to invariance under this generalised structure. The $\mathrm{U}(1)_z$-invariant sector of the putative spectrum selects all states annihilated by $Q_z$ in (\ref{eq:Q_zCharge}). The resulting excitations are universally valid for every quotient $\Sigma_2 = H^2/\Gamma$ and are independent of the details of the Riemann surface, including its genus and the choice of discrete group $\Gamma$. The aim of this paper is to characterise this universal subset of the KK spectrum of the BBBW solutions for all values of the twist parameter $z$.

In \cite{Bhattacharya:2026ypg}, we first diagonalised the trombone-enhanced ExFT KK mass matrices \cite{Pico:2026kch} to obtain putative spectra and only then selected the physical modes above the MN1 background. This computationally demanding procedure can be made substantially more efficient by reversing the order of these steps: here, we first construct the $\mathrm{U}(1)_z$-invariant mass matrices and then diagonalise them. The spectra obtained in this way are guaranteed to contain only physical modes. We implement this approach in appendix~\ref{app:InvariantMassMatrices}, by constructing the action of $Q_z$ in (\ref{eq:Q_zCharge}) on the generic KK towers. Solving the corresponding zero-$Q_z$-charge conditions gives bases of $\mathrm{U}(1)_z$-invariant tensors, which we organise by their total component flavour and R-charges. Finally, we contract the KK trombone mass matrices of \cite{Bhattacharya:2026ypg,Pico:2026kch} with these tensors to obtain  $\mathrm{U}(1)_z$-invariant mass matrices (\ref{eq:InvariantMassGravitonNew}). The trombone contributions do not, however, necessarily drop out of these invariant mass matrices. For example, they disappear from the $\mathrm{U}(1)_z$-invariant graviton mass matrix but remain in the corresponding scalar mass matrix.

As emphasised above, the mass matrices (\ref{eq:InvariantMassGravitonNew}) contain only modes that are constant on $\Sigma_2$. Determining additional modes with non-trivial dependence on the Riemann surface requires solving the corresponding spectral problem on $\Sigma_2$. These include modes obtained by allowing the coefficients in our $\mathrm{U}(1)_z$-invariant sectors to vary over the surface, rather than holding them constant as we have done throughout this paper. The complete BBBW spectrum also contains modes that are not invariant under $\mathrm{U}(1)_z$. These can nevertheless give rise to globally well-defined fluctuations, provided that their transformations between coordinate patches of the $S^4$ fibration over $\Sigma_2$ are compensated by those of their $\Sigma_2$-dependent coefficients. In fact, some of these additional families may themselves have universal spectral data, as occurs in the MN1 case \cite{Bhattacharya:2026ypg}. Our U$(1)_z$-invariant construction therefore captures a useful universal subset of the physical KK spectrum, but neither gives the complete spectrum nor exhausts all universal modes. Further discussion on the scope of our results can be found in section \ref{sec:Discussion}.


\section{\texorpdfstring{Universal light operator spectrum of $\cN=1$ class $\cS$ SCFTs}{Universal light operator spectrum of N=1 class S SCFTs}}
\label{sec:UniversalSpectrum}


We now determine holographically the light operator spectrum of the four-dimensional $\cN=1$ class $\cS$ SCFTs of BBBW \cite{Bah:2011vv,Bah:2012dg} within the universal sector under consideration, for all values of the twist parameter $z$. To this end, we diagonalise the $\mathrm{U}(1)_z$-invariant, $\Sigma_2$-constant mass matrices specified in section \ref{sec:U1zMassMatrices}, and organise the resulting eigenstates in supermultiplets of the $\cN=1$ superconformal group, $\textrm{SU}(2,2|1)$, with definite U$(1)_+$ flavour.


\subsection{General structure}
\label{sec:SpectrumGeneralStructure}

Let $(\widehat p,\widehat q)$ be coprime twisting data satisfying
$\widehat p+\widehat q>0$, and define
\begin{equation}
\label{eq:SpectrumTwistData1}
z=\frac{\widehat p-\widehat q}{\widehat p+\widehat q} \;  ,
\end{equation}
without loss of generality, along with
\begin{equation}
\label{eq:SpectrumTwistData2}
f_z \equiv \widehat p+\widehat q\, ,\qquad
\delta_z \equiv \widehat p-\widehat q\, ,\qquad
\lambda_z \equiv |\delta_z|\, ,\qquad
\mu_z \equiv |\widehat p|+|\widehat q|\, ,
\end{equation}
so that $z=\delta_z/f_z$, $|z|=\lambda_z/f_z$ and
$\mu_z=\max(f_z,\lambda_z)$. The hatted integers in
(\ref{eq:SpectrumTwistData1}) are obtained from the line-bundle
degrees $p,q$ in (\ref{eq:IntroTwistingData}) by setting
$\widehat p=p/d$ and $\widehat q=q/d$, with $d=\gcd(|p|,|q|)$.
The original degrees $p,q$ need not be coprime, and no such
restriction is imposed in \cite{Bah:2011vv,Bah:2012dg}.
Removing their common factor $d$ ensures that the positive integer
label $\nu$ introduced below enumerates all allowed nonzero flavour-charge magnitudes. Thus, the twist parameter $z$ in (\ref{eq:SpectrumTwistData1}) is
the same as $z$ in (\ref{eq:IntroTwistingData}). Since $p+q=2g-2>0$, the hatted integers automatically satisfy
$\widehat p+\widehat q>0$. Since $z$ fixes the ratio $\widehat p/\widehat q=(1+z)/(1-z)$, coprimality and the condition $\widehat p+\widehat q>0$ uniquely determine 
$(\widehat p,\widehat q)$. Thus all quantities in (\ref{eq:SpectrumTwistData2}) depend on $z$ alone, a dependence that we have indicated with a subindex $z$. As we will see, the quantities
(\ref{eq:SpectrumTwistData2}) specify the physical flavour and
 R-charges, and determine the allowed KK levels
at which flavour-charged modes appear. The mixing parameter $\epsilon (z)$ defined in (\ref{eq:SpectrumExactRCharge}) will also play a significant role below.

With these definitions, we evaluate the $\mathrm{U}(1)_z$-invariant mass matrices constructed in section~\ref{sec:U1zMassMatrices} on the BBBW solution \cite{Bah:2011vv,Bah:2012dg}, using the conventions of \cite{Varela:2025xeb,Bhattacharya:2026ypg}. We work at a generic value of the twist parameter $z$ in (\ref{eq:SpectrumTwistData1}) and at fixed KK level $k=0,1,\ldots$, with explicit calculations extending through $k=20$. The KK label $k$ specifies the degree of the $S^4$ spherical harmonic from which the corresponding mode descends, as explained in appendix~\ref{app:InvariantMassMatrices}. After diagonalising these matrices, we remove the Higgsed and spurious states following the procedure described in \cite{Bhattacharya:2026ypg}. We then translate the remaining masses, normalised to the AdS$_5$ radius, into conformal dimensions $E$ using the standard AdS/CFT dictionary \cite{Maldacena:1997re,Gubser:1998bc,Witten:1998qj}, see {\it e.g.}~(4.15) of \cite{Varela:2025xeb}. We also track the $\mathrm{U}(1)_+$ flavour and $\mathrm{U}(1)_-$ reference R-charges (\ref{eq:R0FfromJpm}) according to the group-theoretical branchings of \cite{Bhattacharya:2026ypg}. These determine the superconformal R-charge $r$ of each mass eigenstate through (\ref{eq:SpectrumExactRCharge}). Finally, we organise the eigenstates into representations of $\mathrm{SU}(2,2|1)\times\mathrm{U}(1)_+$ according to their conformal dimensions $E$, R-charges $r$ and flavour charges $F$. The conformal dimensions $E$ generally depend on $z$, as do the R-charges $r$ and flavour charges $F$. The R-charges inherit this dependence from the mixing (\ref{eq:SpectrumExactRCharge}). The $z$-dependence of the flavour charges arises instead from the $\mathrm{U}(1)_z$-invariance condition: although the flavour generator (\ref{eq:R0FfromJpm}) is independent of $z$, requiring modes to be annihilated by $Q_z$ in (\ref{eq:Q_zCharge}) selects a $z$-dependent subset of flavour charged sectors. 

Following \cite{Cordova:2016emh}, we denote the representations of $\mathrm{SU}(2,2|1)\times\mathrm{U}(1)_+$ by $X\bar{Y}[E;j_1,j_2;r]_F$. Here $E$, $(j_1,j_2)$ and $r$ are the conformal dimension, (half-)integer Lorentz spins and R-charge of the superconformal primary (SCP) of the supermultiplet, while $F$ is the $\mathrm{U}(1)_+$ flavour charge, common to all states in a given multiplet. The labels $X$ and $Y$ specify the shortening conditions, if any. We also use the standard nomenclature whereby an $\mathrm{SU}(2,2|1)$ supermultiplet is called a graviton, gravitino, vector or tensor multiplet according to its characteristic top-spin field. Multiplets with a spin-$\frac12$ top-spin component are referred to as chiral multiplets. For ease of reference, and to facilitate assigning mass eigenstates to supermultiplets, we provide tables of the explicit state content of these representations in appendix~\ref{app:NOneMultipletContent}. 

Let $\textrm{Spec}_k^0(z)$ denote the collection of such $\mathrm{U}(1)_z$-invariant $\mathrm{SU}(2,2|1)\times\mathrm{U}(1)_+$ supermultiplets present at fixed twist parameter $z$ and KK level $k=0,1,2,\ldots$. The universal sector of the BBBW large-$N$ light operator spectrum under consideration is thus obtained by taking the direct sum over all KK levels,
\begin{equation}
\label{eq:FullSpec}
\textrm{Spec}^0(z)
=
\bigoplus_{k=0}^{\infty}\textrm{Spec}_k^0(z)
\; ,
\end{equation}
where the superscript $0$ indicates $\mathrm{U}(1)_z$ invariance. The organisation of (\ref{eq:FullSpec}) by KK level follows directly from our calculation. It is also natural to identify families of supermultiplets that extend across multiple KK levels, with a spacing in $k$ that may depend on $z$. Thus, the $\mathrm{U}(1)_z$-invariant spectrum also admits the useful decomposition
\begin{equation}
\label{eq:SpectrumFourfoldSplit}
\textrm{Spec}^0(z)
=
\cS_{\textrm{$F$-neutral}}
\oplus
\cL_{\textrm{$F$-neutral}}(z)
\oplus
\cS_{\textrm{$F$-charged}}(z)
\oplus
\cL_{\textrm{$F$-charged}}(z)
\; ,
\end{equation}
with each of the four blocks on the r.h.s.~capturing multiplets that arise at different KK levels. In (\ref{eq:SpectrumFourfoldSplit}), $\cS$ and $\cL$ denote collections of short and long multiplets, respectively, while the subscripts $F$-neutral and $F$-charged distinguish the sectors with vanishing and non-vanishing flavour charge. Both descriptions, (\ref{eq:FullSpec}) and (\ref{eq:SpectrumFourfoldSplit}), of $\textrm{Spec}^0(z)$ are complementary. Sections \ref{sec:FlavourNeutralSpectrum} and \ref{sec:FlavourChargedSpectrum} will describe $\textrm{Spec}^0(z)$ for all $z$ following the decomposition (\ref{eq:SpectrumFourfoldSplit}), occasionally referring to KK levels as well. Section~\ref{sec:RepresentativeSpectra} will then use the organisation (\ref{eq:FullSpec}) to display the spectrum at the first few KK levels for selected values of $z$.

Before presenting the details, let us comment on some general features of the structure of the spectrum. Multiplets in $\textrm{Spec}^0(z)$ are either CPT-self-conjugate or occur in CPT-conjugate pairs. Both can appear in isolation or form infinite families indexed, in the description (\ref{eq:SpectrumFourfoldSplit}), by up to four integers $(\kappa, \nu , \eta , a)$. The label $\eta$, which we will refer to as the R-charge offset, takes on the following possible values for each type of multiplet,
\begin{equation} \label{eq:etaDef}
\textrm{Graviton}: \; \eta =0 \; , \quad 
\textrm{Gravitino}: \; \eta = \pm1  \; , \quad 
\textrm{Vector}: \; \eta = 0,\,  \pm2  \; ,
\end{equation}
regardless of whether these are long or short, and is undefined for the chiral and tensor multiplets in the protected spectrum. The labels $\kappa$ and $\nu$ are non-negative, with typically infinite ranges
\begin{equation}
\label{eq:kappanuRange}
\kappa=\kappa_{\textrm{min}} \, ,\;
\kappa_{\textrm{min}}+1 \, ,\;
\kappa_{\textrm{min}}+2 \, ,\;
\ldots \; ,
\qquad
\nu=\nu_{\textrm{min}} \, ,\;
\nu_{\textrm{min}}+1 \, ,\;
\nu_{\textrm{min}}+2 \, ,\;
\ldots \; .
\end{equation}
The non-negative lower limits $\kappa_{\textrm{min}}$, $\nu_{\textrm{min}}$ will be specified on a case-by-case basis. Both may depend on $z$, while $\kappa_{\textrm{min}}$ may also depend on $\nu_{\textrm{min}}$. At each value of $\kappa$ and $\nu$, there may be finitely many multiplets or CPT pairs, distinguished by an additional index $a$. The finite range of $a$ depends on the multiplet and on $\kappa$ through the integers
\begin{equation}
\label{eq:NeutralParityDefinitions}
\check{\kappa} =\left\lfloor\frac{\kappa}{2}\right\rfloor \, ,\qquad
\hat{\kappa}  =\left\lceil\frac{\kappa}{2}\right\rceil \, ,
\end{equation}
where $\lfloor x\rfloor$ and $\lceil x\rceil$ denote the usual floor and ceiling functions. From the point of view of the description (\ref{eq:FullSpec}), the KK level $k$ at which a multiplet arises becomes fixed, at each value of $z$, by the labels $(\kappa,\nu,\eta, a)$ of the description (\ref{eq:SpectrumFourfoldSplit}). More concretely, $k$ is independent of $\eta$ and $a$, and depends on $\kappa$, $\nu$ and possibly $z$ as 
\begin{equation}
\label{eq:GeneralKKLevel}
k=\alpha\kappa+\beta\mu_z\nu+\gamma \; .
\end{equation}
Here, $\alpha,\beta\in\{0,1\}$ and $\gamma$ is an integer, with all three coefficients fixed separately for each multiplet. Per (\ref{eq:GeneralKKLevel}), when $\beta =1$, increasing $\nu$ by one at fixed $\kappa$ shifts the KK level by $\mu_z$, defined in (\ref{eq:SpectrumTwistData2}), so the spacing along the family typically depends on $z$. Isolated multiplets arise at fixed KK level (\ref{eq:GeneralKKLevel}) with $\alpha=\beta=0$ and $\gamma\geq0$.

The conformal dimensions of the multiplets in $\textrm{Spec}^0(z)$ generally depend on $z$ and are determined by the roots of the polynomial families constructed in appendix~\ref{app:Spectral polynomials}. Each polynomial is specified by the labels $(\kappa,\nu,\eta)$, while its roots are indexed by $a$. Comparing the resulting dimensions with the shortening conditions summarised in appendix~\ref{app:NOneMultipletContent} determines which roots correspond to short or long multiplets. Separating these contributions leads to the family-dependent lower limits $\kappa_{\textrm{min}}$ and $\nu_{\textrm{min}}$ in (\ref{eq:kappanuRange}), specified in sections~\ref{sec:FlavourNeutralSpectrum}~and~\ref{sec:FlavourChargedSpectrum}. Thus, these limits can differ from those used to define the polynomial families in appendix~\ref{app:Spectral polynomials}, whose roots describe both types of multiplets, long and short. The range of $a$ is determined in a similar way. In all cases that we have studied, the polynomials have as many real roots, counted with multiplicity, as their respective degrees in (\ref{eq:xdegCharged}). Before distinguishing short and long multiplets, $a$ therefore runs over the full range allowed by the degree of the corresponding polynomial. For the long spectrum, however, roots that determine short-multiplet dimensions must be excluded. These exclusions account for the restricted ranges of $a$ given in sections~\ref{sec:FlavourNeutralSpectrum}~and~\ref{sec:FlavourChargedSpectrum} on a case-by-case basis.

The flavour-neutral short spectrum, $\cS_{\textrm{$F$-neutral}}$ in (\ref{eq:SpectrumFourfoldSplit}), consists entirely of isolated multiplets that are present for all $z$, with $z$-independent dimensions, R-charges and flavour charges. The long neutral spectrum, $\cL_{\textrm{$F$-neutral}}(z)$, consists of infinite families of graviton, gravitino and vector multiplets indexed by $\kappa$ and $a$, at the corresponding value of $\eta$. Likewise, these families are present for all $z$, {\it i.e.}~the supermultiplet content of this sector is independent of $z$. The flavour and R-charges also stay $z$-independent. With a few exceptions, however, the conformal dimensions do vary with $z$ in this sector, and this dependence enters exclusively through the square of the R-symmetry mixing parameter, $\epsilon(z)^2$. 
%
The flavour-charged spectrum, $\cS_{\textrm{$F$-charged}}(z)\oplus\cL_{\textrm{$F$-charged}}(z)$, also contains infinite families and, at certain values of $z$, isolated multiplets. The conformal dimensions, flavour and R-charges are $z$-dependent.  The multiplet content also depends on $z$. Within the interior and exterior regions (\ref{eq:zRegions}), the values
\begin{equation}
\label{eq:Distinguishedz}
\begin{aligned}
\textrm{distinguished interior sequence:}\quad
& |z|=\frac{j-1}{j+1} \; ,\\[5pt]
\textrm{distinguished exterior sequence:}\quad
& |z|=\frac{j+1}{j-1} \; ,
\end{aligned}
\qquad j=2,3,4,\ldots \; ,
\end{equation}
play a distinct role. In the interior, flavour-charged short multiplets occur only at these values. In the exterior, there are infinite multiplet families at every $z$, but the lower limits in (\ref{eq:kappanuRange}) of some families differ at the distinguished points (\ref{eq:Distinguishedz}).

Our spectrum is invariant under the reflection $z\mapsto-z$,
\begin{equation}
\label{eq:FullSpecInv}
\textrm{Spec}^0(-z)\cong\textrm{Spec}^0(z)
\; ,
\end{equation}
with flavour charge $F$ at $z$ identified with $-F$ at $-z$. In the decomposition (\ref{eq:SpectrumFourfoldSplit}), CPT-conjugate multiplets have equal SCP dimensions, opposite flavour and SCP R-charges, and exchanged Lorentz spins and shortening labels. The reflection instead preserves the dimension, SCP R-charge, Lorentz spins and shortening labels, while reversing only the flavour charge. Each of the four summands in (\ref{eq:SpectrumFourfoldSplit}) is therefore separately invariant: the neutral sectors are preserved, while the positive- and negative-flavour subsectors are exchanged within each charged sector. This symmetry reflects the exchange of the BBBW line bundles $\cL_p$ and $\cL_q$, which sends $z$ to $-z$ and yields an equivalent SCFT.


Finally, only a handful of multiplets in $\textrm{Spec}^0(z)$ have previously been identified at generic values of the twist parameter $z$. The existence of the stress-tensor multiplet, which belongs to $\cS_{\textrm{$F$-neutral}}$ for all $z$, follows from the general conjecture of \cite{Gauntlett:2007ma}, proved in \cite{Cassani:2019vcl}. This multiplet was explicitly obtained in \cite{Szepietowski:2012tb}, together with the flavour-neutral long vector multiplet in $\cL_{\textrm{$F$-neutral}}(z)$ listed at $\kappa=2$, $a=1$ in table~\ref{tab:FirstZ-neutral-long}. This long vector remains an $\mathrm{SU}(2)_+$ singlet in the MN1 limit, $z=0$, where it was also found in \cite{Faedo:2019cvr,Cassani:2020cod}. The $\mathrm{U}(1)_+$ flavour-current multiplet at generic $z$ was also obtained in \cite{Cassani:2020cod}. More generally, we recently determined the $\mathrm{U}(1)_0$-invariant spectrum of MN1 in \cite{Bhattacharya:2026ypg}, which we recover in section~\ref{sec:RepresentativeMN1} by specialising our general-$z$ results to $z=0$. In \cite{Bhattacharya:2024tjw} we obtained the putative $|z|=1$, MN2 spectrum, and in section~\ref{sec:RepresentativeMN2} below we extract its globally defined $\mathrm{U}(1)_1$-invariant multiplets. All other multiplets reported for general $z$ in sections~\ref{sec:FlavourNeutralSpectrum}~and~\ref{sec:FlavourChargedSpectrum} are new.


\subsection{$\mathrm{U}(1)_z$-invariant flavour-neutral spectrum}
\label{sec:FlavourNeutralSpectrum}


Let us start by describing the two flavour-neutral contributions to
(\ref{eq:SpectrumFourfoldSplit}), namely the protected spectrum
$\cS_{\textrm{$F$-neutral}}$ and the long spectrum $\cL_{\textrm{$F$-neutral}} (z)$ at a fixed value of $z$. Since $F=0$, the SCP R-charge is $r=R_0$, independent of $z$ by (\ref{eq:SpectrumExactRCharge}),  and does not
mix with flavour. The U$(1)_z$-invariance condition (\ref{eq:Q_zCharge}) at $F=0$ is also $z$-independent. Consequently, the allowed charges and the number of multiplets are the same at every BBBW background. Only the dimensions of the long multiplets vary with $z$.

\subsubsection{Protected spectrum}
\label{sec:FlavourNeutralProtected}

%
%

The protected flavour-neutral contribution $\cS_{\textrm{$F$-neutral}}$ to (\ref{eq:SpectrumFourfoldSplit}) is independent of $z$. Accordingly, these multiplets are present for all BBBW backgrounds with $z$-independent dimensions and SCP R-charges related to the labels $\eta$ in (\ref{eq:etaDef}). The set $\cS_{\textrm{$F$-neutral}}$ is composed of a handful of isolated multiplets and, for this reason, it is convenient to characterise it by KK level (\ref{eq:GeneralKKLevel}) with $\alpha=\beta=0$, rather than using the generic labels $(\kappa, \nu, \eta,a)$.  At level $k=0$, the short flavour-neutral spectrum contains, as expected, the stress-tensor multiplet, $\AOneAOne[3;\half,\half;0]_0$, and the U$(1)_+$ flavour-current multiplet, $\ATwoATwo[2;0,0;0]_0$.  At level $k=1$ there is one CPT pair of short gravitino multiplets. This content is summarised in table \ref{tab:FirstZ-neutral-short}. There are no further protected flavour-neutral
multiplets at higher levels $k \geq 2$.
\begin{table}[H]
\centering
\caption{Protected flavour-neutral spectrum
$\cS_{\textrm{$F$-neutral}}$. 
}
\label{tab:FirstZ-neutral-short}
\small\renewcommand{\arraystretch}{1.25}\setlength{\tabcolsep}{5pt}
\begin{tabular}{@{}l|l| l@{}}
\hline
Type & Multiplet  & Level \\
\hline
\hline
Graviton
& $\AOneAOne[3;\half,\half;0]_0$ & $k=0$ \\[3pt]
\hline
Gravitino
& $\LAOne[\tfrac92;0,\half;1]_0\oplus\AOneL[\tfrac92;\half,0;-1]_0$
& $k=1$ \\[3pt]
\hline
Vector
& $\ATwoATwo[2;0,0;0]_0$ & $k=0$ \\
\hline
\end{tabular}
\end{table}
\noindent At the MN1 point, $z=0$, U$(1)_+$ flavour-neutral and charged multiplets
organise into representations of the enhanced $\mathrm{SU}(2)_+$
flavour symmetry, while the stress-tensor multiplet remains an $\mathrm{SU}(2)_+$ singlet. At the MN2 endpoints, $|z|=1$, they instead assemble into
multiplets of the enhanced $\mathcal N=2$ superconformal algebra;
in particular, the $\mathcal N=1$ flavour-current multiplet
becomes part of the $\mathcal N=2$ stress-tensor multiplet.
Further details are given in sections~\ref{sec:RepresentativeMN1} and~\ref{sec:RepresentativeMN2}.

\subsubsection{Long spectrum}
\label{sec:FlavourNeutralLong}

The long, flavour-neutral contribution $\cL_{\textrm{$F$-neutral}}(z)$ to the $\mathrm{U}(1)_z$-invariant spectrum (\ref{eq:SpectrumFourfoldSplit}) at fixed $z$ is summarised in table~\ref{tab:FirstZ-neutral-long}. The multiplicities and SCP R-charges of the multiplets contained in $\cL_{\textrm{$F$-neutral}}(z)$ are independent of $z$, whereas the dimensions do generally vary with $z$. This dependence enters exclusively through the square of the R-symmetry mixing parameter $\epsilon=\epsilon(z)$ in (\ref{eq:SpectrumExactRCharge}). With the index ranges given in the table, all multiplets remain long for all $z$: their dimensions lie strictly above the corresponding unitarity bounds, and none undergoes shortening.
\begin{table}[H]
\centering
\caption{Long flavour-neutral spectrum
$\cL_{\textrm{$F$-neutral}}(z)$.}
\label{tab:FirstZ-neutral-long}
\fontsize{9}{11}\selectfont
\renewcommand{\arraystretch}{1.3}
\setlength{\tabcolsep}{3pt}
\resizebox{\linewidth}{!}{%
\begin{tabular}{@{}l|l|l|l@{}}
\hline
Type & Multiplet & Label ranges & Level \\
\hline
\hline
Graviton
& $\LL[1+\sqrt{4+g_{\kappa a, z}};\half,\half;0]_0$
& \makecell[l]{$\kappa\geq1$\\$a=1,\ldots,\check{\kappa}+1$}
& $k=\kappa$ \\[4pt]
\hline
Gravitino
& $\LL[1+\sqrt{\tfrac14+h_{\kappa a, z}};0,\half;\sigma_{\kappa a}]_0\oplus\LL[1+\sqrt{\tfrac14+h_{\kappa a, z}};\half,0;-\sigma_{\kappa a}]_0$
& \makecell[l]{$\kappa\geq2$\\$a=1,\ldots,2\hat{\kappa}-1$}
& $k=\kappa$ \\[4pt]
\hline
Vector
& $\LL[1+\sqrt{7+18\epsilon^2};0,0;0]_0$
& \makecell[l]{$\kappa=2$\\$a=1$}
& $k=\kappa-2=0$ \\[4pt]
\hline
Vector
& $\LL[1+\sqrt{1+p_{\kappa a, z}};0,0;0]_0$
& \makecell[l]{$\kappa\geq3$\\$a=1,\ldots,2\check{\kappa}$}
& $k=\kappa-2$ \\[4pt]
\hline
Vector
& $\LL[1+\sqrt{1+q_{\kappa a, z}};0,0;2]_0\oplus\LL[1+\sqrt{1+q_{\kappa a, z}};0,0;-2]_0$
& \makecell[l]{$\kappa\geq4$\\$a=1,\ldots,\check{\kappa}-1$}
& $k=\kappa$ \\[4pt]
\hline
\end{tabular}%
}
\end{table}

This sector contains four infinite families, labelled by $\kappa$ in (\ref{eq:kappanuRange}), with lower limits $\kappa_{\textrm{min}}$ specified in table~\ref{tab:FirstZ-neutral-long}. Two families consist of CPT-self-conjugate graviton and vector multiplets, respectively, while the other two contain CPT-conjugate pairs of gravitino and vector multiplets. At fixed $\kappa$, the KK level is given by (\ref{eq:GeneralKKLevel}) with $\alpha=1$, $\beta=0$ and a family-dependent value of $\gamma$. The SCP dimensions are governed by the polynomial families $G_\kappa(x;z)$, $H_\kappa(x;z)$, $P_\kappa(x;z)$ and $Q_\kappa(x;z)$ constructed in appendix~\ref{app:NeutralPolynomials}. These are polynomials in the auxiliary variable $x$, whose coefficients depend on $z$ exclusively through polynomial expressions in the square, $\epsilon(z)^2$, of the R-symmetry mixing parameter (\ref{eq:SpectrumExactRCharge}). Their roots are denoted by $g_{\kappa a,z}$, $h_{\kappa a,z}$, $p_{\kappa a,z}$ and $q_{\kappa a,z}$, with the index $a$ used separately for each family. The roots, and hence the SCP dimensions, are generally complicated functions of $\epsilon(z)^2$, determined implicitly by relations such as
\begin{equation}
G_\kappa(g_{\kappa a,z};z)=0
\; ,\qquad a=1,\ldots,\check{\kappa}+1
\; ,\qquad \kappa=1,2,\ldots
\; ,
\end{equation}
and similarly for $h_{\kappa a,z}$, $p_{\kappa a,z}$ and $q_{\kappa a,z}$.

\begin{figure}
\centering
\captionsetup[subfigure]{position=top}
    \begin{subfigure}[t]{.48\textwidth}
        \centering
        \caption{$\kappa=0$}
        \label{fig:BBBWGlobalSpectralFlowKappa0}
        \includegraphics[width=\linewidth]{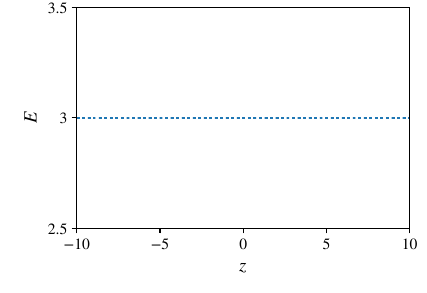}
    \end{subfigure}%
    \hfill
    \begin{subfigure}[t]{.48\textwidth}
        \centering
        \caption{$\kappa=1$}
        \label{fig:BBBWGlobalSpectralFlowKappa1}
        \includegraphics[width=\linewidth]{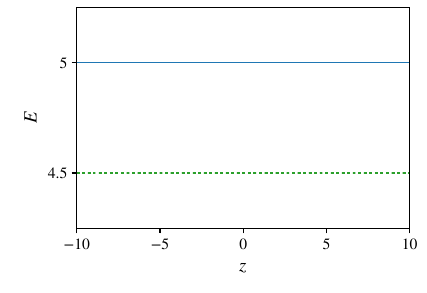}
    \end{subfigure}\\[8pt]
    \begin{subfigure}[t]{.48\textwidth}
        \centering
        \caption{$\kappa=2$}
        \label{fig:BBBWGlobalSpectralFlowKappa2}
        \includegraphics[width=\linewidth]{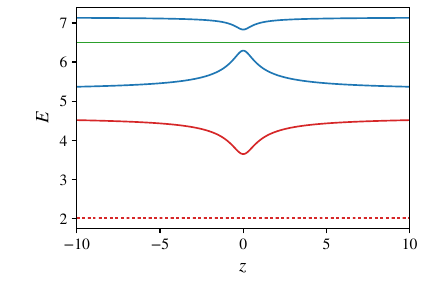}
    \end{subfigure}%
    \hfill
    \begin{subfigure}[t]{.48\textwidth}
        \centering
        \caption{$\kappa=3$}
        \label{fig:BBBWGlobalSpectralFlowKappa3}
        \includegraphics[width=\linewidth]{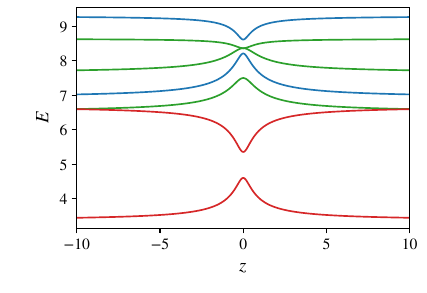}
    \end{subfigure}\\[8pt]
    \begin{subfigure}[t]{.48\textwidth}
        \centering
        \caption{$\kappa=4$}
        \label{fig:BBBWGlobalSpectralFlowKappa4}
        \includegraphics[width=\linewidth]{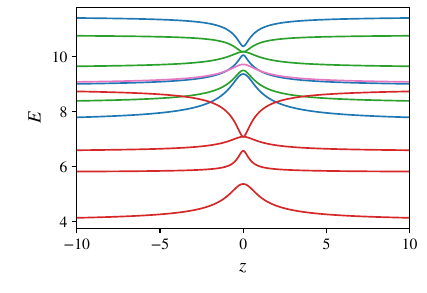}
    \end{subfigure}%
    \hfill
    \begin{subfigure}[t]{.48\textwidth}
        \centering
        \caption{$\kappa=5$}
        \label{fig:BBBWGlobalSpectralFlowKappa5}
        \includegraphics[width=\linewidth]{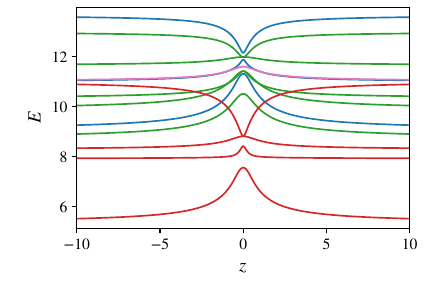}
    \end{subfigure}
    \caption{Dimensions $E$ as functions of $z$, with $-10\leq z\leq10$, of the flavour-neutral short (dotted lines) and long (solid lines) multiplets listed in tables~\ref{tab:FirstZ-neutral-short} and~\ref{tab:FirstZ-neutral-long}, for $\kappa=0, \ldots , 5$. Graviton and gravitino dimensions, governed by $G_\kappa$, $H_\kappa$, are blue and green, respectively. Vector multiplets governed by $P_\kappa$ and $Q_\kappa$ are red and pink. The protected stress-tensor, gravitino and flavour-current multiplets occur at $\kappa=0,1,2$, respectively,~{\it i.e.}~KK levels $k=0,1,0$. Coincident curves represent multiplets with equal dimensions.}
    \label{fig:BBBWF0SpecKappa}
\end{figure}

The finite range of $a$ specified in the table depends on $\kappa$ through the integers (\ref{eq:NeutralParityDefinitions}) and, with one exception, spans all roots allowed by the corresponding polynomial degree (\ref{eq:xdeg}). The exception occurs in the $\eta=0$ vector family at $\kappa=2$, for which only $a=1$ is retained. Although listed separately in table~\ref{tab:FirstZ-neutral-long}, this multiplet belongs to the family governed by the polynomial family $P_\kappa(x;z)$ constructed in appendix~\ref{app:NeutralMatrixVectorP}. Of the two roots of $P_2(x;z)$ in (\ref{eq:NeutralPFirstMembers}), $p_{21,z}=6+18\epsilon^2$ determines this long vector multiplet, while $p_{22,z}=0$ describes the protected $\mathrm{U}(1)_+$ flavour-current multiplet listed in table~\ref{tab:FirstZ-neutral-short}. For $\kappa\geq3$, all roots of $P_\kappa(x;z)$ correspond to long vector multiplets. A similar restriction applies to the graviton family: although $G_\kappa(x;z)$ is defined in appendix~\ref{app:NeutralMatrixGraviton} for all $\kappa\geq0$, its $\kappa=0$ member, $G_0(x;z)$ in (\ref{eq:NeutralGFirstMembers}), describes the protected stress-tensor multiplet in table~\ref{tab:FirstZ-neutral-short}. Likewise, the root of $H_1(x;z)$ in (\ref{eq:NeutralHFirstMembers}) gives the short gravitino pair of table~\ref{tab:FirstZ-neutral-short}. These exclusions illustrate the label restrictions discussed in section~\ref{sec:SpectrumGeneralStructure}.

Interestingly, the polynomials $G_1(x;z)$ in (\ref{eq:NeutralGFirstMembers}) and $H_2(x;z)$ in (\ref{eq:NeutralHFirstMembers}) have $\epsilon$-\-in\-de\-pend\-ent coefficients. Their roots, $g_{11,z}=12$ and $h_{21,z}=30$, therefore give long graviton and gravitino multiplets with dimensions that remain constant for all $z$:
\begin{equation}
\LL[5;\tfrac12,\tfrac12;0]_0
\; ,\qquad
\LL[\tfrac{13}{2};0,\half;1]_0
\oplus
\LL[\tfrac{13}{2};\half,0;-1]_0
\; .
\end{equation}
The graviton multiplet has $\kappa=1$, $a=1$ and occurs at KK level $k=1$, while the gravitino pair has $\kappa=2$, $a=1$ and occurs at $k=2$, as indicated in table~\ref{tab:FirstZ-neutral-long}. All other multiplets in that table have SCP dimensions that depend on $\epsilon^2$. In contrast, the SCP R-charges remain $z$-independent for all multiplets in this sector. The CPT-self-conjugate graviton and vector multiplets have R-neutral SCPs, with $\eta=0$, while the vector CPT pairs have SCP R-charges $\eta= \pm2$. Each gravitino CPT pair has SCP R-charges $\eta=\pm1$. In table~\ref{tab:FirstZ-neutral-long}, we write these as $\sigma_{\kappa a}$ and $-\sigma_{\kappa a}$, where $\sigma_{\kappa a} = \pm1$ is the R-charge of the SCP with Lorentz spins $(0,\frac12)$. Its sign depends on the root $h_{\kappa a,z}$ of $H_\kappa(x;z)$, as discussed in appendix~\ref{app:NeutralMatrixGravitino}. At fixed $\kappa\geq2$, the sequence $\sigma_{\kappa a}$, with $a=1,\ldots,2\hat{\kappa}-1$, always contains exactly $\hat{\kappa}$ entries equal to $+1$ and $\hat{\kappa}-1$ entries equal to $-1$.

As already emphasised, the coefficients of all four polynomial families are polynomials in $\epsilon^2$ with rational numerical coefficients. Together with $\epsilon(-z)=-\epsilon(z)$ from (\ref{eq:SpectrumExactRCharge}), this establishes the invariance of $\cL_{\textrm{$F$-neutral}}(z)$ under $z\mapsto-z$ stated in section~\ref{sec:SpectrumGeneralStructure}. Figure~\ref{fig:BBBWF0SpecKappa} displays the SCP dimensions $E$ in the full U$(1)_z$-invariant flavour-neutral spectrum, $\cS_{\textrm{$F$-neutral}}\oplus\cL_{\textrm{$F$-neutral}}(z)$, as functions of $z$ for the first few values of $\kappa$. The plots exhibit this reflection symmetry, together with stationary behaviour at the MN1 point, $z=0$.


\subsection{$\mathrm{U}(1)_z$-invariant flavour-charged spectrum}
\label{sec:FlavourChargedSpectrum}


We now turn to the last two contributions to (\ref{eq:SpectrumFourfoldSplit}): the protected, $\cS_{\textrm{$F$-charged}}(z)$, and long, $\cL_{\textrm{$F$-charged}}(z)$, sectors of the $\mathrm{U}(1)_z$-invariant, flavour-charged BBBW spectrum. All multiplets in these sectors occur in CPT-conjugate pairs. Their SCP dimensions generally vary with $z$, as in the long flavour-neutral sector. Here, however, this dependence enters through both $\epsilon(z)$ and the flavour and SCP R-charges, which themselves depend on the twist parameter:
\begin{equation}
\label{eq:ProtectedMasterCharge}
F_{\nu,z} =- s \nu f_z
\; (z\neq0) \; \textrm{with $s \equiv \operatorname{sgn}(z)$},\qquad
r_{\nu \eta ,z} = \rho_{\nu,z} + \eta, \; \textrm{with $\rho_{\nu,z}=\nu\big(\lambda_z+|\epsilon(z)|f_z\big)$},
\end{equation}
with $F_{\nu,0}= 2\nu$, $\rho_{\nu,0} =0$ at MN1. The charges thus inherit their $z$-dependence from the twisting data $f_z$ and $\lambda_z$ defined in (\ref{eq:SpectrumTwistData2}) and the mixing parameter $\epsilon(z)$ in (\ref{eq:SpectrumExactRCharge}). Both charges (\ref{eq:ProtectedMasterCharge}) are labelled by the integer $\nu$ in (\ref{eq:kappanuRange}), while the SCP R-charge also involves the constant offset $\eta$ introduced in (\ref{eq:etaDef}). Finally, the multiplet content itself varies with $z$, with distinct contributions in the interior and exterior regions (\ref{eq:zRegions}) and at the MN1 and MN2 points. We therefore organise the discussion of the protected and long flavour-charged spectra in sections~\ref{sec:FlavourChargedProtected} and~\ref{sec:FlavourChargedLong}, respectively, according to these regions and special values of $z$.

\subsubsection{Protected spectrum}
\label{sec:FlavourChargedProtected}

In $0\leq|z|<1$, isolated protected multiplets occur only at selected values of $z$. For $|z|\geq1$, the spectrum $\cS_{\textrm{$F$-charged}}(z)$ also contains infinite discrete towers. We describe all possible cases below, with the corresponding multiplet content summarised in tables \ref{tab:MN1ProtectedMultiplets}--\ref{tab:ExteriorProtectedMultiplets}.

\newpage

\paragraph{The MN1 point, $z=0$. }
\label{sec:FlavourChargedProtectedMN1}

The $\textrm{SU}(2,2|1) \times \textrm{U}(1)_+$ multiplet content of the protected flavour-charged spectrum over the MN1 solution is summarised in table \ref{tab:MN1ProtectedMultiplets}. There are only isolated multiplets which we choose, as in section \ref{sec:FlavourNeutralProtected}, to label by KK level (\ref{eq:GeneralKKLevel}) with $\alpha=\beta=0$ and, as it turns out, $\gamma=0$ or $\gamma=1$. These include a pair of vector multiplets at $k=0$, and two pairs of gravitino multiplets at $k=1$.
\begin{table}[H]
\centering
\caption{Protected flavour-charged spectrum
$\cS_{\textrm{$F$-charged}}(0)$ at MN1, $z=0$.}
\label{tab:MN1ProtectedMultiplets}
\small\renewcommand{\arraystretch}{1.25}\setlength{\tabcolsep}{5pt}
\begin{tabular}{@{}l|l|l@{}}
\hline
Type & Multiplet & Level \\
\hline
\hline
Gravitino
& $\LAOne[\tfrac92;0,\half;1]_{2}\oplus\AOneL[\tfrac92;\half,0;-1]_{-2}$
& $k=1$ \\[3pt]
\hline
Gravitino
& $\AOneL[\tfrac92;\half,0;-1]_{2}\oplus\LAOne[\tfrac92;0,\half;1]_{-2}$
& $k=1$ \\[3pt]
\hline
Vector
& $\ATwoATwo[2;0,0;0]_{2}\oplus\ATwoATwo[2;0,0;0]_{-2}$
& $k=0$ \\
\hline
\end{tabular}
\end{table}
\noindent Together with the U$(1)_+$ flavour current multiplet in table \ref{tab:FirstZ-neutral-short}, the vector multiplets in table \ref{tab:MN1ProtectedMultiplets} sit in a triplet of the enhanced SU$(2)_+$ flavour symmetry of MN1. The $k=1$ flavour-charged gravitino multiplets also form SU$(2)_+$ triplets with their flavour-neutral counterparts. See section \ref{sec:RepresentativeMN1} for further details.

\paragraph{The interior, $0<|z|<1$.}

In this regime, we find short multiplets at the distinguished values of $z$ defined in (\ref{eq:Distinguishedz}), and only at these values. As for MN1, the short spectrum is also isolated in these cases. It contains a pair of gravitino multiplets at KK level $k=j$ and a pair of vector multiplets at $k=j-1$ with the quantum numbers indicated in table~\ref{tab:InteriorProtectedMultiplets}. 
\begin{table}[H]
\centering
\caption{Protected flavour-charged spectrum
$\cS_{\textrm{$F$-charged}}(z)$ at interior values of $z$: $0<|z|<1$.}
\label{tab:InteriorProtectedMultiplets}
\small\renewcommand{\arraystretch}{1.25}\setlength{\tabcolsep}{5pt}
\begin{tabular}{@{}l|l|l@{}}
\hline
Type & Multiplet & Level \\
\hline
\hline
Gravitino
& $\LAOne[\tfrac92+\tfrac32\rho_j;0,\half;\rho_j+1]_{F_j}
\oplus\AOneL[\tfrac92+\tfrac32\rho_j;\half,0;-\rho_j-1]_{-F_j}$
& $k=j$ \\[3pt]
\hline
Vector
& $\LATwo[2+\tfrac32\rho_j;0,0;\rho_j]_{F_j}
\oplus\ATwoL[2+\tfrac32\rho_j;0,0;-\rho_j]_{-F_j}$
& $k=j-1$ \\
\hline
\end{tabular}
\end{table}
\noindent In the table, we have defined
\begin{equation}
 F_j=- \operatorname{sgn}(z) \, (j+1),
 \qquad
 \rho_j=j-1+(j+1)|\epsilon(z)|
       =j-1+\frac{j^2-1}{j+1+2\sqrt{j^2-j+1}} .
\end{equation}
These expressions follow from (\ref{eq:ProtectedMasterCharge}) with $\nu=1$ and $z$ evaluated at its distinguished interior values in (\ref{eq:Distinguishedz}), for $j=2,3, \ldots$. 

\paragraph{The MN2 points, $|z|=1$.}

Table~\ref{tab:MN2ProtectedMultiplets} lists the $\mathrm{U}(1)_1$-invariant flavour-charged short multiplets of $\mathrm{SU}(2,2|1)\times\mathrm{U}(1)_+$ present in the MN2 spectrum. As in the range $0\leq|z|<1$, the protected content at the MN2 points includes isolated multiplets, in this case a pair of gravitino multiplets at KK level $k=0$. The main difference from the preceding cases is the appearance of infinite towers, which now account for most of the protected spectrum. These towers comprise pairs of graviton, gravitino and vector multiplets, as well as chiral multiplets, labelled with the integer $\nu$ introduced in (\ref{eq:kappanuRange}) with the lower limit indicated in the table for each multiplet. The towers arise at KK levels (\ref{eq:GeneralKKLevel}) with $\mu_z=1$ for MN2 and $\alpha=0$, $\beta=1$ and various family-dependent values of $\gamma$, also indicated in the table. 

\begin{table}
\centering
\caption{Protected flavour-charged spectrum
$\cS_{\textrm{$F$-charged}}(s)$ at MN2, 
with $s\equiv\operatorname{sgn}(z)=\pm1$.}
\label{tab:MN2ProtectedMultiplets}
\fontsize{9}{11}\selectfont
\renewcommand{\arraystretch}{1.3}
\setlength{\tabcolsep}{3pt}
\resizebox{\linewidth}{!}{%
\begin{tabular}{@{}l|l|l|l@{}}
\hline
Type & Multiplet & Label ranges & Level \\
\hline
\hline
Graviton
& $\LAOne[2\nu+3;\half,\half;\tfrac{4\nu}{3}]_{-s\nu}\oplus\AOneL[2\nu+3;\half,\half;-\tfrac{4\nu}{3}]_{s\nu}$
& $\nu\geq1$
& $k=\nu$ \\[4pt]
\hline
Gravitino
& $A_1\bar A_2[\tfrac52;\half,0;\tfrac13]_{-s}\oplus A_2\bar A_1[\tfrac52;0,\half;-\tfrac13]_{s}$
& $-$
& $k=0$ \\[4pt]
\hline
Gravitino
& $\LATwo[2\nu+\tfrac12;\half,0;\tfrac{4\nu-3}{3}]_{-s\nu}\oplus\ATwoL[2\nu+\tfrac12;0,\half;-\tfrac{4\nu-3}{3}]_{s\nu}$
& $\nu\geq2$
& $k=\nu-1$ \\[4pt]
\hline
Gravitino
& $\LAOne[2\nu+\tfrac92;0,\half;\tfrac{4\nu+3}{3}]_{-s\nu}\oplus\AOneL[2\nu+\tfrac92;\half,0;-\tfrac{4\nu+3}{3}]_{s\nu}$
& $\nu\geq1$
& $k=\nu+1$ \\[4pt]
\hline
Vector
& $\LATwo[2\nu+2;0,0;\tfrac{4\nu}{3}]_{-s\nu}\oplus\ATwoL[2\nu+2;0,0;-\tfrac{4\nu}{3}]_{s\nu}$
& $\nu\geq1$
& $k=\nu$ \\[4pt]
\hline
Vector
& $\LATwo[2\nu+2;0,0;\tfrac{4\nu}{3}]_{-s\nu}\oplus\ATwoL[2\nu+2;0,0;-\tfrac{4\nu}{3}]_{s\nu}$
& $\nu\geq1$
& $k=\nu-1$ \\[4pt]
\hline
Chiral
& $\LBone[2\nu;0,0;\tfrac{4\nu}{3}]_{-s\nu}\oplus\BOneL[2\nu;0,0;-\tfrac{4\nu}{3}]_{s\nu}$
& $\nu\geq2$
& $k=\nu-2$ \\[4pt]
\hline
\end{tabular}%
}
\end{table}
\noindent At either MN2 endpoint, $f_z = \lambda_z =1$ and the definitions (\ref{eq:ProtectedMasterCharge}) give \(F_{\nu,z}=-s\nu\) and \(\rho_{\nu,z}=4\nu/3\). 
 Together with the flavour-neutral protected sector, these multiplets provide the \(\mathcal N=1\) constituents of the spectrum with enhanced \(\mathcal N=2\) supersymmetry discussed in section~\ref{sec:RepresentativeMN2}.

\paragraph{The exterior, $|z|>1$.}
There are no isolated flavour-charged short multiplets in this
region: all occur in infinite towers labelled by the integer $\nu$ in (\ref{eq:kappanuRange}), with $z$-dependent charges determined by
(\ref{eq:ProtectedMasterCharge}). As at MN2, there are graviton,
gravitino, vector and chiral towers, now accompanied by tensor
multiplets. Their content is summarised in
table~\ref{tab:ExteriorProtectedMultiplets}.

The integer $\nu$ ranges as in (\ref{eq:kappanuRange}) with $\nu_{\min} =1$ in all cases, except for the tower of vector multiplets with $\eta =\pm2$ and the tower of chiral multiplets, for which $\nu_{\min} =2$ if $z$ takes any of the distinguished exterior values defined in (\ref{eq:Distinguishedz}). For $|z| >1$, $\mu_z = \lambda_z$, and the multiplets in each tower occur at KK levels (\ref{eq:GeneralKKLevel}) with $\alpha=0$, $\beta=1$ and the family-dependent values of $\gamma$ indicated in the table. For example, $z=3$ has $\lambda_z=3$ and belongs to the
distinguished exterior sequence (\ref{eq:Distinguishedz}) with $j=2$.
Its short graviton tower occurs at KK levels $k=3,6,9,\ldots$, while
its chiral tower starts at $\nu_{\min} =2$ and occurs at levels
$k=2,5,8,\ldots$.
By contrast, $z=5$ has $\lambda_z=5$ and lies outside the distinguished sequence (\ref{eq:Distinguishedz}), so all towers start at $\nu_{\min} =1$.
The short graviton and chiral towers then occur at levels 
$k=5,10,15,\ldots$ and $k=1,6,11,\ldots$, respectively.

\begin{table}
\centering
\caption{Protected flavour-charged spectrum
$\cS_{\textrm{$F$-charged}}(z)$ at exterior values of $z$: $|z| >1$.}
\label{tab:ExteriorProtectedMultiplets}
\fontsize{9}{11}\selectfont
\renewcommand{\arraystretch}{1.3}
\setlength{\tabcolsep}{3pt}
\resizebox{\linewidth}{!}{%
\begin{tabular}{@{}l|l|l|l@{}}
\hline
Type & Multiplet & Label ranges & Level \\
\hline
\hline
Graviton
& $\LAOne[3+\tfrac32\rho_{\nu,z};\half,\half;\rho_{\nu,z}]_{F_{\nu,z}}\oplus\AOneL[3+\tfrac32\rho_{\nu,z};\half,\half;-\rho_{\nu,z}]_{-F_{\nu,z}}$
& $\nu\geq1$
& $k=\nu\lambda_z$ \\[4pt]
\hline
Gravitino
& $\LATwo[2+\tfrac32(\rho_{\nu,z}-1);\half,0;\rho_{\nu,z}-1]_{F_{\nu,z}}\oplus\ATwoL[2+\tfrac32(\rho_{\nu,z}-1);0,\half;-\rho_{\nu,z}+1]_{-F_{\nu,z}}$
& $\nu\geq1$
& $k=\nu\lambda_z-2$ \\[4pt]
\hline
Gravitino
& $\LATwo[2+\tfrac32(\rho_{\nu,z}-1);\half,0;\rho_{\nu,z}-1]_{F_{\nu,z}}\oplus\ATwoL[2+\tfrac32(\rho_{\nu,z}-1);0,\half;-\rho_{\nu,z}+1]_{-F_{\nu,z}}$
& $\nu\geq1$
& $k=\nu\lambda_z-1$ \\[4pt]
\hline
Gravitino
& $\LAOne[3+\tfrac32(\rho_{\nu,z}+1);0,\half;\rho_{\nu,z}+1]_{F_{\nu,z}}\oplus\AOneL[3+\tfrac32(\rho_{\nu,z}+1);\half,0;-\rho_{\nu,z}-1]_{-F_{\nu,z}}$
& $\nu\geq1$
& $k=\nu\lambda_z+1$ \\[4pt]
\hline
Vector
& $\LATwo[2+\tfrac32(\rho_{\nu,z}-2);0,0;\rho_{\nu,z}-2]_{F_{\nu,z}}\oplus\ATwoL[2+\tfrac32(\rho_{\nu,z}-2);0,0;-\rho_{\nu,z}+2]_{-F_{\nu,z}}$
& \makecell[l]{$\nu\geq2$: distinguished $z$\\
               $\nu\geq1$: otherwise}
& $k=\nu\lambda_z-3$ \\[4pt]
\hline
Vector
& $\LATwo[2+\tfrac32\rho_{\nu,z};0,0;\rho_{\nu,z}]_{F_{\nu,z}}\oplus\ATwoL[2+\tfrac32\rho_{\nu,z};0,0;-\rho_{\nu,z}]_{-F_{\nu,z}}$
& $\nu\geq1$
& $k=\nu\lambda_z$ \\[4pt]
\hline
Tensor
& $\LBone[\tfrac32(\rho_{\nu,z}-1);\half,0;\rho_{\nu,z}-1]_{F_{\nu,z}}\oplus\BOneL[\tfrac32(\rho_{\nu,z}-1);0,\half;-\rho_{\nu,z}+1]_{-F_{\nu,z}}$
& $\nu\geq1$
& $k=\nu\lambda_z-3$ \\[4pt]
\hline
Chiral
& $\LBone[\tfrac32(\rho_{\nu,z}-2);0,0;\rho_{\nu,z}-2]_{F_{\nu,z}}\oplus\BOneL[\tfrac32(\rho_{\nu,z}-2);0,0;-\rho_{\nu,z}+2]_{-F_{\nu,z}}$
& \makecell[l]{$\nu\geq2$: distinguished $z$\\
               $\nu\geq1$: otherwise}
& $k=\nu\lambda_z-4$ \\[4pt]
\hline
\end{tabular}%
}
\end{table}

Several exterior protected multiplets in table~\ref{tab:ExteriorProtectedMultiplets} formally reproduce MN2 multiplets in table~\ref{tab:MN2ProtectedMultiplets}. Setting $f_z=\lambda_z=1$, $\rho_{\nu,z}=4\nu/3$ and $F_{\nu,z}=-s\nu$, with $s=\operatorname{sgn}(z)$, in table~\ref{tab:ExteriorProtectedMultiplets} reproduces the graviton and first vector towers of table~\ref{tab:MN2ProtectedMultiplets} at $k=\nu$, together with its gravitino towers at $k=\nu-1,\nu+1$. At $\nu=1$, the gravitino at $k=\nu-1$ undergoes additional shortening to the isolated $A_1\bar A_2\oplus A_2\bar A_1$ pair at $k=0$. The exterior gravitino at $k=\nu\lambda_z-2$ gives the same formal MN2 representation as the branch at $k=\nu\lambda_z-1$, but occurs one KK level lower. It therefore constitutes an additional exterior occurrence and does not reproduce an MN2 multiplet at the same level. The tensor and chiral towers, as well as the $\eta = \pm 2$ vector tower have no counterparts in table~\ref{tab:MN2ProtectedMultiplets} with matching dimensions, spins and charges, even allowing a change of KK level. These are new protected families for $|z| >1$, with no immediate counterpart at MN2. 

\subsubsection{Long spectrum}
\label{sec:FlavourChargedLong}

We finally describe the long, flavour-charged contribution $\cL_{\textrm{$F$-charged}}(z)$ to the $\mathrm{U}(1)_z$-invariant spectrum (\ref{eq:SpectrumFourfoldSplit}). This sector contains infinite families of CPT pairs of graviton, gravitino and vector multiplets, typically labelled by the non-negative integers $(\kappa , \nu , a )$ introduced in section~\ref{sec:SpectrumGeneralStructure}, along with $\eta$ in~(\ref{eq:etaDef}). As in $\cS_{\textrm{$F$-charged}}(z)$, everything in $\cL_{\textrm{$F$-charged}}(z)$ depends on $z$: the SCP conformal dimensions, the flavour and R-charges (\ref{eq:ProtectedMasterCharge}), and the multiplet content itself. The latter differs between the interior and exterior regions (\ref{eq:zRegions}) and the MN1 and MN2 points. For this reason, the discussion below will be organised by these regions and special values of $z$, as in section~\ref{sec:FlavourChargedProtected}.

The SCP dimensions are determined by the roots $g_{\kappa\nu a,z}$, $h_{\kappa\nu\eta a,z}$, $p_{\kappa\nu a,z}$ and $q_{\kappa\nu\eta a,z}$ of the four polynomial families $G_{\kappa\nu}(x;z)$, $H_{\kappa\nu\eta}(x;z)$, $P_{\kappa\nu}(x;z)$ and $Q_{\kappa\nu\eta}(x;z)$ constructed in appendix~\ref{app:ChargedPolynomialsNew}. The first two govern graviton and gravitino multiplets, respectively, while the last two are associated with vector multiplets. These are polynomials in the auxiliary variable $x$, with degrees (\ref{eq:xdegCharged}) and coefficients that depend on $z$ through the mixing parameter $\epsilon(z)$ in (\ref{eq:SpectrumExactRCharge}) and the charges (\ref{eq:ProtectedMasterCharge}). Each family is labelled by the non-negative integers $\kappa$, $\nu$ in (\ref{eq:kappanuRange}) and the offset $\eta$ in (\ref{eq:etaDef}). For $G_{\kappa\nu}(x;z)$ and $P_{\kappa\nu}(x;z)$, the latter is fixed to $\eta=0$ and is suppressed in the notation. The roots are generally complicated functions of $z$, defined implicitly by their polynomial equations. For example,
\begin{equation}
\label{eq:FCRoots}
H_{\kappa\nu\eta}(h_{\kappa\nu\eta a,z};z)=0
\; ,
\end{equation}
and similarly for $g_{\kappa\nu a,z}$, $p_{\kappa\nu a,z}$ and $q_{\kappa\nu\eta a,z}$. To describe the long spectrum, the typically infinite ranges of $\kappa$ and $\nu$ may need to be restricted in the tables below relative to those used to define the polynomial families in appendix~\ref{app:ChargedPolynomialsNew}. Likewise, the finite range of $a$ need not include every root allowed by the corresponding degree (\ref{eq:xdegCharged}). These restrictions exclude roots that describe the short multiplets already reported in section~\ref{sec:FlavourChargedProtected}.

The charges (\ref{eq:ProtectedMasterCharge}), as well as the dimensions through (\ref{eq:FCRoots}), etc., inherit the discontinuous, non-monotonic dependence on $z$ from the twisting data (\ref{eq:SpectrumTwistData2}). This is illustrated in figure~\ref{fig:BBBWGravitonCloudInterior} below. A simpler behaviour nevertheless emerges in the large-charge limit, $\nu\rightarrow\infty$: ratios formed from the SCP dimensions $E_{\kappa\nu\eta a,z}$, flavour charges $F_{\nu ,z}$ and SCP R-charges $r_{\nu \eta ,z}$ approach smooth functions of $z$. These limiting ratios generally take different forms in the regions (\ref{eq:zRegions}), but are universal across all multiplet types and independent of the remaining labels $\kappa$, $\eta$, $a$. More explicitly, for any given multiplet in $\cL_{\textrm{$F$-charged}}(z)$, the ratio of the SCP dimension to the magnitude of the flavour charge tends to
\begin{equation}
\label{eq:ChargedLongLargeNuEF}
\frac{E_{\kappa\nu\eta a,z}}{|F_{\nu ,z}|}
\xrightarrow[\nu\to\infty]{}
\begin{cases}
\displaystyle \frac32\, , & z=0\, ,\\[4pt]
\displaystyle 1+\frac12\sqrt{1+3z^2}\, , & 0<|z|<1\, ,\\[4pt]
2\, , & |z|=1\, ,\\[4pt]
\displaystyle \frac32\bigl(|z|+|\epsilon(z)|\bigr)\, , & |z|>1\, .
\end{cases}
\end{equation}
For the ratio of the SCP dimension to the magnitude of its R-charge, we find
\begin{equation}
\label{eq:ChargedLongLargeNuER}
\frac{E_{\kappa\nu\eta a,z}}{|r_{\nu \eta ,z}|}
\xrightarrow[\nu\to\infty]{}
\begin{cases}
\displaystyle \frac{1}{2|\epsilon(z)|}
=\frac{1+\sqrt{1+3z^2}}{2|z|}\, , & 0<|z|<1\, ,\\[6pt]
\displaystyle \frac32\, , & |z|\geq1\, .
\end{cases}
\end{equation}
Thus, all long families have the same limiting ratio $3/2$ at MN2 and throughout the exterior. This common limit does not imply shortening: the difference between the SCP dimension and the corresponding unitarity bound need not tend to zero. Finally, the ratio of the flavour-charge magnitude to the SCP R-charge magnitude obeys
\begin{equation}
\label{eq:ChargedLongLargeNuFR}
\frac{|F_{\nu ,z}|}{|r_{\nu \eta ,z}|}
\xrightarrow[\nu\to\infty]{}
\frac{1}{|z|+|\epsilon(z)|}\, ,\qquad z\neq0\, .
\end{equation}
This expression applies in both the interior and exterior regions  (\ref{eq:zRegions}), and gives $3/4$ at MN2. At MN1, the SCP R-charges $r$ remain fixed at $\eta=0,\pm1,\pm2$, while the dimensions $E$ and flavour-charge magnitudes $|F|$ grow with $\nu$. Consequently, $E/|r|$ and $|F|/|r|$ are undefined for R-neutral SCPs and diverge for those with nonzero R-charge.


\paragraph{The MN1 point, $z=0$.}

The spectrum $\cL_{\textrm{$F$-charged}}(0)$ is summarised in table~\ref{tab:NewChargedMN1Long}. At MN1, all multiplets carry flavour charges $F=\pm2\nu$, with $\nu\geq1$ as in (\ref{eq:kappanuRange}). Their SCP R-charges are independent of the labels  $(\kappa , \nu , a )$: as for $\cL_{\textrm{$F$-neutral}}(z)$, they are $\eta = 0$ for the graviton family and some vector families, $\eta = \pm1$ for the gravitino multiplets, and $\eta = \pm2$ for the remaining vector families. The SCP conformal dimensions are still governed by the polynomials constructed in appendix~\ref{app:ChargedPolynomialsNew}. At this point, however, they admit the following simpler expressions in terms of the KK level $k$, (\ref{eq:GeneralKKLevel}) with $\mu_z=2$, and a label $\ell$ assigned to each CPT pair:
\begin{equation}
\label{eq:MN1SCPDimensions}
\begin{aligned}
E^{(\mathrm{grav})}_{k\ell}
 &=1+\sqrt{4+3k(k+3)-3\ell(\ell+1)},
&\quad
E^{(\mathrm{gino})}_{k\ell}
 &=1+\sqrt{\tfrac{25}{4}+3k(k+3)-3\ell(\ell+1)},\\[4pt]
E^{(\mathrm{vec},0)}_{k\ell}
 &=1+\sqrt{7+3k(k+3)-3\ell(\ell+1)},
&
E^{(\mathrm{vec},2)}_{k\ell}
 &=1+\sqrt{10+3k(k+3)-3\ell(\ell+1)}.
\end{aligned}
\end{equation}
Here $\ell$ is specified in terms of $\nu$ and an independent integer $b$, which replaces the generic polynomial-root index $a$. The finite range of $b$ is determined by $\kappa$ through the quantities (\ref{eq:NeutralParityDefinitions}), and $\ell$ is the spin of the SU$(2)_+$ flavour group of MN1, see section~\ref{sec:RepresentativeMN1}.

\begin{table}
\centering
\caption{Long flavour-charged spectrum
$\cL_{\textrm{$F$-charged}}(0)$ at MN1, $z=0$. }
\label{tab:NewChargedMN1Long}
\fontsize{9}{11}\selectfont
\renewcommand{\arraystretch}{1.3}
\setlength{\tabcolsep}{3pt}
\resizebox{\linewidth}{!}{%
\begin{tabular}{@{}l|l|l|l@{}}
\hline
Type & Multiplet & Label ranges & Level and spin \\
\hline
\hline
Graviton
& $\LL[E^{(\mathrm{grav})}_{k\ell};\half,\half;0]_{2\nu}\oplus\LL[E^{(\mathrm{grav})}_{k\ell};\half,\half;0]_{-2\nu}$
& \makecell[l]{$\nu\geq1,\quad\kappa\geq0$\\
               $b=0,\ldots,\check{\kappa}$}
& \makecell[l]{$k=2\nu+\kappa$\\
               $\ell=\nu+b$} \\[4pt]
\hline
Gravitino
& $\LL[E^{(\mathrm{gino})}_{k\ell};0,\half;1]_{2\nu}\oplus\LL[E^{(\mathrm{gino})}_{k\ell};\half,0;-1]_{-2\nu}$
& \makecell[l]{$\nu\geq1,\quad\kappa\geq1$\\
               $(\nu,\kappa)\neq(1,1)$\\
               $b=0,\ldots,\hat{\kappa}-1$}
& \makecell[l]{$k=2\nu+\kappa-2$\\
               $\ell=\nu+b$} \\[4pt]
\hline
Gravitino
& $\LL[E^{(\mathrm{gino})}_{k\ell};0,\half;1]_{-2\nu}\oplus\LL[E^{(\mathrm{gino})}_{k\ell};\half,0;-1]_{2\nu}$
& \makecell[l]{$\nu\geq1,\quad\kappa\geq1$\\
               $(\nu,\kappa)\neq(1,1)$\\
               $b=0,\ldots,\hat{\kappa}-1$}
& \makecell[l]{$k=2\nu+\kappa-2$\\
               $\ell=\nu+b$} \\[4pt]
\hline
Gravitino
& $\LL[E^{(\mathrm{gino})}_{k\ell};0,\half;-1]_{2\nu}\oplus\LL[E^{(\mathrm{gino})}_{k\ell};\half,0;1]_{-2\nu}$
& \makecell[l]{$\nu\geq1,\quad\kappa\geq3$\\
               $b=0,\ldots,\hat{\kappa}-2$}
& \makecell[l]{$k=2\nu+\kappa-2$\\
               $\ell=\nu+b$} \\[4pt]
\hline
Gravitino
& $\LL[E^{(\mathrm{gino})}_{k\ell};0,\half;-1]_{-2\nu}\oplus\LL[E^{(\mathrm{gino})}_{k\ell};\half,0;1]_{2\nu}$
& \makecell[l]{$\nu\geq1,\quad\kappa\geq3$\\
               $b=0,\ldots,\hat{\kappa}-2$}
& \makecell[l]{$k=2\nu+\kappa-2$\\
               $\ell=\nu+b$} \\[4pt]
\hline
Vector
& $d_{\kappa b}\bigl(\LL[E^{(\mathrm{vec},0)}_{k\ell};0,0;0]_{2\nu}\oplus\LL[E^{(\mathrm{vec},0)}_{k\ell};0,0;0]_{-2\nu}\bigr)$
& \makecell[l]{$\nu\geq1,\quad\kappa\geq0$\\
               $(\nu,\kappa)\neq(1,0)$\\
               $b=0,\ldots,\check{\kappa}$}
& \makecell[l]{$k=2\nu+\kappa-2$\\
               $\ell=\nu+b$} \\[4pt]
\hline
Vector
& $\LL[E^{(\mathrm{vec},2)}_{k\ell};0,0;2]_{2}\oplus\LL[E^{(\mathrm{vec},2)}_{k\ell};0,0;-2]_{-2}$
& \makecell[l]{$\nu=1,\quad\kappa\geq4$\\
               $b=0,\ldots,\check{\kappa}-2$}
& \makecell[l]{$k=\kappa$\\
               $\ell=2+b$} \\[4pt]
\hline
Vector
& $\LL[E^{(\mathrm{vec},2)}_{k\ell};0,0;2]_{2\nu}\oplus\LL[E^{(\mathrm{vec},2)}_{k\ell};0,0;-2]_{-2\nu}$
& \makecell[l]{$\nu\geq2,\quad\kappa\geq4$\\
               $b=0,\ldots,\check{\kappa}-2$}
& \makecell[l]{$k=2\nu+\kappa-4$\\
               $\ell=\nu+b$} \\[4pt]
\hline
Vector
& $\LL[E^{(\mathrm{vec},2)}_{k\ell};0,0;-2]_{2}\oplus\LL[E^{(\mathrm{vec},2)}_{k\ell};0,0;2]_{-2}$
& \makecell[l]{$\nu=1,\quad\kappa\geq4$\\
               $b=0,\ldots,\check{\kappa}-2$}
& \makecell[l]{$k=\kappa$\\
               $\ell=2+b$} \\[4pt]
\hline
Vector
& $\LL[E^{(\mathrm{vec},2)}_{k\ell};0,0;-2]_{2\nu}\oplus\LL[E^{(\mathrm{vec},2)}_{k\ell};0,0;2]_{-2\nu}$
& \makecell[l]{$\nu\geq2,\quad\kappa\geq4$\\
               $b=0,\ldots,\check{\kappa}-2$}
& \makecell[l]{$k=2\nu+\kappa-4$\\
               $\ell=\nu+b$} \\[4pt]
\hline
\end{tabular}%
}
\end{table}

All CPT pairs have multiplicity one, except those in one vector family. The multiplicity of each CPT pair in this family at fixed $\kappa$ and $b=0,\ldots,\check{\kappa}$ is
\begin{equation}
\label{eq:MN1VectorMultiplicity}
d_{\kappa b}=2-\delta_{b,\check{\kappa}-1}-\delta_{b,\check{\kappa}} \; .
\end{equation}
The departure from the usual unit multiplicity arises from repeated roots of the corresponding polynomials $P_{\kappa\nu}(x;0)$. The label ranges also exclude $(\nu,\kappa)=(1,0)$ for this vector family and $(\nu,\kappa)=(1,1)$ for certain gravitino families, as indicated in the table. At these particular combinations of labels, the corresponding multiplets shorten and are already included in the protected spectrum of table~\ref{tab:MN1ProtectedMultiplets}. With these exclusions, all multiplets listed in table~\ref{tab:NewChargedMN1Long} are strictly long.

\paragraph{The interior, $0<|z|<1$.}

Table~\ref{tab:NewChargedInteriorLong} lists the multiplets present in $\cL_{\textrm{$F$-charged}}(z)$ for interior $z$. The table contains infinite families of graviton, gravitino and vector multiplets labelled with $(\kappa , \nu , a )$. As remarked at the beginning of this subsection, the SCP dimensions are governed by the roots, $g_{\kappa\nu a, z}$, $h_{\kappa\nu\eta a, z}$, $p_{\kappa\nu a, z}$ and $q_{\kappa\nu\eta a, z}$, of the four polynomial families $G_{\kappa\nu}(x;z)$, $H_{\kappa\nu\eta}(x;z)$, $P_{\kappa\nu}(x;z)$ and $Q_{\kappa\nu\eta}(x;z)$ constructed in appendix~\ref{app:ChargedPolynomialsNew}. The flavour and SCP R-charges (\ref{eq:ProtectedMasterCharge}) are also $z$-dependent. For the gravitino multiplet with SCP Lorentz spins $(0 , \frac12)$, we have defined the R-charge shorthands
\begin{eqnarray} \label{eq:GinoRCharge}
\sigma_{\kappa \nu \eta a} = s_{\kappa \nu \eta a} (\rho_{\nu , z} + \eta) \; , \qquad
s_{\kappa \nu \eta a} = \pm 1 \; ,
\end{eqnarray}
with the sequence $s_{\kappa \nu \eta a}$ of $+$ or $-$ signs specified for each root $h_{\kappa\nu\eta a, z}$ in appendix~\ref{app:NewChargedGravitino}.

\begin{table}
\centering
\caption{Long flavour-charged spectrum
$\cL_{\textrm{$F$-charged}}(z)$ at interior values of $z$: $0 < |z| <1$.}
\label{tab:NewChargedInteriorLong}
\fontsize{9}{11}\selectfont
\renewcommand{\arraystretch}{1.3}
\setlength{\tabcolsep}{3pt}
\resizebox{\linewidth}{!}{%
\begin{tabular}{@{}l|l|l|l@{}}
\hline
Type & Multiplet & Label ranges & Level \\
\hline
\hline
Graviton
& $\LL[1+\sqrt{4+g_{\kappa\nu a, z}};\half,\half;\rho_{\nu,z}]_{F_{\nu,z}}
   \oplus
   \LL[1+\sqrt{4+g_{\kappa\nu a, z}};\half,\half;-\rho_{\nu,z}]_{-F_{\nu,z}}$
& \makecell[l]{$\nu\geq1,\quad\kappa\geq0$\\
               $a=1,\ldots,\check{\kappa}+1$}
& $k=\nu f_z+\kappa$ \\[4pt]
\hline
Gravitino
& $\LL[1+\sqrt{\tfrac14+h_{\kappa\nu\eta a, z}};0,\half;\sigma_{\kappa\nu\eta a}]_{s_{\kappa\nu\eta a}F_{\nu,z}}
   \oplus
   \LL[1+\sqrt{\tfrac14+h_{\kappa\nu\eta a, z}};\half,0;-\sigma_{\kappa\nu\eta a}]_{-s_{\kappa\nu\eta a}F_{\nu,z}}$
& \makecell[l]{$\eta=-1,\quad\nu\geq1$\\
               $\kappa\geq1$\\
               $a=1,\ldots,2\hat{\kappa}-1$}
& $k=\nu f_z+\kappa-2$ \\[4pt]
\hline
Gravitino
& $\LL[1+\sqrt{\tfrac14+h_{\kappa\nu\eta a, z}};0,\half;\sigma_{\kappa\nu\eta a}]_{s_{\kappa\nu\eta a}F_{1,z}}
   \oplus
   \LL[1+\sqrt{\tfrac14+h_{\kappa\nu\eta a, z}};\half,0;-\sigma_{\kappa\nu\eta a}]_{-s_{\kappa\nu\eta a}F_{1,z}}$
& \makecell[l]{distinguished $z$\\
               $\eta=1,\quad\nu=1$\\
               $\kappa\geq2$\\
               $a=1,\ldots,2\hat{\kappa}-1$}
& $k=f_z+\kappa-2$ \\[4pt]
\hline
Gravitino
& $\LL[1+\sqrt{\tfrac14+h_{\kappa\nu\eta a, z}};0,\half;\sigma_{\kappa\nu\eta a}]_{s_{\kappa\nu\eta a}F_{1,z}}
   \oplus
   \LL[1+\sqrt{\tfrac14+h_{\kappa\nu\eta a, z}};\half,0;-\sigma_{\kappa\nu\eta a}]_{-s_{\kappa\nu\eta a}F_{1,z}}$
& \makecell[l]{other interior $z$\\
               $\eta=1,\quad\nu=1$\\
               $\kappa\geq1$\\
               $a=1,\ldots,2\hat{\kappa}-1$}
& $k=f_z+\kappa-2$ \\[4pt]
\hline
Gravitino
& $\LL[1+\sqrt{\tfrac14+h_{\kappa\nu\eta a, z}};0,\half;\sigma_{\kappa\nu\eta a}]_{s_{\kappa\nu\eta a}F_{\nu,z}}
   \oplus
   \LL[1+\sqrt{\tfrac14+h_{\kappa\nu\eta a, z}};\half,0;-\sigma_{\kappa\nu\eta a}]_{-s_{\kappa\nu\eta a}F_{\nu,z}}$
& \makecell[l]{$\eta=1,\quad\nu\geq2$\\
               $\kappa\geq1$\\
               $a=1,\ldots,2\hat{\kappa}-1$}
& $k=\nu f_z+\kappa-2$ \\[4pt]
\hline
Vector
& $\LL[1+\sqrt{1+p_{\kappa\nu a, z}};0,0;\rho_{\nu,z}]_{F_{\nu,z}}
   \oplus
   \LL[1+\sqrt{1+p_{\kappa\nu a, z}};0,0;-\rho_{\nu,z}]_{-F_{\nu,z}}$
& \makecell[l]{distinguished $z$\\
               $\nu\geq1$\\
               $\kappa\geq1$ if $\nu=1$\\
               $\kappa\geq0$ if $\nu\geq2$\\
               $a=1$ for $\kappa=0,1$\\
               $a=1,\ldots,2\check{\kappa}$ for $\kappa\geq2$}
& $k=\nu f_z+\kappa-2$ \\[4pt]
\hline
Vector
& $\LL[1+\sqrt{1+p_{\kappa\nu a, z}};0,0;\rho_{\nu,z}]_{F_{\nu,z}}
   \oplus
   \LL[1+\sqrt{1+p_{\kappa\nu a, z}};0,0;-\rho_{\nu,z}]_{-F_{\nu,z}}$
& \makecell[l]{other interior $z$\\
               $\nu\geq1,\quad\kappa\geq0$\\
               $a=1$ for $\kappa=0,1$\\
               $a=1,\ldots,2\check{\kappa}$ for $\kappa\geq2$}
& $k=\nu f_z+\kappa-2$ \\[4pt]
\hline
Vector
& $\LL[1+\sqrt{1+q_{\kappa\nu\eta a, z}};0,0;\rho_{\nu,z}-2]_{F_{\nu,z}}
   \oplus
   \LL[1+\sqrt{1+q_{\kappa\nu\eta a, z}};0,0;-\rho_{\nu,z}+2]_{-F_{\nu,z}}$
& \makecell[l]{$\eta=-2,\quad\nu\geq1$\\
               $\kappa\geq4$\\
               $a=1,\ldots,\check{\kappa}-1$}
& $k=\nu f_z+\kappa-4$ \\[4pt]
\hline
Vector
& $\LL[1+\sqrt{1+q_{\kappa\nu\eta a, z}};0,0;\rho_{1,z}+2]_{F_{1,z}}
   \oplus
   \LL[1+\sqrt{1+q_{\kappa\nu\eta a, z}};0,0;-\rho_{1,z}-2]_{-F_{1,z}}$
& \makecell[l]{distinguished $z$\\
               $\eta=2,\quad\nu=1$\\
               $\kappa\geq4$\\
               $a=1,\ldots,\check{\kappa}-1$}
& $k=f_z+\kappa-2$ \\[4pt]
\hline
Vector
& $\LL[1+\sqrt{1+q_{\kappa\nu\eta a, z}};0,0;\rho_{1,z}+2]_{F_{1,z}}
   \oplus
   \LL[1+\sqrt{1+q_{\kappa\nu\eta a, z}};0,0;-\rho_{1,z}-2]_{-F_{1,z}}$
& \makecell[l]{other interior $z$\\
               $\eta=2,\quad\nu=1$\\
               $\kappa\geq4$\\
               $a=1,\ldots,\check{\kappa}-1$}
& $k=f_z+\kappa-4$ \\[4pt]
\hline
Vector
& $\LL[1+\sqrt{1+q_{\kappa\nu\eta a, z}};0,0;\rho_{\nu,z}+2]_{F_{\nu,z}}
   \oplus
   \LL[1+\sqrt{1+q_{\kappa\nu\eta a, z}};0,0;-\rho_{\nu,z}-2]_{-F_{\nu,z}}$
& \makecell[l]{$\eta=2,\quad\nu\geq2$\\
               $\kappa\geq4$\\
               $a=1,\ldots,\check{\kappa}-1$}
& $k=\nu f_z+\kappa-4$ \\[4pt]
\hline
\end{tabular}%
}
\end{table}

The ranges of the labels $(\kappa , \nu , a )$ in the table reflect the existence of the corresponding polynomials and, where appropriate, exclude roots that hit the unitarity bound and become short. At the distinguished interior values of $z$ in (\ref{eq:Distinguishedz}), the $\eta=1$ gravitino family with $\nu=1$ starts at $\kappa_{\textrm{min}}=2$ instead of $\kappa_{\textrm{min}}=1$, while the central-vector family with $\nu=1$ starts at $\kappa_{\textrm{min}}=1$ instead of $\kappa_{\textrm{min}}=0$. These excluded multiplets are those already listed in table~\ref{tab:InteriorProtectedMultiplets}. The separate $\eta=2$, $\nu=1$ vector cases account for a change in the KK-level offset at distinguished $z$ and do not exclude any protected roots. With the label ranges indicated in the table, all multiplets given therein are strictly long. Finally, the KK level $k$ is determined by $\kappa$ and $\nu$ through (\ref{eq:GeneralKKLevel}) with $\mu_z=f_z$, $\alpha=\beta=1$ and the multiplet-dependent offset $\gamma$ indicated in the table.

\begin{figure}
\centering
\includegraphics[width=\linewidth]{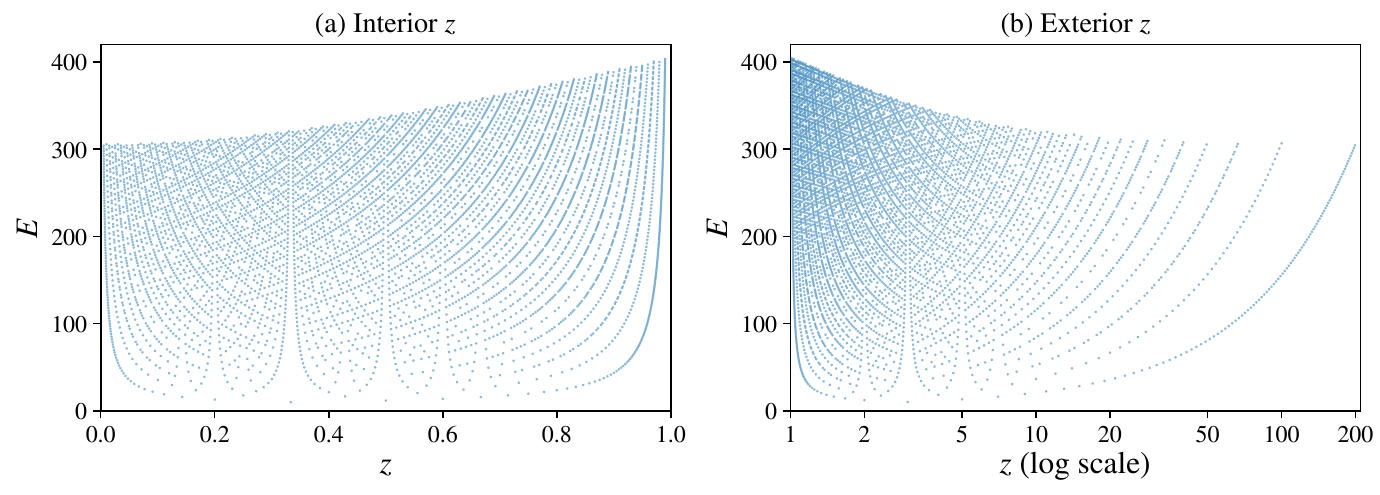}
\caption{SCP dimensions (\ref{eq:EzGrav}) of the long flavour-charged graviton multiplet listed in tables~\ref{tab:NewChargedInteriorLong} (Panel~(a), $0<z<1$) and~\ref{tab:NewChargedExteriorLong} (Panel~(b), $z>1$) at $\nu=\kappa=a=1$. Each panel contains $6115$ BBBW backgrounds, with one blue point per CPT  graviton multiplet pair and background. All points are unconnected. The horizontal axis is linear in~(a) and logarithmic in~(b), and the panels share the same vertical scale.}
\label{fig:BBBWGravitonCloudInterior}
\label{fig:BBBWGravitonCloudExterior}
\end{figure}

Figure~\ref{fig:BBBWGravitonCloudInterior}(a) shows the SCP dimension
\begin{equation} \label{eq:EzGrav}
E(z)=1+\sqrt{4+g_{111, z}}
\end{equation}
of the graviton CPT pair with $\nu=\kappa=a=1$ listed in
table~\ref{tab:NewChargedInteriorLong}, as a function of $z$. The survey in the plot includes all $6115$ BBBW AdS$_5$ backgrounds with interior $z$ satisfying 
$f_z\leq B$ with $B=200$. This encompasses both distinguished values of
$z$ in the sense of (\ref{eq:Distinguishedz}) and generic ones in the range 
$1/199\leq z\leq99/100$. Since $k=f_z+1$ for this family and
these labels, the corresponding KK levels range from $4$ to $201$. Other choices of $(\nu,\kappa,a)$ for the graviton multiplets in table~\ref{tab:NewChargedInteriorLong}
yield dimension plots with the same qualitative dependence on $z$. A similar qualitative behaviour also occurs for all other multiplets in table~\ref{tab:NewChargedInteriorLong}, when their dimensions are plotted as functions of $z$ at fixed values of $(\nu,\kappa,a)$.

\begin{figure}
\centering
\includegraphics[width=\linewidth]{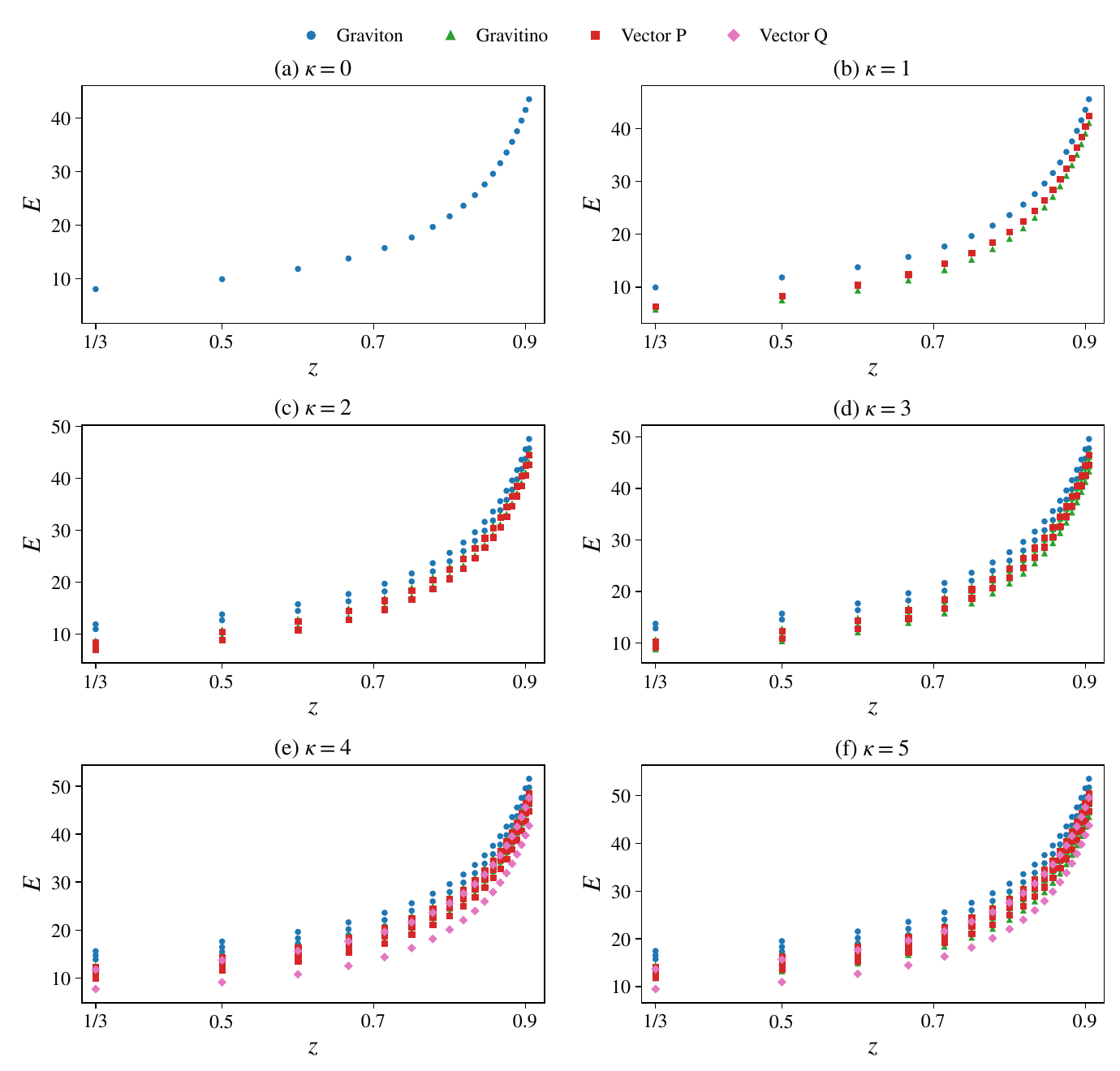}
\caption{The long flavour-charged dimensions $E$ as functions of $z$ on the distinguished interior sequence (\ref{eq:Distinguishedz}), at $\nu=1$, $\kappa=0,\ldots,5$, and every allowed value of the index $a$ and offset $\eta$ from table~\ref{tab:NewChargedInteriorLong}. One point is plotted per CPT pair and background. Distinct pairs with identical dimensions give overlapping points.}
\label{fig:BBBWChargedDistinguishedInteriorNuOne}
\end{figure}

The survey in figure~\ref{fig:BBBWGravitonCloudInterior}(a) illustrates the discontinuous, non-monotonic dependence
of these graviton SCP dimensions on $z$. Although $E(z)$ is uniquely determined
at each admissible $z$, nearby rational values of $z$ can have very
different primitive twisting data (\ref{eq:SpectrumTwistData2}).
These data enter the coefficients of $G_{11}(x;z)$ in (\ref{eq:ChargedGFirstMembers}) and hence determine
its root $g_{111, z}$. For example, $z=\tfrac13$ and
$z=\tfrac{33}{100}$ differ by only $\tfrac1{300}$, but their primitive
twisting pairs $(\widehat p, \widehat q)$ are $(2,1)$ and $(133,67)$, giving $f_z=3$ and $f_z=200$,
respectively. The corresponding dimensions are
\begin{equation}
E(\tfrac13)=1+\sqrt{28+30\sqrt3}\simeq9.9\,,\qquad E(\tfrac{33}{100})=1+\tfrac1{33}\sqrt{57985387+462134\sqrt{13267}}\simeq320.6\,.
\end{equation}
This example also explains the sparse valleys around certain values of $z$: the dimension is small at the central value, whereas nearby distinct backgrounds require much larger $f_z$ and hence much larger dimensions. 
Increasing the cutoff $B$ includes additional AdS$_5$ backgrounds
and extends the range of KK levels. Both the density of plotted
points and the upper scale of the dimensions therefore depend on
the cutoff. In particular, the point density does not represent
a density of states within a single background. The survey compares
different backgrounds at fixed integer labels, while the flavour
charge and KK level vary with $z$; it therefore does not trace a
smooth continuation at fixed charge.

Although figure~\ref{fig:BBBWGravitonCloudInterior}(a) keeps $\nu=1$ fixed, the large-charge behaviour (\ref{eq:ChargedLongLargeNuEF}) is already visible in the upper limiting curve $B\left(1+\frac12\sqrt{1+3z^2}\right)$. For $B=200$, this curve rises from approximately $300$ to $400$, accounting for the upper edge of the plot. In deriving (\ref{eq:ChargedLongLargeNuEF}), we approached the large flavour-charge limit by taking $\nu$ large at fixed $z$. Since $|F|=\nu f_z$, however, the same regime can also be reached at fixed $\nu$ by sampling backgrounds with increasingly large values of $f_z$. The figure illustrates this second route to the limit, in which the large charge arises from the twisting data of the sampled BBBW backgrounds.

A simpler dependence emerges upon restricting the survey to the distinguished interior sequence (\ref{eq:Distinguishedz}). Along this sequence, the primitive twisting data $(\widehat p,\widehat q)=(j,1)$, $j=2,3,  \ldots$ vary in a correlated manner, and the dimensions of the multiplets of table~\ref{tab:NewChargedInteriorLong} increase monotonically with $j$, and hence with $z_j$. This behaviour reflects this particular choice of BBBW backgrounds, and is not a consequence of the sampled $z$ values alone. Figure \ref{fig:BBBWChargedDistinguishedInteriorNuOne} shows the evolution of the dimensions along the distinguished sequence of all available multiplets in table~\ref{tab:NewChargedInteriorLong} at $\nu=1$ and $\kappa=0 , \ldots , 5$. Each panel at fixed $\kappa$ has as many dots of the same colour, corresponding to the same type of multiplet, as the range of the index $a$ allows. The divergence of the dimensions as $z_j\to1^-$ reflects the simultaneous growth of the flavour charge and KK level. Indeed, along this sequence,
\begin{equation}
\label{eq:DistinguishedInteriorChargeGrowth}
f_{z_j}=j+1=\frac{2}{1-z_j}\xrightarrow[j\to\infty]{}\infty.
\end{equation}
and $k=f_{z_j}+\kappa+\gamma$ diverges as well. At the MN2 endpoint $z=1$, the primitive twisting pair is instead $(\widehat p,\widehat q)=(1,0)$, and the dimensions at fixed KK level remain finite.

\paragraph{The MN2 endpoints, $|z|=1$.} 

The $\textrm{SU}(2,2|1) \times \textrm{U}(1)_+$ content of the long, flavour-charged, U$(1)_1$-invariant spectrum of MN2 is displayed in table~\ref{tab:NewChargedMN2Long}. At $|z|=1$, the roots of the polynomials constructed in appendix~\ref{app:ChargedPolynomialsNew} yield the following explicit expressions for the SCP conformal dimensions:
\begin{equation}
\label{eq:NewChargedMN2Dimensions}
 D_{kb}=\sqrt{4+2k(k+3)+2(k-2b)(k-2b+1)} \; .
\end{equation}
Here, $k$ is the KK label obtained from the usual labels  $\kappa$, $\nu$ through (\ref{eq:GeneralKKLevel}), and the independent integer $b$ replaces the
generic polynomial-root index $a$. The finite range of $b$ is still determined by
$\kappa$ through the quantities (\ref{eq:NeutralParityDefinitions}) for each type of multiplet listed in table~\ref{tab:NewChargedMN2Long}. Branches in the table without $b$ contain a single occurrence at $\kappa$, $\nu$ fixed. The index $b$ fixes the SU$(2)$ R-symmetry spin $\ell=k-2b$ of the parent $\mathcal N=2$ representation whose branching contains the relevant $\mathcal N=1$ multiplet. See section~\ref{sec:RepresentativeMN2} for details. As usual, all multiplets in table~\ref{tab:NewChargedMN2Long} are strictly long: protected occurrences have been listed separately in table~\ref{tab:MN2ProtectedMultiplets}.

\begin{table}[H]
\centering
\caption{Long flavour-charged spectrum
$\cL_{\textrm{$F$-charged}}(s)$ at MN2, 
with $s\equiv\operatorname{sgn}(z)=\pm1$.}
\label{tab:NewChargedMN2Long}
\fontsize{9}{11}\selectfont
\renewcommand{\arraystretch}{1.3}
\setlength{\tabcolsep}{3pt}
\resizebox{\linewidth}{!}{%
\begin{tabular}{@{}l|l|l|l@{}}
\hline
Type & Multiplet & Label ranges & Level \\
\hline
\hline
Graviton
& $\LL[2k+3;\half,\half;\tfrac{4\nu}{3}]_{-s\nu}\oplus\LL[2k+3;\half,\half;-\tfrac{4\nu}{3}]_{s\nu}$
& $\nu\geq1,\quad\kappa\geq1$
& $k=\nu+\kappa$ \\[4pt]
\hline
Graviton
& $\LL[D_{kb}+1;\half,\half;\tfrac{4\nu}{3}]_{-s\nu}\oplus\LL[D_{kb}+1;\half,\half;-\tfrac{4\nu}{3}]_{s\nu}$
& \makecell[l]{$\nu\geq1,\quad\kappa\geq1$\\
               $b=1,\ldots,\check{\kappa}$}
& $k=\nu+\kappa$ \\[4pt]
\hline
Gravitino
& $\LL[2k+\tfrac52;0,\half;\tfrac{4\nu+3}{3}]_{-s\nu}\oplus\LL[2k+\tfrac52;\half,0;-\tfrac{4\nu+3}{3}]_{s\nu}$
& $\nu\geq1,\quad\kappa\geq2$
& $k=\nu+\kappa$ \\[4pt]
\hline
Gravitino
& $\LL[2k+\tfrac52;0,\half;-\tfrac{4\nu-3}{3}]_{s\nu}\oplus\LL[2k+\tfrac52;\half,0;\tfrac{4\nu-3}{3}]_{-s\nu}$
& $\nu\geq1,\quad\kappa\geq2$
& $k=\nu+\kappa-2$ \\[4pt]
\hline
Gravitino
& $\LL[D_{kb}+\half;0,\half;\tfrac{4\nu+3}{3}]_{-s\nu}\oplus\LL[D_{kb}+\half;\half,0;-\tfrac{4\nu+3}{3}]_{s\nu}$
& \makecell[l]{$\nu\geq1,\quad\kappa\geq2$\\
               $b=1,\ldots,\hat{\kappa}-1$}
& $k=\nu+\kappa$ \\[4pt]
\hline
Gravitino
& $\LL[D_{kb}+\half;0,\half;-\tfrac{4\nu-3}{3}]_{s\nu}\oplus\LL[D_{kb}+\half;\half,0;\tfrac{4\nu-3}{3}]_{-s\nu}$
& \makecell[l]{$\nu\geq1,\quad\kappa\geq2$\\
               $b=1,\ldots,\hat{\kappa}-1$}
& $k=\nu+\kappa-2$ \\[4pt]
\hline
Gravitino
& $\LL[D_{kb}+\tfrac32;0,\half;-\tfrac{4\nu+3}{3}]_{s\nu}\oplus\LL[D_{kb}+\tfrac32;\half,0;\tfrac{4\nu+3}{3}]_{-s\nu}$
& \makecell[l]{$\nu\geq1,\quad\kappa\geq2$\\
               $b=1,\ldots,\hat{\kappa}-1$}
& $k=\nu+\kappa$ \\[4pt]
\hline
Gravitino
& $\LL[D_{kb}+\tfrac32;0,\half;\tfrac{4\nu-3}{3}]_{-s\nu}\oplus\LL[D_{kb}+\tfrac32;\half,0;-\tfrac{4\nu-3}{3}]_{s\nu}$
& \makecell[l]{$\nu\geq1,\quad\kappa\geq2$\\
               $b=1,\ldots,\hat{\kappa}-1$}
& $k=\nu+\kappa-2$ \\[4pt]
\hline
Vector
& $\LL[2k+4;0,0;\tfrac{4\nu}{3}]_{-s\nu}\oplus\LL[2k+4;0,0;-\tfrac{4\nu}{3}]_{s\nu}$
& $\nu\geq1,\quad\kappa\geq2$
& $k=\nu+\kappa-2$ \\[4pt]
\hline
Vector
& $\LL[2k+2;0,0;\tfrac{4\nu}{3}]_{-s\nu}\oplus\LL[2k+2;0,0;-\tfrac{4\nu}{3}]_{s\nu}$
& $\nu\geq1,\quad\kappa\geq3$
& $k=\nu+\kappa-2$ \\[4pt]
\hline
Vector
& $\LL[D_{kb};0,0;\tfrac{4\nu}{3}]_{-s\nu}\oplus\LL[D_{kb};0,0;-\tfrac{4\nu}{3}]_{s\nu}$
& \makecell[l]{$\nu\geq1,\quad\kappa\geq2$\\
               $b=1,\ldots,\check{\kappa}-1$}
& $k=\nu+\kappa-2$ \\[4pt]
\hline
Vector
& $\LL[D_{kb}+2;0,0;\tfrac{4\nu}{3}]_{-s\nu}\oplus\LL[D_{kb}+2;0,0;-\tfrac{4\nu}{3}]_{s\nu}$
& \makecell[l]{$\nu\geq1,\quad\kappa\geq2$\\
               $b=1,\ldots,\check{\kappa}-1$}
& $k=\nu+\kappa-2$ \\[4pt]
\hline
Vector
& $\LL[D_{kb}+1;0,0;\tfrac{4\nu+6}{3}]_{-s\nu}\oplus\LL[D_{kb}+1;0,0;-\tfrac{4\nu+6}{3}]_{s\nu}$
& \makecell[l]{$\nu\geq1,\quad\kappa\geq4$\\
               $b=1,\ldots,\check{\kappa}-1$}
& $k=\nu+\kappa$ \\[4pt]
\hline
Vector
& $\LL[D_{kb}+1;0,0;-\tfrac23]_{-s}\oplus\LL[D_{kb}+1;0,0;\tfrac23]_{s}$
& \makecell[l]{$\nu=1,\quad\kappa\geq4$\\
               $b=1,\ldots,\check{\kappa}-1$}
& $k=\kappa-1$ \\[4pt]
\hline
Vector
& $\LL[D_{kb}+1;0,0;\tfrac{4\nu-6}{3}]_{-s\nu}\oplus\LL[D_{kb}+1;0,0;-\tfrac{4\nu-6}{3}]_{s\nu}$
& \makecell[l]{$\nu\geq2,\quad\kappa\geq4$\\
               $b=1,\ldots,\check{\kappa}-1$}
& $k=\nu+\kappa-4$ \\[4pt]
\hline
\end{tabular}%
}
\end{table}


\paragraph{The exterior, $|z|>1$.}

\begin{figure}
\centering
\includegraphics[width=\linewidth]{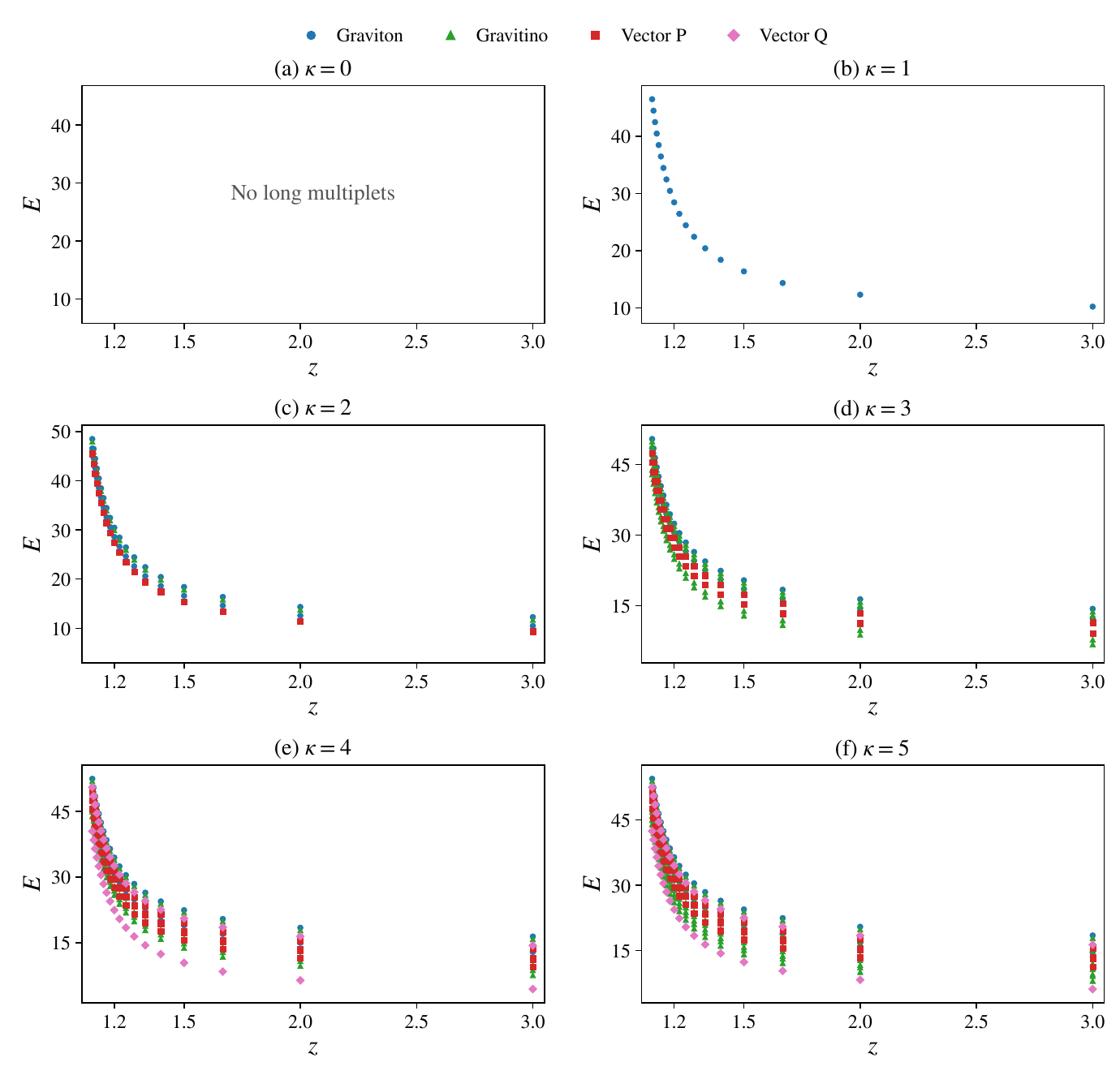}
\caption{The long flavour-charged dimensions $E$ as functions of $z$ on the distinguished exterior sequence (\ref{eq:Distinguishedz}), at $\nu=1$, $\kappa=0,\ldots,5$, and every allowed value of the index $a$ and offset $\eta$ from table~\ref{tab:NewChargedExteriorLong}. One point is plotted per CPT pair and background. Distinct pairs with identical dimensions give overlapping points.}
\label{fig:BBBWChargedDistinguishedExteriorNuOne}
\end{figure}

The situation is again generic in this region: the graviton, gravitino and vector CPT pairs collected in
table~\ref{tab:NewChargedExteriorLong} retain the generic labels
$(\kappa,\nu,a)$, along with $\eta$ as appropriate. Their SCP dimensions follow from the roots of
$G_{\kappa\nu}$, $H_{\kappa\nu\eta}$, $P_{\kappa\nu}$ and
$Q_{\kappa\nu\eta}$ in appendix~\ref{app:ChargedPolynomialsNew},
with charges specified by (\ref{eq:ProtectedMasterCharge}) and the
gravitino sign assignments (\ref{eq:GinoRCharge}). The KK level formula (\ref{eq:GeneralKKLevel}) applies with $\mu_z=\lambda_z$, $\alpha=\beta=1$ and the offsets $\gamma$ listed in the table. The label ranges account for both
the existence of physical multiplets and the removal of protected occurrences,
which belong to table~\ref{tab:ExteriorProtectedMultiplets}. For example, for the $\eta=-2$ vector family, the first long member has $\kappa=4$
when $\nu=1$ at a distinguished exterior value of $z$ in
(\ref{eq:Distinguishedz}). In all other cases this family starts at
$\kappa=6$: the $\kappa=4$ physical vector pair is absent and the $\kappa=5$ root is
protected, as explained in appendix~\ref{app:NewChargedOuterVector}.
For the vector multiplet with zero SCP R-charge offset, 
$\delta_{\kappa2}$ reduces the
root count $2\check{\kappa}$ by one only at $\kappa=2$. All entries
in the table are strictly long.

\begin{table}[H]
\centering
\caption{Long flavour-charged spectrum
$\cL_{\textrm{$F$-charged}}(z)$ at exterior values of $z$: $|z| > 1$.}
\label{tab:NewChargedExteriorLong}
\fontsize{9}{11}\selectfont
\renewcommand{\arraystretch}{1.3}
\setlength{\tabcolsep}{3pt}
\resizebox{\linewidth}{!}{%
\begin{tabular}{@{}l|l|l|l@{}}
\hline
Type & Multiplet & Label ranges & Level \\
\hline
\hline
Graviton
& $\LL[1+\sqrt{4+g_{\kappa\nu a, z}};\half,\half;\rho_{\nu,z}]_{F_{\nu,z}}\oplus \LL[1+\sqrt{4+g_{\kappa\nu a, z}};\half,\half;-\rho_{\nu,z}]_{-F_{\nu,z}}$
& \makecell[l]{$\nu\geq1,\quad\kappa\geq1$\\
               $a=1,\ldots,\check{\kappa}+1$}
& $k=\nu\lambda_z+\kappa$ \\[4pt]
\hline
Gravitino
& $\LL[1+\sqrt{\tfrac14+h_{\kappa\nu\eta a, z}};0,\half;\sigma_{\kappa\nu\eta a}]_{s_{\kappa\nu\eta a}F_{\nu,z}}\oplus \LL[1+\sqrt{\tfrac14+h_{\kappa\nu\eta a, z}};\half,0;-\sigma_{\kappa\nu\eta a}]_{-s_{\kappa\nu\eta a}F_{\nu,z}}$
& \makecell[l]{$\eta=1,\quad\nu\geq1$\\
               $\kappa\geq2$\\
               $a=1,\ldots,2\hat{\kappa}-1$}
& $k=\nu\lambda_z+\kappa$ \\[4pt]
\hline
Gravitino
& $\LL[1+\sqrt{\tfrac14+h_{\kappa\nu\eta a, z}};0,\half;\sigma_{\kappa\nu\eta a}]_{s_{\kappa\nu\eta a}F_{\nu,z}}\oplus \LL[1+\sqrt{\tfrac14+h_{\kappa\nu\eta a, z}};\half,0;-\sigma_{\kappa\nu\eta a}]_{-s_{\kappa\nu\eta a}F_{\nu,z}}$
& \makecell[l]{$\eta=-1,\quad\nu\geq1$\\
               $\kappa=3,\quad a=1,2$}
& $k=\nu\lambda_z-1$ \\[4pt]
\hline
Gravitino
& $\LL[1+\sqrt{\tfrac14+h_{\kappa\nu\eta a, z}};0,\half;\sigma_{\kappa\nu\eta a}]_{s_{\kappa\nu\eta a}F_{\nu,z}}\oplus \LL[1+\sqrt{\tfrac14+h_{\kappa\nu\eta a, z}};\half,0;-\sigma_{\kappa\nu\eta a}]_{-s_{\kappa\nu\eta a}F_{\nu,z}}$
& \makecell[l]{$\eta=-1,\quad\nu\geq1$\\
               $\kappa\geq4$\\
               $a=1,\ldots,2\hat{\kappa}-1$}
& $k=\nu\lambda_z+\kappa-4$ \\[4pt]
\hline
Vector
& $\LL[1+\sqrt{1+p_{\kappa\nu a, z}};0,0;\rho_{\nu,z}]_{F_{\nu,z}}\oplus \LL[1+\sqrt{1+p_{\kappa\nu a, z}};0,0;-\rho_{\nu,z}]_{-F_{\nu,z}}$
& \makecell[l]{$\nu\geq1,\quad\kappa\geq2$\\
               $a=1,\ldots,2\check{\kappa}-\delta_{\kappa2}$}
& $k=\nu\lambda_z+\kappa-2$ \\[4pt]
\hline
Vector
& $\LL[1+\sqrt{1+q_{\kappa\nu\eta a, z}};0,0;\rho_{\nu,z}+2]_{F_{\nu,z}}\oplus \LL[1+\sqrt{1+q_{\kappa\nu\eta a, z}};0,0;-\rho_{\nu,z}-2]_{-F_{\nu,z}}$
& \makecell[l]{$\eta=2,\quad\nu\geq1$\\
               $\kappa\geq4$\\
               $a=1,\ldots,\check{\kappa}-1$}
& $k=\nu\lambda_z+\kappa$ \\[4pt]
\hline
Vector
& $\LL[1+\sqrt{1+q_{\kappa\nu\eta a, z}};0,0;\rho_{1,z}-2]_{F_{1,z}}\oplus \LL[1+\sqrt{1+q_{\kappa\nu\eta a, z}};0,0;-\rho_{1,z}+2]_{-F_{1,z}}$
& \makecell[l]{distinguished $z$\\
               $\eta=-2,\quad\nu=1$\\
               $\kappa\geq4$\\
               $a=1,\ldots,\check{\kappa}-1$}
& $k=\lambda_z+\kappa-6$ \\[4pt]
\hline
Vector
& $\LL[1+\sqrt{1+q_{\kappa\nu\eta a, z}};0,0;\rho_{1,z}-2]_{F_{1,z}}\oplus \LL[1+\sqrt{1+q_{\kappa\nu\eta a, z}};0,0;-\rho_{1,z}+2]_{-F_{1,z}}$
& \makecell[l]{other exterior $z$\\
               $\eta=-2,\quad\nu=1$\\
               $\kappa\geq6$\\
               $a=1,\ldots,\check{\kappa}-1$}
& $k=\lambda_z+\kappa-8$ \\[4pt]
\hline
Vector
& $\LL[1+\sqrt{1+q_{\kappa\nu\eta a, z}};0,0;\rho_{\nu,z}-2]_{F_{\nu,z}}\oplus \LL[1+\sqrt{1+q_{\kappa\nu\eta a, z}};0,0;-\rho_{\nu,z}+2]_{-F_{\nu,z}}$
& \makecell[l]{$\eta=-2,\quad\nu\geq2$\\
               $\kappa\geq6$\\
               $a=1,\ldots,\check{\kappa}-1$}
& $k=\nu\lambda_z+\kappa-8$ \\[4pt]
\hline
\end{tabular}%
}
\end{table}

Figure~\ref{fig:BBBWGravitonCloudExterior}(b) shows the SCP dimension
(\ref{eq:EzGrav}) of the long graviton CPT pair in table~\ref{tab:NewChargedExteriorLong} with $\nu=\kappa=a=1$
over the $6115$ exterior backgrounds satisfying $\lambda_z\leq B=200$.
The sample spans $100/99\leq z\leq199$, with $k=\lambda_z+1\leq201$,
and is displayed on a logarithmic horizontal axis. As for the interior $z$ case of figure~\ref{fig:BBBWGravitonCloudExterior}(a), the non-monotonic dependence and sparse valleys reflect the discontinuity of the twisting data (\ref{eq:SpectrumTwistData2}) with $z$. The leading upper scale,
$\frac{3B}{2}(1+|\epsilon(z)|/z)$ in agreement with (\ref{eq:ChargedLongLargeNuEF}), decreases from approximately $400$
near $z=1$ towards $300$ at large $z$.
Figure~\ref{fig:BBBWChargedDistinguishedExteriorNuOne} instead shows all
available long multiplets in table~\ref{tab:NewChargedExteriorLong} at
$\nu=1$ and $\kappa=0,\ldots,5$, including every allowed $a$ and $\eta$,
on the distinguished exterior sequence with $j=2,\ldots,20$. Here,
$(\widehat p,\widehat q)=(j,-1)$ gives $f_{z_j}=j-1$ and $\lambda_{z_j}=j+1$.
Along this sequence, the dimensions increase with $j$, and hence as $z_j$
decreases towards $1$. Their growth, $E \approx 2j$, reflects increasing
charge and KK level, while the MN2 dimensions at fixed KK level remain finite.

\section{Spectra at representative values of the twist parameter}
\label{sec:RepresentativeSpectra}

We now discuss the spectrum at selected values of $z$, following the organisation by KK level in (\ref{eq:FullSpec}). The symmetry (\ref{eq:FullSpecInv}) allows us to restrict attention to non-negative $z$. Our examples cover the four categories obtained by distinguishing between interior and exterior regions, (\ref{eq:zRegions}), and between distinguished and generic values of $z$, (\ref{eq:Distinguishedz}). We choose one fixed twist parameter from each category and comment separately on the special MN1 and MN2 cases. We also examine the large-$|z|$ limit at fixed KK level. For each BBBW background and level $k$ under consideration, we assemble the multiplets in the four contributions to (\ref{eq:SpectrumFourfoldSplit}) determined in sections~\ref{sec:FlavourNeutralSpectrum} and~\ref{sec:FlavourChargedSpectrum}. This includes every isolated multiplet arising at that level, together with all allowed members of the infinite families. For the latter, we select the values of $(\kappa,\nu,a)$, and of $\eta$ where appropriate, that yield the required $k$ through (\ref{eq:GeneralKKLevel}) and the specific level formula in the corresponding table of section~\ref{sec:UniversalSpectrum}. The numerical spectra in sections~\ref{sec:RepresentativeOneFifth},~\ref{sec:RepresentativeOneThird},~\ref{sec:RepresentativeThree}--\ref{sec:RepresentativeLargeZ} are displayed at $k=0,\ldots,5$. The enhanced MN1 and MN2 cases are treated separately in sections~\ref{sec:RepresentativeMN1} and \ref{sec:RepresentativeMN2}.

In order to save space, tables \ref{tab:RepresentativeOneFifth}--\ref{tab:RepresentativeLargeZ} below use some shorthands. For long graviton ($j=\frac12$) and vector ($j=0$) multiplets, the notation $\LL[E;j,j;\pm r]_{\pm F}$ stands for two, not four, multiplets, namely,
\begin{equation}
\LL[E;j,j;\pm r]_{\pm F} \equiv \LL[E;j,j;r]_F \oplus \LL[E;j,j;-r]_{-F} \; .
\end{equation}
For entries whose CPT conjugate changes the shortening labels or
Lorentz spins, $\oplus\,\mathrm{CPT}$ denotes the conjugate obtained by
$X\bar Y\mapsto Y\bar X$, $(j_1,j_2)\mapsto(j_2,j_1)$ and
$(r,F)\mapsto(-r,-F)$.
A multiplicity multiplying such an entry multiplies the entire pair.
CPT-self-conjugate flavour-neutral multiplets are displayed once. Numerical dimensions and R-charges are rounded to three decimal places. Some flavour-charged multiplet families first appear above this displayed range and are therefore absent from the corresponding tables. Their onset levels are discussed for the exterior examples in sections~\ref{sec:RepresentativeThree} and~\ref{sec:RepresentativeFive}.

\subsection{\texorpdfstring{$z=0$}{z=0}: the MN1 endpoint}
\label{sec:RepresentativeMN1}

At the MN1 point, the flavour symmetry enhances to
$\mathrm{SU}(2)_+$, with $\mathrm{U}(1)_+$ its Cartan subgroup,
and the $\mathrm{SU}(2,2|1)\times\mathrm{U}(1)_+$ multiplets of
section~\ref{sec:UniversalSpectrum} reassemble at each KK level
into representations of
$\mathrm{SU}(2,2|1)\times\mathrm{SU}(2)_+$.
In our charge normalisation, the spin-$\ell$ representation of $\mathrm{SU}(2)_+$
branches as
\begin{equation}
\bm{2 \ell +1} \rightarrow \oplus_{m=-\ell}^{\ell}\mathbf{1}_{2m} \; .
\end{equation}
Accordingly, $\mathrm{SU}(2,2|1)$ multiplets with identical superconformal quantum numbers and $\mathrm{U}(1)_+$ flavour charges $F=-2\ell,-2\ell+2,\ldots,2\ell$  combine into the $\bm{2 \ell +1}$ representation of  $\mathrm{SU}(2)_+$. For example, the protected multiplets in
tables~\ref{tab:FirstZ-neutral-short}
and~\ref{tab:MN1ProtectedMultiplets} give the flavour-current
triplet at $k=0$ and a CPT pair of short-gravitino triplets at
$k=1$, while the stress-tensor multiplet remains an $\mathrm{SU}(2)_+$ singlet.
Further examples follow by combining the long neutral and
charged sectors in tables~\ref{tab:FirstZ-neutral-long}
and~\ref{tab:NewChargedMN1Long}.
For example, at $k=2$, the graviton multiplets with $E=1+2\sqrt7$ and
$F=0,\pm2$ form
$\LL[1+2\sqrt7;\half,\half;0]\otimes[1]$,
whereas the vector multiplets with $E=1+\sqrt{19}$ and
$F=0,\pm2,\pm4$ form
$\LL[1+\sqrt{19};0,0;0]\otimes[2]$. The U$(1)_0$-invariant spectrum of MN1 discussed in section 4.2 of \cite{Bhattacharya:2026ypg} is thus recovered.

\subsection{\texorpdfstring{$z=\frac15$}{z=1/5}: a generic interior value}
\label{sec:RepresentativeOneFifth}

\begin{table}[!t]
\centering
\begingroup
\fontsize{8}{10}\selectfont
\thinmuskip=2mu\medmuskip=3mu\thickmuskip=4mu
\renewcommand{\arraystretch}{1.08}
\setlength{\tabcolsep}{4pt}
\setlength{\doublerulesep}{0pt}
\caption{The $z=\frac15$ spectrum at $k=0,\ldots,5$.}\label{tab:RepresentativeOneFifth}
\begin{tabular}{@{}p{0.105\textwidth}|p{\dimexpr0.895\textwidth-8.4pt\relax}@{}}
\hline\hline
\textbf{Multiplet} & $z=\frac15$ \\
\hline
\multicolumn{2}{l}{\rule{0pt}{13pt}$k=0$}\\*
Graviton & \raggedright \mbox{\(A_1\bar A_1[3.000;\tfrac{1}{2},\tfrac{1}{2};0]_{0}\)} \arraybackslash \\*[2pt]
Vector & \raggedright \mbox{\(A_2\bar A_2[2.000;0,0;0]_{0}\)} \allowbreak
\mbox{\(\oplus\,L\bar L[3.678;0,0;0]_{0}\)} \arraybackslash \\[3pt]
\hline\hline
\multicolumn{2}{l}{\rule{0pt}{13pt}$k=1$}\\*
Graviton & \raggedright \mbox{\(L\bar L[5.000;\tfrac{1}{2},\tfrac{1}{2};0]_{0}\)} \arraybackslash \\*[2pt]
Gravitino & \raggedright \mbox{\((L\bar A_1[4.500;0,\tfrac{1}{2};1]_{0}\oplus\mathrm{CPT})\)} \arraybackslash \\*[2pt]
Vector & \raggedright \mbox{\(L\bar L[4.521;0,0;0]_{0}\)} \allowbreak
\mbox{\(\oplus\,L\bar L[5.447;0,0;0]_{0}\)} \arraybackslash \\[3pt]
\hline\hline
\multicolumn{2}{l}{\rule{0pt}{13pt}$k=2$}\\*
Graviton & \raggedright \mbox{\(L\bar L[6.250;\tfrac{1}{2},\tfrac{1}{2};0]_{0}\)} \allowbreak
\mbox{\(\oplus\,L\bar L[6.854;\tfrac{1}{2},\tfrac{1}{2};0]_{0}\)} \arraybackslash \\*[2pt]
Gravitino & \raggedright \mbox{\((L\bar L[6.500;0,\tfrac{1}{2};1]_{0}\oplus\mathrm{CPT})\)} \arraybackslash \\*[2pt]
Vector & \raggedright \mbox{\(L\bar L[5.325;0,0;0]_{0}\)} \allowbreak
\mbox{\(\oplus\,L\bar L[6.435;0,0;0]_{0}\)} \allowbreak
\mbox{\(\oplus\,L\bar L[7.067;0,0;0]_{0}\)} \allowbreak
\mbox{\(\oplus\,L\bar L[7.240;0,0;0]_{0}\)} \arraybackslash \\[3pt]
\hline\hline
\multicolumn{2}{l}{\rule{0pt}{13pt}$k=3$}\\*
Graviton & \raggedright \mbox{\(L\bar L[8.130;\tfrac{1}{2},\tfrac{1}{2};0]_{0}\)} \allowbreak
\mbox{\(\oplus\,L\bar L[8.681;\tfrac{1}{2},\tfrac{1}{2};0]_{0}\)} \arraybackslash \\*[2pt]
Gravitino & \raggedright \mbox{\((L\bar L[7.465;0,\tfrac{1}{2};1]_{0}\oplus\mathrm{CPT})\)} \allowbreak
\mbox{\(\oplus\,(L\bar L[8.346;0,\tfrac{1}{2};-1]_{0}\oplus\mathrm{CPT})\)} \allowbreak
\mbox{\(\oplus\,(L\bar L[8.381;0,\tfrac{1}{2};1]_{0}\oplus\mathrm{CPT})\)} \arraybackslash \\*[2pt]
Vector & \raggedright \mbox{\(L\bar L[7.475;0,0;0]_{0}\)} \allowbreak
\mbox{\(\oplus\,L\bar L[8.252;0,0;0]_{0}\)} \allowbreak
\mbox{\(\oplus\,L\bar L[8.795;0,0;0]_{0}\)} \allowbreak
\mbox{\(\oplus\,L\bar L[9.045;0,0;0]_{0}\)} \allowbreak
\mbox{\(\oplus\,L\bar L[7.097;0,0;\pm 1.486]_{\mp 5}\)} \arraybackslash \\[3pt]
\hline\hline
\multicolumn{2}{l}{\rule{0pt}{13pt}$k=4$}\\*
Graviton & \raggedright \mbox{\(L\bar L[9.315;\tfrac{1}{2},\tfrac{1}{2};0]_{0}\)} \allowbreak
\mbox{\(\oplus\,L\bar L[9.937;\tfrac{1}{2},\tfrac{1}{2};0]_{0}\)} \allowbreak
\mbox{\(\oplus\,L\bar L[10.503;\tfrac{1}{2},\tfrac{1}{2};0]_{0}\)} \arraybackslash \\*[2pt]
Gravitino & \raggedright \mbox{\((L\bar L[9.440;0,\tfrac{1}{2};1]_{0}\oplus\mathrm{CPT})\)} \allowbreak
\mbox{\(\oplus\,(L\bar L[10.163;0,\tfrac{1}{2};-1]_{0}\oplus\mathrm{CPT})\)} \allowbreak
\mbox{\(\oplus\,(L\bar L[10.222;0,\tfrac{1}{2};1]_{0}\oplus\mathrm{CPT})\)} \allowbreak
\mbox{\(\oplus\,(L\bar L[8.854;\tfrac{1}{2},0;0.486]_{-5}\oplus\mathrm{CPT})\)} \allowbreak
\mbox{\(\oplus\,(L\bar L[9.437;0,\tfrac{1}{2};2.486]_{-5}\oplus\mathrm{CPT})\)} \arraybackslash \\*[2pt]
Vector & \raggedright \mbox{\(L\bar L[8.368;0,0;0]_{0}\)} \allowbreak
\mbox{\(\oplus\,L\bar L[9.411;0,0;0]_{0}\)} \allowbreak
\mbox{\(\oplus\,L\bar L[9.698;0,0;\pm 2]_{0}\)} \allowbreak
\mbox{\(\oplus\,L\bar L[10.046;0,0;0]_{0}\)} \allowbreak
\mbox{\(\oplus\,L\bar L[10.173;0,0;0]_{0}\)} \allowbreak
\mbox{\(\oplus\,L\bar L[10.541;0,0;0]_{0}\)} \allowbreak
\mbox{\(\oplus\,L\bar L[10.853;0,0;0]_{0}\)} \allowbreak
\mbox{\(\oplus\,L\bar L[9.231;0,0;\pm 1.486]_{\mp 5}\)} \arraybackslash \\[3pt]
\hline\hline
\multicolumn{2}{l}{\rule{0pt}{13pt}$k=5$}\\*
Graviton & \raggedright \mbox{\(L\bar L[11.208;\tfrac{1}{2},\tfrac{1}{2};0]_{0}\)} \allowbreak
\mbox{\(\oplus\,L\bar L[11.720;\tfrac{1}{2},\tfrac{1}{2};0]_{0}\)} \allowbreak
\mbox{\(\oplus\,L\bar L[12.323;\tfrac{1}{2},\tfrac{1}{2};0]_{0}\)} \allowbreak
\mbox{\(\oplus\,L\bar L[11.002;\tfrac{1}{2},\tfrac{1}{2};\pm 1.486]_{\mp 5}\)} \arraybackslash \\*[2pt]
Gravitino & \raggedright \mbox{\((L\bar L[10.447;0,\tfrac{1}{2};1]_{0}\oplus\mathrm{CPT})\)} \allowbreak
\mbox{\(\oplus\,(L\bar L[11.324;0,\tfrac{1}{2};1]_{0}\oplus\mathrm{CPT})\)} \allowbreak
\mbox{\(\oplus\,(L\bar L[11.357;0,\tfrac{1}{2};-1]_{0}\oplus\mathrm{CPT})\)} \allowbreak
\mbox{\(\oplus\,(L\bar L[11.964;0,\tfrac{1}{2};-1]_{0}\oplus\mathrm{CPT})\)} \allowbreak
\mbox{\(\oplus\,(L\bar L[12.049;0,\tfrac{1}{2};1]_{0}\oplus\mathrm{CPT})\)} \allowbreak
\mbox{\(\oplus\,(L\bar L[10.854;\tfrac{1}{2},0;0.486]_{-5}\oplus\mathrm{CPT})\)} \allowbreak
\mbox{\(\oplus\,(L\bar L[11.437;0,\tfrac{1}{2};2.486]_{-5}\oplus\mathrm{CPT})\)} \arraybackslash \\*[2pt]
Vector & \raggedright \mbox{\(L\bar L[10.450;0,0;0]_{0}\)} \allowbreak
\mbox{\(\oplus\,L\bar L[11.254;0,0;0]_{0}\)} \allowbreak
\mbox{\(\oplus\,L\bar L[11.567;0,0;\pm 2]_{0}\)} \allowbreak
\mbox{\(\oplus\,L\bar L[11.841;0,0;0]_{0}\)} \allowbreak
\mbox{\(\oplus\,L\bar L[11.922;0,0;0]_{0}\)} \allowbreak
\mbox{\(\oplus\,L\bar L[12.306;0,0;0]_{0}\)} \allowbreak
\mbox{\(\oplus\,L\bar L[12.664;0,0;0]_{0}\)} \allowbreak
\mbox{\(\oplus\,L\bar L[10.036;0,0;\pm 1.486]_{\mp 5}\)} \allowbreak
\mbox{\(\oplus\,L\bar L[10.804;0,0;\mp 0.514]_{\mp 5}\)} \allowbreak
\mbox{\(\oplus\,L\bar L[11.228;0,0;\pm 1.486]_{\mp 5}\)} \allowbreak
\mbox{\(\oplus\,L\bar L[11.784;0,0;\pm 3.486]_{\mp 5}\)} \arraybackslash \\[3pt]
\hline\hline
\end{tabular}
\endgroup
\end{table}

The spectrum at the generic, non-distinguished interior value $z=\frac15$ at levels $k=0, \ldots, 5$ is listed in table \ref{tab:RepresentativeOneFifth}. This BBBW background has $\epsilon=(5-2\sqrt7)/3$ and primitive twisting data $(\widehat p,\widehat q)=(3,2)$, giving $\mu_z = f_z=5$, $\lambda_z=1$. The only protected multiplets are the isolated flavour-neutral instances of table~\ref{tab:FirstZ-neutral-short}. All other multiplets are long, and come from contributions from tables~\ref{tab:FirstZ-neutral-long} and \ref{tab:NewChargedInteriorLong}. In particular, the first flavour-charged $\eta=0$ vector multiplets have labels $(\kappa,\nu,a)=(0,1,1)$ and occur at $k=3$. The first two flavour-charged gravitino pairs have $(\kappa,\nu,a)=(1,1,1)$ and $\eta=\pm1$, and occur at $k=4$. The first flavour-charged graviton multiplets have $(\kappa,\nu,a)=(0,1,1)$ and occur at $k=5$. All these multiplets have flavour charges $F=\pm5$. All flavour-charged entries in table~\ref{tab:RepresentativeOneFifth} have $\nu=1$.



\subsection{\texorpdfstring{$z=\frac13$}{z=1/3}: a distinguished interior value}
\label{sec:RepresentativeOneThird}

\begin{table}[!t]
\centering
\begingroup
\fontsize{8}{10}\selectfont
\thinmuskip=2mu\medmuskip=3mu\thickmuskip=4mu
\renewcommand{\arraystretch}{1.08}
\setlength{\tabcolsep}{4pt}
\setlength{\doublerulesep}{0pt}
\caption{The $z=\frac13$ spectrum at $k=0,\ldots,5$.}\label{tab:RepresentativeOneThird}
\begin{tabular}{@{}p{0.105\textwidth}|p{\dimexpr0.895\textwidth-8.4pt\relax}@{}}
\hline\hline
\textbf{Multiplet} & $z=\frac13$ \\
\hline
\multicolumn{2}{l}{\rule{0pt}{13pt}$k=0$}\\*
Graviton & \raggedright \mbox{\(A_1\bar A_1[3.000;\tfrac{1}{2},\tfrac{1}{2};0]_{0}\)} \arraybackslash \\*[2pt]
Vector & \raggedright \mbox{\(A_2\bar A_2[2.000;0,0;0]_{0}\)} \allowbreak
\mbox{\(\oplus\,L\bar L[3.726;0,0;0]_{0}\)} \arraybackslash \\[3pt]
\hline\hline
\multicolumn{2}{l}{\rule{0pt}{13pt}$k=1$}\\*
Graviton & \raggedright \mbox{\(L\bar L[5.000;\tfrac{1}{2},\tfrac{1}{2};0]_{0}\)} \arraybackslash \\*[2pt]
Gravitino & \raggedright \mbox{\((L\bar A_1[4.500;0,\tfrac{1}{2};1]_{0}\oplus\mathrm{CPT})\)} \arraybackslash \\*[2pt]
Vector & \raggedright \mbox{\((L\bar A_2[4.196;0,0;1.464]_{-3}\oplus\mathrm{CPT})\)} \allowbreak
\mbox{\(\oplus\,L\bar L[4.416;0,0;0]_{0}\)} \allowbreak
\mbox{\(\oplus\,L\bar L[5.557;0,0;0]_{0}\)} \arraybackslash \\[3pt]
\hline\hline
\multicolumn{2}{l}{\rule{0pt}{13pt}$k=2$}\\*
Graviton & \raggedright \mbox{\(L\bar L[6.192;\tfrac{1}{2},\tfrac{1}{2};0]_{0}\)} \allowbreak
\mbox{\(\oplus\,L\bar L[6.883;\tfrac{1}{2},\tfrac{1}{2};0]_{0}\)} \arraybackslash \\*[2pt]
Gravitino & \raggedright \mbox{\((L\bar A_1[6.696;0,\tfrac{1}{2};2.464]_{-3}\oplus\mathrm{CPT})\)} \allowbreak
\mbox{\(\oplus\,(L\bar L[6.500;0,\tfrac{1}{2};1]_{0}\oplus\mathrm{CPT})\)} \allowbreak
\mbox{\(\oplus\,(L\bar L[5.768;\tfrac{1}{2},0;0.464]_{-3}\oplus\mathrm{CPT})\)} \arraybackslash \\*[2pt]
Vector & \raggedright \mbox{\(L\bar L[5.271;0,0;0]_{0}\)} \allowbreak
\mbox{\(\oplus\,L\bar L[6.308;0,0;0]_{0}\)} \allowbreak
\mbox{\(\oplus\,L\bar L[7.043;0,0;0]_{0}\)} \allowbreak
\mbox{\(\oplus\,L\bar L[7.408;0,0;0]_{0}\)} \allowbreak
\mbox{\(\oplus\,L\bar L[6.398;0,0;\pm 1.464]_{\mp 3}\)} \arraybackslash \\[3pt]
\hline\hline
\multicolumn{2}{l}{\rule{0pt}{13pt}$k=3$}\\*
Graviton & \raggedright \mbox{\(L\bar L[8.032;\tfrac{1}{2},\tfrac{1}{2};0]_{0}\)} \allowbreak
\mbox{\(\oplus\,L\bar L[8.753;\tfrac{1}{2},\tfrac{1}{2};0]_{0}\)} \allowbreak
\mbox{\(\oplus\,L\bar L[8.002;\tfrac{1}{2},\tfrac{1}{2};\pm 1.464]_{\mp 3}\)} \arraybackslash \\*[2pt]
Gravitino & \raggedright \mbox{\((L\bar L[7.414;0,\tfrac{1}{2};1]_{0}\oplus\mathrm{CPT})\)} \allowbreak
\mbox{\(\oplus\,(L\bar L[8.316;0,\tfrac{1}{2};-1]_{0}\oplus\mathrm{CPT})\)} \allowbreak
\mbox{\(\oplus\,(L\bar L[8.403;0,\tfrac{1}{2};1]_{0}\oplus\mathrm{CPT})\)} \allowbreak
\mbox{\(\oplus\,(L\bar L[7.768;\tfrac{1}{2},0;0.464]_{-3}\oplus\mathrm{CPT})\)} \allowbreak
\mbox{\(\oplus\,(L\bar L[8.696;0,\tfrac{1}{2};2.464]_{-3}\oplus\mathrm{CPT})\)} \arraybackslash \\*[2pt]
Vector & \raggedright \mbox{\(L\bar L[7.342;0,0;0]_{0}\)} \allowbreak
\mbox{\(\oplus\,L\bar L[8.145;0,0;0]_{0}\)} \allowbreak
\mbox{\(\oplus\,L\bar L[8.772;0,0;0]_{0}\)} \allowbreak
\mbox{\(\oplus\,L\bar L[9.265;0,0;0]_{0}\)} \allowbreak
\mbox{\(\oplus\,L\bar L[7.033;0,0;\pm 1.464]_{\mp 3}\)} \allowbreak
\mbox{\(\oplus\,L\bar L[7.695;0,0;\mp 0.536]_{\mp 3}\)} \allowbreak
\mbox{\(\oplus\,L\bar L[8.395;0,0;\pm 1.464]_{\mp 3}\)} \arraybackslash \\[3pt]
\hline\hline
\multicolumn{2}{l}{\rule{0pt}{13pt}$k=4$}\\*
Graviton & \raggedright \mbox{\(L\bar L[9.234;\tfrac{1}{2},\tfrac{1}{2};0]_{0}\)} \allowbreak
\mbox{\(\oplus\,L\bar L[9.827;\tfrac{1}{2},\tfrac{1}{2};0]_{0}\)} \allowbreak
\mbox{\(\oplus\,L\bar L[10.621;\tfrac{1}{2},\tfrac{1}{2};0]_{0}\)} \allowbreak
\mbox{\(\oplus\,L\bar L[9.942;\tfrac{1}{2},\tfrac{1}{2};\pm 1.464]_{\mp 3}\)} \arraybackslash \\*[2pt]
Gravitino & \raggedright \mbox{\((L\bar L[9.359;0,\tfrac{1}{2};1]_{0}\oplus\mathrm{CPT})\)} \allowbreak
\mbox{\(\oplus\,(L\bar L[10.138;0,\tfrac{1}{2};-1]_{0}\oplus\mathrm{CPT})\)} \allowbreak
\mbox{\(\oplus\,(L\bar L[10.279;0,\tfrac{1}{2};1]_{0}\oplus\mathrm{CPT})\)} \allowbreak
\mbox{\(\oplus\,(L\bar L[8.775;\tfrac{1}{2},0;0.464]_{-3}\oplus\mathrm{CPT})\)} \allowbreak
\mbox{\(\oplus\,(L\bar L[9.372;0,\tfrac{1}{2};2.464]_{-3}\oplus\mathrm{CPT})\)} \allowbreak
\mbox{\(\oplus\,(L\bar L[9.677;\tfrac{1}{2},0;0.464]_{-3}\oplus\mathrm{CPT})\)} \allowbreak
\mbox{\(\oplus\,(L\bar L[9.684;0,\tfrac{1}{2};0.464]_{-3}\oplus\mathrm{CPT})\)} \allowbreak
\mbox{\(\oplus\,(L\bar L[10.310;\tfrac{1}{2},0;2.464]_{-3}\oplus\mathrm{CPT})\)} \allowbreak
\mbox{\(\oplus\,(L\bar L[10.634;0,\tfrac{1}{2};2.464]_{-3}\oplus\mathrm{CPT})\)} \arraybackslash \\*[2pt]
Vector & \raggedright \mbox{\(L\bar L[8.290;0,0;0]_{0}\)} \allowbreak
\mbox{\(\oplus\,L\bar L[9.216;0,0;0]_{0}\)} \allowbreak
\mbox{\(\oplus\,L\bar L[9.668;0,0;\pm 2]_{0}\)} \allowbreak
\mbox{\(\oplus\,L\bar L[9.986;0,0;0]_{0}\)} \allowbreak
\mbox{\(\oplus\,L\bar L[10.107;0,0;0]_{0}\)} \allowbreak
\mbox{\(\oplus\,L\bar L[10.540;0,0;0]_{0}\)} \allowbreak
\mbox{\(\oplus\,L\bar L[11.124;0,0;0]_{0}\)} \allowbreak
\mbox{\(\oplus\,L\bar L[9.122;0,0;\pm 1.464]_{\mp 3}\)} \allowbreak
\mbox{\(\oplus\,L\bar L[9.488;0,0;\mp 0.536]_{\mp 3}\)} \allowbreak
\mbox{\(\oplus\,L\bar L[10.325;0,0;\pm 1.464]_{\mp 3}\)} \allowbreak
\mbox{\(\oplus\,L\bar L[9.013;0,0;\pm 2.928]_{\mp 6}\)} \arraybackslash \\[3pt]
\hline\hline
\multicolumn{2}{l}{\rule{0pt}{13pt}$k=5$}\\*
Graviton & \raggedright \mbox{\(L\bar L[11.071;\tfrac{1}{2},\tfrac{1}{2};0]_{0}\)} \allowbreak
\mbox{\(\oplus\,L\bar L[11.625;\tfrac{1}{2},\tfrac{1}{2};0]_{0}\)} \allowbreak
\mbox{\(\oplus\,L\bar L[12.487;\tfrac{1}{2},\tfrac{1}{2};0]_{0}\)} \allowbreak
\mbox{\(\oplus\,L\bar L[10.914;\tfrac{1}{2},\tfrac{1}{2};\pm 1.464]_{\mp 3}\)} \allowbreak
\mbox{\(\oplus\,L\bar L[11.842;\tfrac{1}{2},\tfrac{1}{2};\pm 1.464]_{\mp 3}\)} \arraybackslash \\*[2pt]
Gravitino & \raggedright \mbox{\((L\bar L[10.365;0,\tfrac{1}{2};1]_{0}\oplus\mathrm{CPT})\)} \allowbreak
\mbox{\(\oplus\,(L\bar L[11.232;0,\tfrac{1}{2};1]_{0}\oplus\mathrm{CPT})\)} \allowbreak
\mbox{\(\oplus\,(L\bar L[11.286;0,\tfrac{1}{2};-1]_{0}\oplus\mathrm{CPT})\)} \allowbreak
\mbox{\(\oplus\,(L\bar L[11.958;0,\tfrac{1}{2};-1]_{0}\oplus\mathrm{CPT})\)} \allowbreak
\mbox{\(\oplus\,(L\bar L[12.147;0,\tfrac{1}{2};1]_{0}\oplus\mathrm{CPT})\)} \allowbreak
\mbox{\(\oplus\,(L\bar L[10.720;\tfrac{1}{2},0;0.464]_{-3}\oplus\mathrm{CPT})\)} \allowbreak
\mbox{\(\oplus\,(L\bar L[11.355;0,\tfrac{1}{2};2.464]_{-3}\oplus\mathrm{CPT})\)} \allowbreak
\mbox{\(\oplus\,(L\bar L[11.509;0,\tfrac{1}{2};0.464]_{-3}\oplus\mathrm{CPT})\)} \allowbreak
\mbox{\(\oplus\,(L\bar L[11.557;\tfrac{1}{2},0;0.464]_{-3}\oplus\mathrm{CPT})\)} \allowbreak
\mbox{\(\oplus\,(L\bar L[12.201;\tfrac{1}{2},0;2.464]_{-3}\oplus\mathrm{CPT})\)} \allowbreak
\mbox{\(\oplus\,(L\bar L[12.543;0,\tfrac{1}{2};2.464]_{-3}\oplus\mathrm{CPT})\)} \allowbreak
\mbox{\(\oplus\,(L\bar L[10.500;\tfrac{1}{2},0;1.928]_{-6}\oplus\mathrm{CPT})\)} \allowbreak
\mbox{\(\oplus\,(L\bar L[11.428;0,\tfrac{1}{2};3.928]_{-6}\oplus\mathrm{CPT})\)} \arraybackslash \\*[2pt]
Vector & \raggedright \mbox{\(L\bar L[10.305;0,0;0]_{0}\)} \allowbreak
\mbox{\(\oplus\,L\bar L[11.022;0,0;0]_{0}\)} \allowbreak
\mbox{\(\oplus\,L\bar L[11.542;0,0;\pm 2]_{0}\)} \allowbreak
\mbox{\(\oplus\,L\bar L[11.825;0,0;0]_{0}\)} \allowbreak
\mbox{\(\oplus\,L\bar L[11.837;0,0;0]_{0}\)} \allowbreak
\mbox{\(\oplus\,L\bar L[12.339;0,0;0]_{0}\)} \allowbreak
\mbox{\(\oplus\,L\bar L[12.984;0,0;0]_{0}\)} \allowbreak
\mbox{\(\oplus\,L\bar L[9.947;0,0;\pm 1.464]_{\mp 3}\)} \allowbreak
\mbox{\(\oplus\,L\bar L[10.731;0,0;\mp 0.536]_{\mp 3}\)} \allowbreak
\mbox{\(\oplus\,L\bar L[11.090;0,0;\pm 1.464]_{\mp 3}\)} \allowbreak
\mbox{\(\oplus\,L\bar L[11.262;0,0;\mp 0.536]_{\mp 3}\)} \allowbreak
\mbox{\(\oplus\,L\bar L[11.728;0,0;\pm 3.464]_{\mp 3}\)} \allowbreak
\mbox{\(\oplus\,L\bar L[11.826;0,0;\pm 1.464]_{\mp 3}\)} \allowbreak
\mbox{\(\oplus\,L\bar L[12.229;0,0;\pm 1.464]_{\mp 3}\)} \allowbreak
\mbox{\(\oplus\,L\bar L[11.103;0,0;\pm 2.928]_{\mp 6}\)} \arraybackslash \\[3pt]
\hline\hline
\end{tabular}
\endgroup
\end{table}

We next consider the distinguished value $z=\frac13$, with $j=2$ in (\ref{eq:Distinguishedz}), where $(\widehat p,\widehat q)=(2,1)$, $\mu_z = f_z=3$, $\lambda_z=1$ and  $\epsilon=1-2\sqrt3/3$. The spectrum now contains the isolated protected contributions from tables~\ref{tab:FirstZ-neutral-short}~and~\ref{tab:InteriorProtectedMultiplets}. All other multiplets are long and come from tables~\ref{tab:FirstZ-neutral-long} and \ref{tab:NewChargedInteriorLong}. At $k=4$, the long flavour-charged vector spectrum with $\eta=0$ contains three CPT pairs, with labels 
$
 (\kappa,\nu,a)=(3,1,1), \; (3,1,2),\; (0,2,1)
$
subject to $3\nu+\kappa-2=4$. The first two pairs have flavour $F=\pm3$, and the third has $F=\pm6$. This illustrates how multiplets with different labels in (\ref{eq:SpectrumFourfoldSplit}) contribute to the same KK level in (\ref{eq:FullSpec}).


\begingroup
\makeatletter
\setlength{\@fptop}{0pt}
\makeatother
\clearpage
\endgroup

\subsection{\texorpdfstring{$z=1$}{z=1}: the MN2 endpoint}
\label{sec:RepresentativeMN2}

At the MN2 point, $z=1$, the $\mathrm{SU}(2,2|1) \times \mathrm{U}(1)_+$ multiplets listed in tables~\ref{tab:FirstZ-neutral-short}, \ref{tab:FirstZ-neutral-long}, \ref{tab:MN2ProtectedMultiplets} and~\ref{tab:NewChargedMN2Long} reassemble at each KK level into representations of the enhanced $\cN=2$ superconformal group, $\mathrm{SU}(2,2|2)$. The resulting contributions to the spectrum (\ref{eq:FullSpec}) are, in $\cN=2$ notation,
\begin{equation}
\label{eq:MN2ReassembledSpectrum}
\begin{split}
\mathrm{Spec}_k^0(1)={}& A_2\bar A_2[2k+2;0,0;k,0]
 \oplus B_1\bar B_1[2k+4;0,0;k+2,0]
\\
&\oplus\bigoplus_{b=1}^{\lfloor k/2\rfloor}
 L\bar L[D_{kb};0,0;k-2b,0]\; ,\qquad k=0,1,\ldots\; .
\end{split}
\end{equation}
Here, we follow again \cite{Cordova:2016emh} and denote an $\cN=2$ multiplet by $X\bar Y[E;j_1,j_2;\ell,n]$, where $E$, $(j_1,j_2)$, $\ell$ and $n$ are the dimension, Lorentz spins, SU$(2)$ R-symmetry spin and U$(1)$ R-charge of its SCP. The long sum in the second line of (\ref{eq:MN2ReassembledSpectrum}) is empty at $k=0,1$, and its first member is $L\bar L[2\sqrt6;0,0;0,0]$ at $k=2$. The dimensions 
$D_{kb}$ are given in (\ref{eq:NewChargedMN2Dimensions}). Each multiplet in (\ref{eq:MN2ReassembledSpectrum}) occurs once.

The putative MN2 spectrum was given in (24) of \cite{Bhattacharya:2024tjw}. The globally defined $\mathrm{U}(1)_1$-invariant spectrum (\ref{eq:MN2ReassembledSpectrum}) above consists of a subset of the multiplets listed in \cite{Bhattacharya:2024tjw}. In particular, the only U$(1)_1$-invariant short multiplets are those that appear in the first line of (\ref{eq:MN2ReassembledSpectrum}). The long multiplet dimensions, $D_{kb}$ in (\ref{eq:NewChargedMN2Dimensions}), are related to the dimensions $E_{k \ell n j}$ in \cite{Bhattacharya:2024tjw} through $D_{kb}=E_{k,k-2b,0,0}$. Appendix~\ref{app:N=2Enhancement} gives a detailed account of how the $\cN=1$ multiplets obtained at MN2 in section~\ref{sec:UniversalSpectrum} reassemble into the $\cN=2$ multiplets (\ref{eq:MN2ReassembledSpectrum}).

\begin{samepage}
\subsection{\texorpdfstring{$z=3$}{z=3}: a distinguished exterior value}
\label{sec:RepresentativeThree}

We now turn to the exterior and first consider the distinguished value $z=3$, corresponding to $j=2$ in (\ref{eq:Distinguishedz}). For this BBBW background, the primitive data are $(\widehat p,\widehat q)=(2,-1)$, so $\mu_z = \lambda_z =3$, $f_z =1$, and $\epsilon=(1-2\sqrt7)/9$. Table~\ref{tab:RepresentativeThree} lists the available multiplets through KK level $k=5$ by collecting all possible contributions from tables~\ref{tab:FirstZ-neutral-short},~\ref{tab:FirstZ-neutral-long},~\ref{tab:ExteriorProtectedMultiplets} and \ref{tab:NewChargedExteriorLong}. As at every distinguished exterior value, the short chiral family and one short vector family have $\nu_{\textrm{min}}=2$, while all other protected families have $\nu_{\textrm{min}}=1$: see table~\ref{tab:ExteriorProtectedMultiplets}. The first chiral multiplet,
$L\bar B_1[\tfrac{17+2\sqrt7}{3};0,0;\tfrac{34+4\sqrt7}{9}]_{-2}$
along with its CPT conjugate, indeed has $\nu_{\textrm{min}}=2$ and appears at $k=2$. The other chiral multiplet in the table arises at $k=5$ with $\nu=3$. As for all other exterior BBBW backgrounds, the $z=3$ protected spectrum contains tensor multiplets. Within the displayed range, these tensor multiplets occur at KK levels $k=0,3$. The long flavour-charged vector family with $\eta=2$ first appears at $k=7$ and is therefore absent from table~\ref{tab:RepresentativeThree}.

\par\end{samepage}

\newpage
\begingroup
\fontsize{8}{10}\selectfont
\thinmuskip=2mu\medmuskip=3mu\thickmuskip=4mu
\renewcommand{\arraystretch}{1.0}
\setlength{\tabcolsep}{4pt}
\setlength{\LTpre}{6pt}
\setlength{\LTcapwidth}{\textwidth}
\setlength{\LTpost}{10pt}
\begin{longtable}{@{}p{0.105\textwidth}|p{\dimexpr0.895\textwidth-8.4pt\relax}@{}}
\caption{The $z=3$ spectrum at $k=0,\ldots,5$.}\label{tab:RepresentativeThree}\\
\hline\hline
\textbf{Multiplet} & $z=3$ \\
\hline
\endfirsthead
\multicolumn{2}{c}{\footnotesize\tablename\ \thetable\ (continued): $z=3$}\\
\hline\hline
\textbf{Multiplet} & Representations \\
\hline
\endhead
\hline\multicolumn{2}{r}{\textit{Continued on the next page}}\\
\endfoot
\hline\hline
\endlastfoot
\multicolumn{2}{l}{\rule{0pt}{13pt}$k=0$}\\*
Graviton & \raggedright \mbox{\(A_1\bar A_1[3.000;\tfrac{1}{2},\tfrac{1}{2};0]_{0}\)} \arraybackslash \\*[1pt]
Vector & \raggedright \mbox{\(A_2\bar A_2[2.000;0,0;0]_{0}\)} \allowbreak
\mbox{\(\oplus\,L\bar L[4.331;0,0;0]_{0}\)} \arraybackslash \\*[1pt]
Tensor & \raggedright \mbox{\((L\bar B_1[3.715;\tfrac{1}{2},0;2.477]_{-1}\oplus\mathrm{CPT})\)} \arraybackslash \\[3pt]
\hline\hline
\multicolumn{2}{l}{\rule{0pt}{13pt}$k=1$}\\*
Graviton & \raggedright \mbox{\(L\bar L[5.000;\tfrac{1}{2},\tfrac{1}{2};0]_{0}\)} \arraybackslash \\*[1pt]
Gravitino & \raggedright \mbox{\((L\bar A_1[4.500;0,\tfrac{1}{2};1]_{0}\oplus\mathrm{CPT})\)} \allowbreak
\mbox{\(\oplus\,(L\bar A_2[5.715;\tfrac{1}{2},0;2.477]_{-1}\oplus\mathrm{CPT})\)} \arraybackslash \\*[1pt]
Vector & \raggedright \mbox{\(L\bar L[3.631;0,0;0]_{0}\)} \allowbreak
\mbox{\(\oplus\,L\bar L[6.401;0,0;0]_{0}\)} \allowbreak
\mbox{\(\oplus\,L\bar L[4.401;0,0;\pm 1.477]_{\mp 1}\)} \arraybackslash \\[3pt]
\hline\hline
\multicolumn{2}{l}{\rule{0pt}{13pt}$k=2$}\\*
Graviton & \raggedright \mbox{\(L\bar L[5.561;\tfrac{1}{2},\tfrac{1}{2};0]_{0}\)} \allowbreak
\mbox{\(\oplus\,L\bar L[7.091;\tfrac{1}{2},\tfrac{1}{2};0]_{0}\)} \arraybackslash \\*[1pt]
Gravitino & \raggedright \mbox{\((L\bar A_2[5.715;\tfrac{1}{2},0;2.477]_{-1}\oplus\mathrm{CPT})\)} \allowbreak
\mbox{\(\oplus\,(L\bar L[6.500;0,\tfrac{1}{2};1]_{0}\oplus\mathrm{CPT})\)} \allowbreak
\mbox{\(\oplus\,(L\bar L[6.799;0,\tfrac{1}{2};2.477]_{-1}\oplus\mathrm{CPT})\)} \allowbreak
\mbox{\(\oplus\,(L\bar L[7.799;\tfrac{1}{2},0;2.477]_{-1}\oplus\mathrm{CPT})\)} \arraybackslash \\*[1pt]
Vector & \raggedright \mbox{\(L\bar L[4.403;0,0;0]_{0}\)} \allowbreak
\mbox{\(\oplus\,L\bar L[5.866;0,0;0]_{0}\)} \allowbreak
\mbox{\(\oplus\,L\bar L[6.705;0,0;0]_{0}\)} \allowbreak
\mbox{\(\oplus\,L\bar L[8.496;0,0;0]_{0}\)} \allowbreak
\mbox{\(\oplus\,L\bar L[6.168;0,0;\pm 1.477]_{\mp 1}\)} \arraybackslash \\*[1pt]
Chiral & \raggedright \mbox{\((L\bar B_1[7.431;0,0;4.954]_{-2}\oplus\mathrm{CPT})\)} \arraybackslash \\[3pt]
\hline\hline
\multicolumn{2}{l}{\rule{0pt}{13pt}$k=3$}\\*
Graviton & \raggedright \mbox{\((L\bar A_1[8.215;\tfrac{1}{2},\tfrac{1}{2};3.477]_{-1}\oplus\mathrm{CPT})\)} \allowbreak
\mbox{\(\oplus\,L\bar L[7.237;\tfrac{1}{2},\tfrac{1}{2};0]_{0}\)} \allowbreak
\mbox{\(\oplus\,L\bar L[9.185;\tfrac{1}{2},\tfrac{1}{2};0]_{0}\)} \arraybackslash \\*[1pt]
Gravitino & \raggedright \mbox{\((L\bar L[6.795;0,\tfrac{1}{2};1]_{0}\oplus\mathrm{CPT})\)} \allowbreak
\mbox{\(\oplus\,(L\bar L[7.881;0,\tfrac{1}{2};-1]_{0}\oplus\mathrm{CPT})\)} \allowbreak
\mbox{\(\oplus\,(L\bar L[8.585;0,\tfrac{1}{2};1]_{0}\oplus\mathrm{CPT})\)} \allowbreak
\mbox{\(\oplus\,(L\bar L[7.583;\tfrac{1}{2},0;2.477]_{-1}\oplus\mathrm{CPT})\)} \allowbreak
\mbox{\(\oplus\,(L\bar L[8.761;0,\tfrac{1}{2};2.477]_{-1}\oplus\mathrm{CPT})\)} \allowbreak
\mbox{\(\oplus\,(L\bar L[9.894;\tfrac{1}{2},0;2.477]_{-1}\oplus\mathrm{CPT})\)} \arraybackslash \\*[1pt]
Vector & \raggedright \mbox{\((L\bar A_2[7.215;0,0;3.477]_{-1}\oplus\mathrm{CPT})\)} \allowbreak
\mbox{\(\oplus\,(L\bar A_2[9.431;0,0;4.954]_{-2}\oplus\mathrm{CPT})\)} \allowbreak
\mbox{\(\oplus\,L\bar L[5.905;0,0;0]_{0}\)} \allowbreak
\mbox{\(\oplus\,L\bar L[7.957;0,0;0]_{0}\)} \allowbreak
\mbox{\(\oplus\,L\bar L[8.446;0,0;0]_{0}\)} \allowbreak
\mbox{\(\oplus\,L\bar L[10.597;0,0;0]_{0}\)} \allowbreak
\mbox{\(\oplus\,L\bar L[6.922;0,0;\pm 1.477]_{\mp 1}\)} \allowbreak
\mbox{\(\oplus\,L\bar L[8.156;0,0;\pm 1.477]_{\mp 1}\)} \allowbreak
\mbox{\(\oplus\,L\bar L[9.342;0,0;\pm 3.477]_{\mp 1}\)} \arraybackslash \\*[1pt]
Tensor & \raggedright \mbox{\((L\bar B_1[8.931;\tfrac{1}{2},0;5.954]_{-2}\oplus\mathrm{CPT})\)} \arraybackslash \\[3pt]
\hline\hline
\multicolumn{2}{l}{\rule{0pt}{13pt}$k=4$}\\*
Graviton & \raggedright \mbox{\(L\bar L[8.140;\tfrac{1}{2},\tfrac{1}{2};0]_{0}\)} \allowbreak
\mbox{\(\oplus\,L\bar L[9.179;\tfrac{1}{2},\tfrac{1}{2};0]_{0}\)} \allowbreak
\mbox{\(\oplus\,L\bar L[11.284;\tfrac{1}{2},\tfrac{1}{2};0]_{0}\)} \allowbreak
\mbox{\(\oplus\,L\bar L[10.215;\tfrac{1}{2},\tfrac{1}{2};\pm 3.477]_{\mp 1}\)} \arraybackslash \\*[1pt]
Gravitino & \raggedright \mbox{\((L\bar A_1[9.715;0,\tfrac{1}{2};4.477]_{-1}\oplus\mathrm{CPT})\)} \allowbreak
\mbox{\(\oplus\,(L\bar A_2[10.931;\tfrac{1}{2},0;5.954]_{-2}\oplus\mathrm{CPT})\)} \allowbreak
\mbox{\(\oplus\,(L\bar L[8.597;0,\tfrac{1}{2};1]_{0}\oplus\mathrm{CPT})\)} \allowbreak
\mbox{\(\oplus\,(L\bar L[9.778;0,\tfrac{1}{2};-1]_{0}\oplus\mathrm{CPT})\)} \allowbreak
\mbox{\(\oplus\,(L\bar L[10.680;0,\tfrac{1}{2};1]_{0}\oplus\mathrm{CPT})\)} \allowbreak
\mbox{\(\oplus\,(L\bar L[8.090;\tfrac{1}{2},0;2.477]_{-1}\oplus\mathrm{CPT})\)} \allowbreak
\mbox{\(\oplus\,(L\bar L[9.178;0,\tfrac{1}{2};2.477]_{-1}\oplus\mathrm{CPT})\)} \allowbreak
\mbox{\(\oplus\,(L\bar L[9.602;\tfrac{1}{2},0;2.477]_{-1}\oplus\mathrm{CPT})\)} \allowbreak
\mbox{\(\oplus\,(L\bar L[10.792;0,\tfrac{1}{2};2.477]_{-1}\oplus\mathrm{CPT})\)} \allowbreak
\mbox{\(\oplus\,(L\bar L[11.993;\tfrac{1}{2},0;2.477]_{-1}\oplus\mathrm{CPT})\)} \arraybackslash \\*[1pt]
Vector & \raggedright \mbox{\(L\bar L[6.941;0,0;0]_{0}\)} \allowbreak
\mbox{\(\oplus\,L\bar L[7.985;0,0;0]_{0}\)} \allowbreak
\mbox{\(\oplus\,L\bar L[9.235;0,0;\pm 2]_{0}\)} \allowbreak
\mbox{\(\oplus\,L\bar L[9.290;0,0;0]_{0}\)} \allowbreak
\mbox{\(\oplus\,L\bar L[10.059;0,0;0]_{0}\)} \allowbreak
\mbox{\(\oplus\,L\bar L[10.369;0,0;0]_{0}\)} \allowbreak
\mbox{\(\oplus\,L\bar L[12.699;0,0;0]_{0}\)} \allowbreak
\mbox{\(\oplus\,L\bar L[8.493;0,0;\pm 1.477]_{\mp 1}\)} \allowbreak
\mbox{\(\oplus\,L\bar L[9.087;0,0;\pm 3.477]_{\mp 1}\)} \allowbreak
\mbox{\(\oplus\,L\bar L[10.188;0,0;\pm 1.477]_{\mp 1}\)} \allowbreak
\mbox{\(\oplus\,L\bar L[11.418;0,0;\pm 3.477]_{\mp 1}\)} \allowbreak
\mbox{\(\oplus\,L\bar L[9.497;0,0;\pm 4.954]_{\mp 2}\)} \allowbreak
\mbox{\(\oplus\,L\bar L[11.477;0,0;\pm 4.954]_{\mp 2}\)} \arraybackslash \\[3pt]
\hline\hline
\multicolumn{2}{l}{\rule{0pt}{13pt}$k=5$}\\*
Graviton & \raggedright \mbox{\(L\bar L[9.651;\tfrac{1}{2},\tfrac{1}{2};0]_{0}\)} \allowbreak
\mbox{\(\oplus\,L\bar L[11.169;\tfrac{1}{2},\tfrac{1}{2};0]_{0}\)} \allowbreak
\mbox{\(\oplus\,L\bar L[13.385;\tfrac{1}{2},\tfrac{1}{2};0]_{0}\)} \allowbreak
\mbox{\(\oplus\,L\bar L[10.474;\tfrac{1}{2},\tfrac{1}{2};\pm 3.477]_{\mp 1}\)} \allowbreak
\mbox{\(\oplus\,L\bar L[12.263;\tfrac{1}{2},\tfrac{1}{2};\pm 3.477]_{\mp 1}\)} \arraybackslash \\*[1pt]
Gravitino & \raggedright \mbox{\((L\bar A_2[10.931;\tfrac{1}{2},0;5.954]_{-2}\oplus\mathrm{CPT})\)} \allowbreak
\mbox{\(\oplus\,(L\bar L[9.263;0,\tfrac{1}{2};1]_{0}\oplus\mathrm{CPT})\)} \allowbreak
\mbox{\(\oplus\,(L\bar L[10.355;0,\tfrac{1}{2};-1]_{0}\oplus\mathrm{CPT})\)} \allowbreak
\mbox{\(\oplus\,(L\bar L[10.578;0,\tfrac{1}{2};1]_{0}\oplus\mathrm{CPT})\)} \allowbreak
\mbox{\(\oplus\,(L\bar L[11.767;0,\tfrac{1}{2};-1]_{0}\oplus\mathrm{CPT})\)} \allowbreak
\mbox{\(\oplus\,(L\bar L[12.780;0,\tfrac{1}{2};1]_{0}\oplus\mathrm{CPT})\)} \allowbreak
\mbox{\(\oplus\,(L\bar L[9.791;\tfrac{1}{2},0;2.477]_{-1}\oplus\mathrm{CPT})\)} \allowbreak
\mbox{\(\oplus\,(L\bar L[10.974;0,\tfrac{1}{2};2.477]_{-1}\oplus\mathrm{CPT})\)} \allowbreak
\mbox{\(\oplus\,(L\bar L[11.654;\tfrac{1}{2},0;2.477]_{-1}\oplus\mathrm{CPT})\)} \allowbreak
\mbox{\(\oplus\,(L\bar L[11.715;0,\tfrac{1}{2};4.477]_{-1}\oplus\mathrm{CPT})\)} \allowbreak
\mbox{\(\oplus\,(L\bar L[12.850;0,\tfrac{1}{2};2.477]_{-1}\oplus\mathrm{CPT})\)} \allowbreak
\mbox{\(\oplus\,(L\bar L[14.095;\tfrac{1}{2},0;2.477]_{-1}\oplus\mathrm{CPT})\)} \allowbreak
\mbox{\(\oplus\,(L\bar L[11.976;0,\tfrac{1}{2};5.954]_{-2}\oplus\mathrm{CPT})\)} \allowbreak
\mbox{\(\oplus\,(L\bar L[12.976;\tfrac{1}{2},0;5.954]_{-2}\oplus\mathrm{CPT})\)} \arraybackslash \\*[1pt]
Vector & \raggedright \mbox{\(L\bar L[8.327;0,0;0]_{0}\)} \allowbreak
\mbox{\(\oplus\,L\bar L[9.951;0,0;0]_{0}\)} \allowbreak
\mbox{\(\oplus\,L\bar L[10.874;0,0;0]_{0}\)} \allowbreak
\mbox{\(\oplus\,L\bar L[11.189;0,0;\pm 2]_{0}\)} \allowbreak
\mbox{\(\oplus\,L\bar L[12.163;0,0;0]_{0}\)} \allowbreak
\mbox{\(\oplus\,L\bar L[12.363;0,0;0]_{0}\)} \allowbreak
\mbox{\(\oplus\,L\bar L[14.802;0,0;0]_{0}\)} \allowbreak
\mbox{\(\oplus\,L\bar L[9.452;0,0;\pm 3.477]_{\mp 1}\)} \allowbreak
\mbox{\(\oplus\,L\bar L[9.478;0,0;\pm 1.477]_{\mp 1}\)} \allowbreak
\mbox{\(\oplus\,L\bar L[10.363;0,0;\pm 1.477]_{\mp 1}\)} \allowbreak
\mbox{\(\oplus\,L\bar L[11.099;0,0;\pm 3.477]_{\mp 1}\)} \allowbreak
\mbox{\(\oplus\,L\bar L[11.562;0,0;\pm 3.477]_{\mp 1}\)} \allowbreak
\mbox{\(\oplus\,L\bar L[12.247;0,0;\pm 1.477]_{\mp 1}\)} \allowbreak
\mbox{\(\oplus\,L\bar L[13.508;0,0;\pm 3.477]_{\mp 1}\)} \allowbreak
\mbox{\(\oplus\,L\bar L[11.396;0,0;\pm 4.954]_{\mp 2}\)} \allowbreak
\mbox{\(\oplus\,L\bar L[13.543;0,0;\pm 4.954]_{\mp 2}\)} \arraybackslash \\*[1pt]
Chiral & \raggedright \mbox{\((L\bar B_1[12.646;0,0;8.431]_{-3}\oplus\mathrm{CPT})\)} \arraybackslash \\[3pt]
\end{longtable}
\endgroup

\begin{table}[!t]
\centering
\begingroup
\fontsize{8}{10}\selectfont
\thinmuskip=2mu\medmuskip=3mu\thickmuskip=4mu
\renewcommand{\arraystretch}{1.08}
\setlength{\tabcolsep}{4pt}
\setlength{\doublerulesep}{0pt}
\caption{The $z=5$ spectrum at $k=0,\ldots,5$.}\label{tab:RepresentativeFive}
\begin{tabular}{@{}p{0.105\textwidth}|p{\dimexpr0.895\textwidth-8.4pt\relax}@{}}
\hline\hline
\textbf{Multiplet} & $z=5$ \\
\hline
\multicolumn{2}{l}{\rule{0pt}{13pt}$k=0$}\\*
Graviton & \raggedright \mbox{\(A_1\bar A_1[3.000;\tfrac{1}{2},\tfrac{1}{2};0]_{0}\)} \arraybackslash \\*[2pt]
Vector & \raggedright \mbox{\(A_2\bar A_2[2.000;0,0;0]_{0}\)} \allowbreak
\mbox{\(\oplus\,L\bar L[4.430;0,0;0]_{0}\)} \arraybackslash \\[3pt]
\hline\hline
\multicolumn{2}{l}{\rule{0pt}{13pt}$k=1$}\\*
Graviton & \raggedright \mbox{\(L\bar L[5.000;\tfrac{1}{2},\tfrac{1}{2};0]_{0}\)} \arraybackslash \\*[2pt]
Gravitino & \raggedright \mbox{\((L\bar A_1[4.500;0,\tfrac{1}{2};1]_{0}\oplus\mathrm{CPT})\)} \arraybackslash \\*[2pt]
Vector & \raggedright \mbox{\(L\bar L[3.532;0,0;0]_{0}\)} \allowbreak
\mbox{\(\oplus\,L\bar L[6.510;0,0;0]_{0}\)} \arraybackslash \\*[2pt]
Chiral & \raggedright \mbox{\((L\bar B_1[5.272;0,0;3.515]_{-1}\oplus\mathrm{CPT})\)} \arraybackslash \\[3pt]
\hline\hline
\multicolumn{2}{l}{\rule{0pt}{13pt}$k=2$}\\*
Graviton & \raggedright \mbox{\(L\bar L[5.457;\tfrac{1}{2},\tfrac{1}{2};0]_{0}\)} \allowbreak
\mbox{\(\oplus\,L\bar L[7.113;\tfrac{1}{2},\tfrac{1}{2};0]_{0}\)} \arraybackslash \\*[2pt]
Gravitino & \raggedright \mbox{\((L\bar L[6.500;0,\tfrac{1}{2};1]_{0}\oplus\mathrm{CPT})\)} \arraybackslash \\*[2pt]
Vector & \raggedright \mbox{\((L\bar A_2[7.272;0,0;3.515]_{-1}\oplus\mathrm{CPT})\)} \allowbreak
\mbox{\(\oplus\,L\bar L[4.253;0,0;0]_{0}\)} \allowbreak
\mbox{\(\oplus\,L\bar L[5.836;0,0;0]_{0}\)} \allowbreak
\mbox{\(\oplus\,L\bar L[6.643;0,0;0]_{0}\)} \allowbreak
\mbox{\(\oplus\,L\bar L[8.628;0,0;0]_{0}\)} \arraybackslash \\*[2pt]
Tensor & \raggedright \mbox{\((L\bar B_1[6.772;\tfrac{1}{2},0;4.515]_{-1}\oplus\mathrm{CPT})\)} \arraybackslash \\[3pt]
\hline\multicolumn{2}{r}{\textit{Continued on the next page}}\\
\end{tabular}
\endgroup
\end{table}

\begin{samepage}
\subsection{\texorpdfstring{$z=5$}{z=5}: a generic exterior value}
\label{sec:RepresentativeFive}

The generic exterior value $z=5$ lies outside the distinguished sequence (\ref{eq:Distinguishedz}). The primitive data are $(\widehat p,\widehat q)=(3,-2)$, giving $\mu_z = \lambda_z =5$, $f_z =1$ and $\epsilon=(1-2\sqrt{19})/15$. Table~\ref{tab:RepresentativeFive} collects all contributing multiplets at KK levels $k=0 , \ldots , 5$ from tables~\ref{tab:FirstZ-neutral-short},~\ref{tab:FirstZ-neutral-long},~\ref{tab:ExteriorProtectedMultiplets} and \ref{tab:NewChargedExteriorLong}. At generic exterior values, all protected families in table~\ref{tab:ExteriorProtectedMultiplets} have $\nu_{\textrm{min}}=1$. In particular, the short chiral and $\eta=-2$ vector families now start at $\nu_{\textrm{min}}=1$, rather than $\nu_{\textrm{min}}=2$ as in the distinguished example $z=3$. For that reason, the first chiral multiplet,
$L\bar B_1[\tfrac{22+\sqrt{19}}{5};0,0;\tfrac{44+2\sqrt{19}}{15}]_{-1}$
along with its CPT conjugate, occurs for $z=5$ already at KK level $k=1$, in contrast to the analogous multiplet at $z=3$, which arises at $k=2$ as discussed in section~\ref{sec:RepresentativeThree}.

For the long vector family with $\eta=-2$ and $\nu=1$, the lower limit is now $\kappa_{\textrm{min}}=6$, as for every generic exterior value of $z$. The formal $\kappa=4$ polynomial does not correspond to a physical vector pair, while the $\kappa=5$ polynomial describes the short vector pair at $k=2$. The first two long pairs correspond to $\kappa=6$ and $a=1,2$, and occur at $k=3$. This differs from the distinguished case $z=3$, where the family starts at $\kappa=4$ and contributes one long pair at $k=1$ in table~\ref{tab:RepresentativeThree}. The $\eta=0$ vector family provides another illustration of the separation into short and long multiplets: at $(\kappa,\nu)=(2,1)$, its two polynomial roots describe one short pair and one long pair, both at $k=5$. Several flavour-charged families first appear beyond the $k=5$ cutoff of table~\ref{tab:RepresentativeFive}. The protected gravitino tower with $k=5\nu+1$ in table~\ref{tab:ExteriorProtectedMultiplets} and the long graviton family in table~\ref{tab:NewChargedExteriorLong} both begin at $k=6$. The long $\eta=1$ gravitino family begins at $k=7$, while the long $\eta=2$ vector family begins at $k=9$.

\par\end{samepage}

\begin{table}[!t]
\centering
\begingroup
\fontsize{8}{10}\selectfont
\thinmuskip=2mu\medmuskip=3mu\thickmuskip=4mu
\renewcommand{\arraystretch}{1.08}
\setlength{\tabcolsep}{4pt}
\setlength{\doublerulesep}{0pt}
\captionsetup{skip=0pt}
\caption*{Table~\ref{tab:RepresentativeFive} (continued): $z=5$}
\begin{tabular}{@{}p{0.105\textwidth}|p{\dimexpr0.895\textwidth-8.4pt\relax}@{}}
\hline\hline
\textbf{Multiplet} & Representations \\
\hline
\multicolumn{2}{l}{\rule{0pt}{13pt}$k=3$}\\*
Graviton & \raggedright \mbox{\(L\bar L[7.122;\tfrac{1}{2},\tfrac{1}{2};0]_{0}\)} \allowbreak
\mbox{\(\oplus\,L\bar L[9.231;\tfrac{1}{2},\tfrac{1}{2};0]_{0}\)} \arraybackslash \\*[2pt]
Gravitino & \raggedright \mbox{\((L\bar A_2[8.772;\tfrac{1}{2},0;4.515]_{-1}\oplus\mathrm{CPT})\)} \allowbreak
\mbox{\(\oplus\,(L\bar L[6.691;0,\tfrac{1}{2};1]_{0}\oplus\mathrm{CPT})\)} \allowbreak
\mbox{\(\oplus\,(L\bar L[7.798;0,\tfrac{1}{2};-1]_{0}\oplus\mathrm{CPT})\)} \allowbreak
\mbox{\(\oplus\,(L\bar L[8.607;0,\tfrac{1}{2};1]_{0}\oplus\mathrm{CPT})\)} \arraybackslash \\*[2pt]
Vector & \raggedright \mbox{\(L\bar L[5.699;0,0;0]_{0}\)} \allowbreak
\mbox{\(\oplus\,L\bar L[7.945;0,0;0]_{0}\)} \allowbreak
\mbox{\(\oplus\,L\bar L[8.387;0,0;0]_{0}\)} \allowbreak
\mbox{\(\oplus\,L\bar L[10.753;0,0;0]_{0}\)} \allowbreak
\mbox{\(\oplus\,L\bar L[7.391;0,0;\pm 3.515]_{\mp 1}\)} \allowbreak
\mbox{\(\oplus\,L\bar L[9.347;0,0;\pm 3.515]_{\mp 1}\)} \arraybackslash \\[3pt]
\hline\hline
\multicolumn{2}{l}{\rule{0pt}{13pt}$k=4$}\\*
Graviton & \raggedright \mbox{\(L\bar L[7.951;\tfrac{1}{2},\tfrac{1}{2};0]_{0}\)} \allowbreak
\mbox{\(\oplus\,L\bar L[9.087;\tfrac{1}{2},\tfrac{1}{2};0]_{0}\)} \allowbreak
\mbox{\(\oplus\,L\bar L[11.355;\tfrac{1}{2},\tfrac{1}{2};0]_{0}\)} \arraybackslash \\*[2pt]
Gravitino & \raggedright \mbox{\((L\bar A_2[8.772;\tfrac{1}{2},0;4.515]_{-1}\oplus\mathrm{CPT})\)} \allowbreak
\mbox{\(\oplus\,(L\bar L[8.484;0,\tfrac{1}{2};1]_{0}\oplus\mathrm{CPT})\)} \allowbreak
\mbox{\(\oplus\,(L\bar L[9.710;0,\tfrac{1}{2};-1]_{0}\oplus\mathrm{CPT})\)} \allowbreak
\mbox{\(\oplus\,(L\bar L[10.726;0,\tfrac{1}{2};1]_{0}\oplus\mathrm{CPT})\)} \allowbreak
\mbox{\(\oplus\,(L\bar L[9.846;0,\tfrac{1}{2};4.515]_{-1}\oplus\mathrm{CPT})\)} \allowbreak
\mbox{\(\oplus\,(L\bar L[10.846;\tfrac{1}{2},0;4.515]_{-1}\oplus\mathrm{CPT})\)} \arraybackslash \\*[2pt]
Vector & \raggedright \mbox{\(L\bar L[6.691;0,0;0]_{0}\)} \allowbreak
\mbox{\(\oplus\,L\bar L[7.854;0,0;0]_{0}\)} \allowbreak
\mbox{\(\oplus\,L\bar L[9.143;0,0;0]_{0}\)} \allowbreak
\mbox{\(\oplus\,L\bar L[9.153;0,0;\pm 2]_{0}\)} \allowbreak
\mbox{\(\oplus\,L\bar L[10.073;0,0;0]_{0}\)} \allowbreak
\mbox{\(\oplus\,L\bar L[10.330;0,0;0]_{0}\)} \allowbreak
\mbox{\(\oplus\,L\bar L[12.881;0,0;0]_{0}\)} \allowbreak
\mbox{\(\oplus\,L\bar L[9.220;0,0;\pm 3.515]_{\mp 1}\)} \allowbreak
\mbox{\(\oplus\,L\bar L[11.446;0,0;\pm 3.515]_{\mp 1}\)} \arraybackslash \\[3pt]
\hline\hline
\multicolumn{2}{l}{\rule{0pt}{13pt}$k=5$}\\*
Graviton & \raggedright \mbox{\((L\bar A_1[11.272;\tfrac{1}{2},\tfrac{1}{2};5.515]_{-1}\oplus\mathrm{CPT})\)} \allowbreak
\mbox{\(\oplus\,L\bar L[9.436;\tfrac{1}{2},\tfrac{1}{2};0]_{0}\)} \allowbreak
\mbox{\(\oplus\,L\bar L[11.099;\tfrac{1}{2},\tfrac{1}{2};0]_{0}\)} \allowbreak
\mbox{\(\oplus\,L\bar L[13.481;\tfrac{1}{2},\tfrac{1}{2};0]_{0}\)} \arraybackslash \\*[2pt]
Gravitino & \raggedright \mbox{\((L\bar L[9.069;0,\tfrac{1}{2};1]_{0}\oplus\mathrm{CPT})\)} \allowbreak
\mbox{\(\oplus\,(L\bar L[10.184;0,\tfrac{1}{2};-1]_{0}\oplus\mathrm{CPT})\)} \allowbreak
\mbox{\(\oplus\,(L\bar L[10.487;0,\tfrac{1}{2};1]_{0}\oplus\mathrm{CPT})\)} \allowbreak
\mbox{\(\oplus\,(L\bar L[11.722;0,\tfrac{1}{2};-1]_{0}\oplus\mathrm{CPT})\)} \allowbreak
\mbox{\(\oplus\,(L\bar L[12.850;0,\tfrac{1}{2};1]_{0}\oplus\mathrm{CPT})\)} \allowbreak
\mbox{\(\oplus\,(L\bar L[10.648;\tfrac{1}{2},0;4.515]_{-1}\oplus\mathrm{CPT})\)} \allowbreak
\mbox{\(\oplus\,(L\bar L[11.821;0,\tfrac{1}{2};4.515]_{-1}\oplus\mathrm{CPT})\)} \allowbreak
\mbox{\(\oplus\,(L\bar L[12.945;\tfrac{1}{2},0;4.515]_{-1}\oplus\mathrm{CPT})\)} \arraybackslash \\*[2pt]
Vector & \raggedright \mbox{\((L\bar A_2[10.272;0,0;5.515]_{-1}\oplus\mathrm{CPT})\)} \allowbreak
\mbox{\(\oplus\,L\bar L[8.022;0,0;0]_{0}\)} \allowbreak
\mbox{\(\oplus\,L\bar L[9.829;0,0;0]_{0}\)} \allowbreak
\mbox{\(\oplus\,L\bar L[10.716;0,0;0]_{0}\)} \allowbreak
\mbox{\(\oplus\,L\bar L[11.123;0,0;\pm 2]_{0}\)} \allowbreak
\mbox{\(\oplus\,L\bar L[12.200;0,0;0]_{0}\)} \allowbreak
\mbox{\(\oplus\,L\bar L[12.348;0,0;0]_{0}\)} \allowbreak
\mbox{\(\oplus\,L\bar L[15.010;0,0;0]_{0}\)} \allowbreak
\mbox{\(\oplus\,L\bar L[9.735;0,0;\pm 3.515]_{\mp 1}\)} \allowbreak
\mbox{\(\oplus\,L\bar L[11.203;0,0;\pm 3.515]_{\mp 1}\)} \allowbreak
\mbox{\(\oplus\,L\bar L[12.394;0,0;\pm 5.515]_{\mp 1}\)} \allowbreak
\mbox{\(\oplus\,L\bar L[13.557;0,0;\pm 3.515]_{\mp 1}\)} \arraybackslash \\[3pt]
\hline\hline
\end{tabular}
\endgroup
\end{table}

\begin{samepage}
\subsection{The large-\texorpdfstring{$|z|$}{|z|} limit at fixed KK level}
\label{sec:RepresentativeLargeZ}

We conclude with the large-$|z|$ limit of the spectrum (\ref{eq:FullSpec}) at fixed KK level. In the exterior, $\mu_z=\lambda_z=|z|f_z$ tends to infinity because $f_z$ is a positive integer. The level formula (\ref{eq:GeneralKKLevel}) therefore sends charged multiplets with fixed allowed labels to increasingly large $k$. Since $\nu\geq1$ and the remaining labels are bounded below, no charged multiplets remain at any fixed $k$ for sufficiently large $|z|$. In conclusion, only the flavour-neutral multiplets of tables~\ref{tab:FirstZ-neutral-short}~and~\ref{tab:FirstZ-neutral-long} survive the large $|z|$, fixed $k$ limit. The protected dimensions of table \ref{tab:FirstZ-neutral-short} are independent of $z$ and thus remain unchanged. The long dimensions of table \ref{tab:FirstZ-neutral-long} generally vary with $z$, but this dependence enters exclusively through $\epsilon(z)^2$. Since $\epsilon(z)^2\to\frac13$ by (\ref{eq:SpectrumExactRCharge}), the polynomials of appendix~\ref{app:NeutralPolynomials} evaluated at $\epsilon^2 = \frac13$ also produce finite, $z$-independent limiting dimensions. This is in agreement with the horizontal asymptotes of the curves in figure~\ref{fig:BBBWF0SpecKappa}. Table~\ref{tab:RepresentativeLargeZ} lists the resulting spectrum at $k=0,\ldots,5$.

\par\end{samepage}

\begin{table}[!t]
\centering
\begingroup
\fontsize{8}{10}\selectfont
\thinmuskip=2mu\medmuskip=3mu\thickmuskip=4mu
\renewcommand{\arraystretch}{1.08}
\setlength{\tabcolsep}{4pt}
\setlength{\doublerulesep}{0pt}
\caption{The fixed-$k$, large-$|z|$ limit of the spectrum at $k=0,\ldots,5$.}\label{tab:RepresentativeLargeZ}
\begin{tabular}{@{}p{0.105\textwidth}|p{\dimexpr0.895\textwidth-8.4pt\relax}@{}}
\hline\hline
\textbf{Multiplet} & $|z|\to\infty$ at fixed $k$ \\
\hline
\multicolumn{2}{l}{\rule{0pt}{13pt}$k=0$}\\*
Graviton & \raggedright \mbox{\(A_1\bar A_1[3.000;\tfrac{1}{2},\tfrac{1}{2};0]_{0}\)} \arraybackslash \\*[2pt]
Vector & \raggedright \mbox{\(A_2\bar A_2[2.000;0,0;0]_{0}\)} \allowbreak
\mbox{\(\oplus\,L\bar L[4.606;0,0;0]_{0}\)} \arraybackslash \\[3pt]
\hline\hline
\multicolumn{2}{l}{\rule{0pt}{13pt}$k=1$}\\*
Graviton & \raggedright \mbox{\(L\bar L[5.000;\tfrac{1}{2},\tfrac{1}{2};0]_{0}\)} \arraybackslash \\*[2pt]
Gravitino & \raggedright \mbox{\((L\bar A_1[4.500;0,\tfrac{1}{2};1]_{0}\oplus\mathrm{CPT})\)} \arraybackslash \\*[2pt]
Vector & \raggedright \mbox{\(L\bar L[3.363;0,0;0]_{0}\)} \allowbreak
\mbox{\(\oplus\,L\bar L[6.694;0,0;0]_{0}\)} \arraybackslash \\[3pt]
\hline\hline
\multicolumn{2}{l}{\rule{0pt}{13pt}$k=2$}\\*
Graviton & \raggedright \mbox{\(L\bar L[5.266;\tfrac{1}{2},\tfrac{1}{2};0]_{0}\)} \allowbreak
\mbox{\(\oplus\,L\bar L[7.148;\tfrac{1}{2},\tfrac{1}{2};0]_{0}\)} \arraybackslash \\*[2pt]
Gravitino & \raggedright \mbox{\((L\bar L[6.500;0,\tfrac{1}{2};1]_{0}\oplus\mathrm{CPT})\)} \arraybackslash \\*[2pt]
Vector & \raggedright \mbox{\(L\bar L[3.986;0,0;0]_{0}\)} \allowbreak
\mbox{\(\oplus\,L\bar L[5.786;0,0;0]_{0}\)} \allowbreak
\mbox{\(\oplus\,L\bar L[6.532;0,0;0]_{0}\)} \allowbreak
\mbox{\(\oplus\,L\bar L[8.847;0,0;0]_{0}\)} \arraybackslash \\[3pt]
\hline\hline
\multicolumn{2}{l}{\rule{0pt}{13pt}$k=3$}\\*
Graviton & \raggedright \mbox{\(L\bar L[6.919;\tfrac{1}{2},\tfrac{1}{2};0]_{0}\)} \allowbreak
\mbox{\(\oplus\,L\bar L[9.305;\tfrac{1}{2},\tfrac{1}{2};0]_{0}\)} \arraybackslash \\*[2pt]
Gravitino & \raggedright \mbox{\((L\bar L[6.500;0,\tfrac{1}{2};1]_{0}\oplus\mathrm{CPT})\)} \allowbreak
\mbox{\(\oplus\,(L\bar L[7.641;0,\tfrac{1}{2};-1]_{0}\oplus\mathrm{CPT})\)} \allowbreak
\mbox{\(\oplus\,(L\bar L[8.641;0,\tfrac{1}{2};1]_{0}\oplus\mathrm{CPT})\)} \arraybackslash \\*[2pt]
Vector & \raggedright \mbox{\(L\bar L[5.341;0,0;0]_{0}\)} \allowbreak
\mbox{\(\oplus\,L\bar L[7.920;0,0;0]_{0}\)} \allowbreak
\mbox{\(\oplus\,L\bar L[8.282;0,0;0]_{0}\)} \allowbreak
\mbox{\(\oplus\,L\bar L[11.012;0,0;0]_{0}\)} \arraybackslash \\[3pt]
\hline\hline
\multicolumn{2}{l}{\rule{0pt}{13pt}$k=4$}\\*
Graviton & \raggedright \mbox{\(L\bar L[7.606;\tfrac{1}{2},\tfrac{1}{2};0]_{0}\)} \allowbreak
\mbox{\(\oplus\,L\bar L[8.922;\tfrac{1}{2},\tfrac{1}{2};0]_{0}\)} \allowbreak
\mbox{\(\oplus\,L\bar L[11.469;\tfrac{1}{2},\tfrac{1}{2};0]_{0}\)} \arraybackslash \\*[2pt]
Gravitino & \raggedright \mbox{\((L\bar L[8.284;0,\tfrac{1}{2};1]_{0}\oplus\mathrm{CPT})\)} \allowbreak
\mbox{\(\oplus\,(L\bar L[9.583;0,\tfrac{1}{2};-1]_{0}\oplus\mathrm{CPT})\)} \allowbreak
\mbox{\(\oplus\,(L\bar L[10.799;0,\tfrac{1}{2};1]_{0}\oplus\mathrm{CPT})\)} \arraybackslash \\*[2pt]
Vector & \raggedright \mbox{\(L\bar L[6.244;0,0;0]_{0}\)} \allowbreak
\mbox{\(\oplus\,L\bar L[7.625;0,0;0]_{0}\)} \allowbreak
\mbox{\(\oplus\,L\bar L[8.870;0,0;0]_{0}\)} \allowbreak
\mbox{\(\oplus\,L\bar L[9.000;0,0;\pm 2]_{0}\)} \allowbreak
\mbox{\(\oplus\,L\bar L[10.083;0,0;0]_{0}\)} \allowbreak
\mbox{\(\oplus\,L\bar L[10.263;0,0;0]_{0}\)} \allowbreak
\mbox{\(\oplus\,L\bar L[13.181;0,0;0]_{0}\)} \arraybackslash \\[3pt]
\hline\hline
\multicolumn{2}{l}{\rule{0pt}{13pt}$k=5$}\\*
Graviton & \raggedright \mbox{\(L\bar L[9.053;\tfrac{1}{2},\tfrac{1}{2};0]_{0}\)} \allowbreak
\mbox{\(\oplus\,L\bar L[10.973;\tfrac{1}{2},\tfrac{1}{2};0]_{0}\)} \allowbreak
\mbox{\(\oplus\,L\bar L[13.637;\tfrac{1}{2},\tfrac{1}{2};0]_{0}\)} \arraybackslash \\*[2pt]
Gravitino & \raggedright \mbox{\((L\bar L[8.712;0,\tfrac{1}{2};1]_{0}\oplus\mathrm{CPT})\)} \allowbreak
\mbox{\(\oplus\,(L\bar L[9.865;0,\tfrac{1}{2};-1]_{0}\oplus\mathrm{CPT})\)} \allowbreak
\mbox{\(\oplus\,(L\bar L[10.323;0,\tfrac{1}{2};1]_{0}\oplus\mathrm{CPT})\)} \allowbreak
\mbox{\(\oplus\,(L\bar L[11.635;0,\tfrac{1}{2};-1]_{0}\oplus\mathrm{CPT})\)} \allowbreak
\mbox{\(\oplus\,(L\bar L[12.965;0,\tfrac{1}{2};1]_{0}\oplus\mathrm{CPT})\)} \arraybackslash \\*[2pt]
Vector & \raggedright \mbox{\(L\bar L[7.488;0,0;0]_{0}\)} \allowbreak
\mbox{\(\oplus\,L\bar L[9.616;0,0;0]_{0}\)} \allowbreak
\mbox{\(\oplus\,L\bar L[10.428;0,0;0]_{0}\)} \allowbreak
\mbox{\(\oplus\,L\bar L[11.000;0,0;\pm 2]_{0}\)} \allowbreak
\mbox{\(\oplus\,L\bar L[12.235;0,0;0]_{0}\)} \allowbreak
\mbox{\(\oplus\,L\bar L[12.338;0,0;0]_{0}\)} \allowbreak
\mbox{\(\oplus\,L\bar L[15.352;0,0;0]_{0}\)} \arraybackslash \\[3pt]
\hline\hline
\end{tabular}
\endgroup
\end{table}

\section{Discussion}
\label{sec:Discussion}

We have determined a sector of the light single-trace operator spectrum at large $N$ that is universally present in the four-dimensional $\cN=1$ class $\cS$ SCFTs dual to the BBBW wrapped-M5 backgrounds~\cite{Bah:2011vv,Bah:2012dg}. This extends our analyses of MN2 \cite{Bhattacharya:2024tjw} and MN1 \cite{Bhattacharya:2026ypg} to general twists $z$. Using the trombone KK operators of~\cite{Pico:2026kch}, we have constructed mass matrices directly on the globally defined $\mathrm{U}(1)_z$-invariant spaces. Here $\mathrm{U}(1)_z$ is the structure group of the generalised structure admitted by the BBBW AdS$_5$ configurations in ExGG \cite{Cassani:2019vcl,Cassani:2020cod}. Within our construction, this invariance selects modes that are constant on the hyperbolic, unpunctured Riemann surface $\Sigma_2$ wrapped by the M5-branes. The resulting spectrum, presented in section~\ref{sec:UniversalSpectrum}, therefore consists of physical modes that are globally defined on the internal six-dimensional $S^4$ fibration over $\Sigma_2$, for all $z$ and all $\Sigma_2$.

Our restriction to expansion coefficients that are constant over $\Sigma_2$ does not necessarily capture the complete universal spectrum over a given BBBW background. Modes with non-trivial dependence on the surface can also have universal dimensions and charges, provided that, at fixed $z$, they exist with these same values for every compact hyperbolic quotient compatible with the twisting data. Their mass eigenfunctions may therefore depend on the choice of $\Sigma_2$, while the corresponding mass eigenvalues remain unchanged. The graviton analyses of MN2 and MN1 in \cite{Bhattacharya:2024tjw,Bhattacharya:2026ypg} illustrate this possibility. In these cases, the dependence on $\Sigma_2$ is governed by a weighted Maass eigenvalue problem. Certain families of Maass eigenvalues are fixed by the weight and discrete quantum numbers, independently of $\Gamma$, and consequently give rise to universal graviton masses despite the non-trivial surface dependence of their eigenfunctions. More generally, the complete KK spectrum on a specified surface $\Sigma_2=H^2/\Gamma$ can, in principle, be determined by solving the full fluctuation equations, including the corresponding spectral problems on $\Sigma_2$. This calculation must be carried out on a case-by-case basis for each quotient and will also include non-universal modes whose mass eigenvalues depend on the particular choice of $\Gamma$.

The additional universal families with nonconstant coefficients on $\Sigma_2$ alluded to above could also contain short multiplets. Therefore, our results do not necessarily exhaust the universal protected sector of the dual SCFTs. The short multiplets that we find in this paper provide candidates for identification with protected operators and determine contributions to the single-particle superconformal index~\cite{Beem:2012yn}. Obtaining the full index would also require the contributing multiplets from the sectors that we have omitted in this work. In particular, the absence of a family of chiral multiplets from our tables does not establish its absence from the SCFT, or even from its universal spectrum, since such a family could arise among modes with non-trivial dependence on $\Sigma_2$. For the same reason, the present results alone do not determine completely the chiral ring or the conformal manifold. It would be interesting to derive the charged shortening conditions directly from the SCFT constructions of~\cite{Benini:2009mz,Bah:2012dg,Beem:2012yn} and identify which protected operators are represented by modes with constant coefficients on $\Sigma_2$ and which require additional sectors. 


\section*{Acknowledgements}


This work was supported by NSF grant PHY-2609785.

\appendix

\addtocontents{toc}{\setcounter{tocdepth}{1}}


\section{\texorpdfstring{ $\mathrm{U}(1)_z$-invariant tensors and mass matrices}{Invariant tensors and U(1)z-invariant mass matrices}}
\label{app:InvariantMassMatrices}

The construction of section~\ref{sec:U1zMassMatrices} requires tensors that are annihilated by the generator $Q_z$ in (\ref{eq:Q_zCharge}). At KK level $k$, we seek these tensors in the products of the five-dimensional field representations $\cR_s$ with the SO$(5)$ representation $[k0]$ of scalar harmonics on $S^4$:
\begin{equation}
\label{eq:SeedTimesHarmonicRepresentations}
\begin{array}{c|c|c|c}
\textrm{field} & s & \cR_s\otimes[k0] & \textrm{invariant tensor} \\ \hline
h_{\mu\nu} & 2 & \bm1\otimes[k0] & K_\alpha{}^\Lambda \\[2pt]
\psi_\mu{}^i & \frac32 & \overline{\bm8}\otimes[k0] & K_\beta{}^{i\Lambda} \\[2pt]
A_\mu{}^{\uM} & 1 & \overline{\bm{27}}\otimes[k0] & K_\gamma{}^{\uM\Lambda} \\[2pt]
B_{\mu\nu\,\uM} & B & \bm{27}\otimes[k0] & \widetilde K_{\gamma\,\uM}{}^\Lambda \\[2pt]
\chi^{ijk} & \frac12 & \overline{\bm{48}}\otimes[k0] & K_\delta{}^{ijk\Lambda} \\[2pt]
\phi^{ijkl} & 0 & \bm{42}\otimes[k0] & K_\zeta{}^{ijkl\Lambda}
\end{array}
\; .
\end{equation}
Here $i,j,\ldots=1,\ldots,8$ are USp$(8)$ indices, $\uM,\uN,\ldots=1,\ldots,27$ are E$_{6(6)}$ indices, and $\Lambda,\Sigma,\ldots$ label a basis of $[k0]$. The labels $\alpha,\beta,\gamma,\delta,\zeta$ run over the independent singlets in the respective products. The vector and two-form spaces have the same dimension, but their tensors carry dual E$_{6(6)}$ indices. We suppress the dependence of the tensors on $k$ and $z$. The putative KK fluctuations lie in the representations listed in (\ref{eq:SeedTimesHarmonicRepresentations}).

\subsection{The generators in the fundamental representations}
\label{sec:InvariantFundamentalGenerators}

We first represent $Q_z=R_0+zF+J_0$ on the fundamental $\bm8$ of USp$(8)$. Its SU$(4)\times$U$(1)_{J_0}$ branching is $\bm4_{+1}\oplus\overline{\bm4}_{-1}$. Further restriction to SU$(2)_+\times$SU$(2)_-$ gives
\begin{equation}
\label{eq:InvariantEightBranching}
\bm8\longrightarrow
\big[(\bm2,\bm1)\oplus(\bm1,\bm2)\big]_{+1}
\oplus
\big[(\bm2,\bm1)\oplus(\bm1,\bm2)\big]_{-1}
\; ,
\end{equation}
where the entries refer to SU$(2)_+\times$SU$(2)_-$ and the subscript is $J_0$. As in the main text, their Cartans are $F$ and $R_0$, respectively, with doublet weights $\pm1$. Choose the order $(F,R_0)=(1,0),(-1,0),(0,1),(0,-1)$ in each of the two $J_0$ sectors, taking $J_0=+1$ first. The three commuting generators are then
\begin{equation}
\label{eq:InvariantEightCartans}
\begin{aligned}
F\big|_{\bm8}&=\operatorname{diag}(1,-1,0,0,1,-1,0,0)\; ,\\
R_0\big|_{\bm8}&=\operatorname{diag}(0,0,1,-1,0,0,1,-1)\; ,\\
J_0\big|_{\bm8}&=\operatorname{diag}(1,1,1,1,-1,-1,-1,-1)\; .
\end{aligned}
\end{equation}
Their combination in (\ref{eq:Q_zCharge}) gives
\begin{equation}
\label{eq:InvariantQEight}
Q_8\equiv Q_z\big|_{\bm8}
=\operatorname{diag}(1+z,1-z,2,0,-1+z,-1-z,0,-2)
\; .
\end{equation}
This is a choice of weight basis for the specified embedding. A compatible symplectic form has $\Omega_{16}=\Omega_{25}=\Omega_{38}=\Omega_{47}=1$, their antisymmetric partners, and all other entries zero. It obeys $Q_8^T\Omega+\Omega Q_8=0$. The conjugate USp$(8)$ representations in (\ref{eq:SeedTimesHarmonicRepresentations}) are identified with the indicated upper-index tensors using this symplectic form.

For the harmonic factor, use Cartesian coordinates $y_a$, $a=1,\ldots,5$, on $\mathbb R^5$, restricted to $\sum_a y_a^2=1$. The generator $J_0$ acts trivially on these coordinates. With $L_{ab}=y_a\partial_b-y_b\partial_a$, the coordinate convention of the calculation is $T_p=-iL_{12}$, $T_q=iL_{34}$. Equation~(\ref{eq:R0FfromJpm}) therefore gives
\begin{equation}
\label{eq:InvariantHarmonicDifferentialCharge}
\widehat Q_z=R_0+zF
=(1+z)T_p+(1-z)T_q
=-i\big[(1+z)L_{12}+(z-1)L_{34}\big]
\; .
\end{equation}
The linear polynomials $y_1+iy_2$, $y_3+iy_4$ and $y_5$ have charges $1+z$, $z-1$ and zero, respectively. On their coefficient vectors, (\ref{eq:InvariantHarmonicDifferentialCharge}) is represented by
\begin{equation}
\label{eq:InvariantQFive}
Q_5=
\begin{pmatrix}
0&-i(1+z)&0&0&0\\
i(1+z)&0&0&0&0\\
0&0&0&-i(z-1)&0\\
0&0&i(z-1)&0&0\\
0&0&0&0&0
\end{pmatrix}
\; .
\end{equation}

\subsection{The induced generators and invariant tensors}
\label{sec:InvariantChargeKernels}

The remaining generators follow by acting on each tensor index. In a USp$(8)$ frame, the vector, spin-$\frac12$ and scalar seeds are antisymmetric, symplectic-traceless tensors of ranks two, three and four, respectively. Their generators act as
\begin{align}
\label{eq:InvariantQTwoIndices}
(Q_{27}A)^{ij}
&=(Q_8)^i{}_{\ell}A^{\ell j}+(Q_8)^j{}_{\ell}A^{i\ell}\; ,\\
\label{eq:InvariantQThreeIndices}
(Q_{48}\chi)^{ijk}
&=(Q_8)^i{}_{\ell}\chi^{\ell jk}
+(Q_8)^j{}_{\ell}\chi^{i\ell k}
+(Q_8)^k{}_{\ell}\chi^{ij\ell}\; ,\\
\label{eq:InvariantQFourIndices}
(Q_{42}\phi)^{ijkl}
&=(Q_8)^i{}_{m}\phi^{mjkl}
+(Q_8)^j{}_{m}\phi^{imkl}
+(Q_8)^k{}_{m}\phi^{ijml}
+(Q_8)^l{}_{m}\phi^{ijkm}\; .
\end{align}
These actions preserve the conditions $\Omega_{ij}A^{ij}=0$, $\Omega_{ij}\chi^{ijk}=0$ and $\Omega_{ij}\phi^{ijkl}=0$. Dressing the vector index with the vacuum coset representative, $A^{ij}=\mathcal V_{\uM}{}^{ij}A^{\uM}$, defines the same generator in the field coordinates of (\ref{eq:SeedTimesHarmonicRepresentations}):
\begin{equation}
\label{eq:InvariantQVectorFrame}
\mathcal V_{\uM}{}^{ij}(Q_{27})^{\uM}{}_{\uN}
=(Q_8)^i{}_{\ell}\mathcal V_{\uN}{}^{\ell j}
+(Q_8)^j{}_{\ell}\mathcal V_{\uN}{}^{i\ell}
\; .
\end{equation}
On a lower two-form index the action is the dual one, $(Q_z B)_{\uM}=-(Q_{27})^{\uN}{}_{\uM}B_{\uN}$. The coset representative and all field-coordinate changes are the same as those used for the parent mass operators.

Similarly, $[k0]=\operatorname{Sym}^k_0\bm5$ is realised by symmetric traceless tensors $H^{a_1\ldots a_k}$. Differentiating the action of an SO$(5)$ matrix on their $k$ indices gives
\begin{equation}
\label{eq:InvariantQHarmonicTensor}
(Q_{[k0]}H)^{a_1\ldots a_k}
=\sum_{r=1}^k(Q_5)^{a_r}{}_{b}
H^{a_1\ldots a_{r-1}b\,a_{r+1}\ldots a_k}
\; .
\end{equation}
This preserves symmetry and tracelessness; the sum is zero at $k=0$. In the chosen harmonic basis, its matrix is fixed by
\begin{equation}
\label{eq:InvariantQHarmonicBasis}
\widehat Q_z\cY_\Sigma(y)
=\cY_\Lambda(y)(Q_{[k0]})^\Lambda{}_{\Sigma}
\; .
\end{equation}
Thus $Q_{[k0]}$ acts on harmonic coefficient columns, with the same convention as $Q_5$ in (\ref{eq:InvariantQFive}).

The desired tensors are now obtained by setting the total charge in each product (\ref{eq:SeedTimesHarmonicRepresentations}) to zero:
\begin{equation}
\label{eq:InvariantKernelGraviton}
\begin{aligned}
0&=(Q_{[k0]})^\Lambda{}_{\Sigma}K_\alpha{}^\Sigma\; ,\\
0&=(Q_8)^i{}_{j}K_\beta{}^{j\Lambda}
+(Q_{[k0]})^\Lambda{}_{\Sigma}K_\beta{}^{i\Sigma}\; ,\\
0&=(Q_{27})^{\uM}{}_{\uN}K_\gamma{}^{\uN\Lambda}
+(Q_{[k0]})^\Lambda{}_{\Sigma}K_\gamma{}^{\uM\Sigma}\; ,\\
0&=-(Q_{27})^{\uN}{}_{\uM}\widetilde K_{\gamma\,\uN}{}^\Lambda
+(Q_{[k0]})^\Lambda{}_{\Sigma}\widetilde K_{\gamma\,\uM}{}^\Sigma\; ,\\
0&=(Q_{48})^{ijk}{}_{\ell mn}K_\delta{}^{\ell mn\Lambda}
+(Q_{[k0]})^\Lambda{}_{\Sigma}K_\delta{}^{ijk\Sigma}\; ,\\
0&=(Q_{42})^{ijkl}{}_{mnpq}K_\zeta{}^{mnpq\Lambda}
+(Q_{[k0]})^\Lambda{}_{\Sigma}K_\zeta{}^{ijkl\Sigma}\; .
\end{aligned}
\end{equation}
The seed and harmonic trace constraints are already incorporated in the representations on which these equations act. A basis of each kernel supplies all its invariant tensors, and its dimension is the corresponding singlet multiplicity. In particular, charged seed components must be retained whenever an opposite harmonic charge is available. One can construct the kernels by pairing explicit charge eigenvectors in the two factors, or by solving the displayed linear systems directly. This construction uses only $Q_z$ and is valid also at values of $z$ where additional singlets appear.

To relate these tensors to the charges (\ref{eq:ProtectedMasterCharge}), choose a basis of each kernel consisting of simultaneous eigenvectors of $F$, $R_0$ and $J_0$. The first two generators act on both the seed and harmonic factors, so their eigenvalues are the total charges; $J_0$ acts only on the seed. Let $j_0$ denote its eigenvalue. The weight lattice obeys $F+R_0+j_0\in2\mathbb Z$. Together with $Q_z=0$, this gives $\widehat q F/f_z\in\mathbb Z$; since $\gcd(\widehat q,f_z)=1$, the allowed flavour charges lie in $f_z\mathbb Z$. For $z\ne0$, choosing the CPT representative $F=F_{\nu,z}=-\operatorname{sgn}(z)\nu f_z$, with $\nu\geq1$, and using $\epsilon(z)=-\operatorname{sgn}(z)|\epsilon(z)|$, we obtain
\begin{equation}
\label{eq:InvariantComponentCharges}
\begin{aligned}
R_0&=-zF_{\nu,z}-j_0=\nu\lambda_z-j_0\; ,\\
r&=R_0+\epsilon(z)F_{\nu,z}=\rho_{\nu,z}-j_0\; .
\end{aligned}
\end{equation}
For a superconformal primary, comparison with (\ref{eq:ProtectedMasterCharge}) identifies $\eta=-j_0$. Here, however, the tensors describe individual field components. Components of one supermultiplet share $F$ and $\nu$, while their R-charges can differ by supercharge shifts, with corresponding changes in $j_0$. The SCP label $\eta$ is therefore assigned when the mass eigenstates are organised into supermultiplets. At MN1, the same weight condition gives $F=\pm2\nu$ and $r=-j_0$; the flavour-neutral sector has $F=0$ and $r=-j_0$ for every $z$. The mixing parameter $\epsilon(z)$ is needed to convert $R_0$ into $r$, but is not needed to construct the invariant tensors.

For example, (\ref{eq:InvariantKernelGraviton}) gives the constant tensor $K=1$ at $k=0$ and $K_1{}^a=\delta^{a5}$ at $k=1$, for $z\ne\pm1$. At $k=2$ and $z\ne0,\pm1$, it gives two tensors,
\begin{equation}
\label{eq:InvariantGravitonQuadraticExample}
K_1{}^{ab}=\frac12\operatorname{diag}(1,1,-1,-1,0)
\; ,\qquad
K_2{}^{ab}=\frac1{\sqrt{20}}\operatorname{diag}(1,1,1,1,-4)
\; .
\end{equation}
Both are traceless and invariant under (\ref{eq:InvariantQHarmonicTensor}). Higher levels are treated in the same way, with the full charge condition retained for the chosen $z$.

The dimension of the graviton kernel can be counted at arbitrary $k$. With the primitive twisting data (\ref{eq:SpectrumTwistData2}), the invariant harmonics have flavour charges $F=0$ or $F=\pm\nu f_z$, with $\nu\geq1$. For $z\ne0$, the graviton seed has $j_0=0$, so the representative with $F=F_{\nu,z}$ has $R_0=\nu\lambda_z$ and $r=\rho_{\nu,z}$ by (\ref{eq:InvariantComponentCharges}). The minimum harmonic degree in each charged sector is $\nu\mu_z$. At degree $k=\nu\mu_z+\kappa$, imposing tracelessness leaves $1+\check\kappa$ independent harmonics at each of the two flavour charges. Including the neutral sector, the number of invariant tensors $K_\alpha{}^\Lambda$ is therefore
\begin{equation}
\label{eq:InvariantGravitonMultiplicity}
N (k,z)
=1+\left\lfloor\frac{k}{2}\right\rfloor
+2\sum_{\nu=1}^{\lfloor k/\mu_z\rfloor}
\left(1+\left\lfloor\frac{k-\nu\mu_z}{2}\right\rfloor\right)
\; ,
\end{equation}
so that $\alpha=1,\ldots,N (k,z)$. The sum is empty for $k<\mu_z$. This expression also applies at $z=0,\pm1$, with the corresponding primitive values of $\mu_z$.

Equation~(\ref{eq:InvariantGravitonMultiplicity}) reproduces the graviton multiplicities in section~\ref{sec:UniversalSpectrum}. The first term counts the flavour-neutral multiplets in tables~\ref{tab:FirstZ-neutral-short}~and~\ref{tab:FirstZ-neutral-long}: the stress-tensor multiplet at $k=0$, and $1+\check\kappa$ long multiplets with $\kappa=k$ for $k\geq1$. Each summand counts the charged CPT pairs at $\kappa=k-\nu\mu_z\geq0$, with the factor of two accounting for their opposite flavour charges. This is the graviton level assignment in (\ref{eq:GeneralKKLevel}); for $\kappa\geq1$, the multiplicity $1+\check\kappa$ also equals the degree of $G_{\kappa\nu}$ in (\ref{eq:xdegCharged}), whose roots determine the long graviton dimensions in tables~\ref{tab:NewChargedMN1Long}--\ref{tab:NewChargedExteriorLong}. At $\kappa=0$, the single charged pair is long for $|z|<1$, as in tables~\ref{tab:NewChargedMN1Long}~and~\ref{tab:NewChargedInteriorLong}, and short for $|z|\geq1$, as in tables~\ref{tab:MN2ProtectedMultiplets}~and~\ref{tab:ExteriorProtectedMultiplets}. Since every $\cN=1$ graviton multiplet contains one spin-$2$ field, the protected and long contributions at each level of (\ref{eq:FullSpec}) together exhaust the invariant graviton space. The count fixes the multiplicities and flavour sectors; their dimensions and shortening are determined by the mass spectrum.

Choose orthonormal bases of the kernels in the seed and harmonic pairings. We denote the dual tensors by raising the invariant label and dualising the field indices. For example,
\begin{equation}
\label{eq:InvariantTensorDuality}
K_\alpha{}^\Lambda K_\Lambda{}^{\alpha'}
=\delta_\alpha{}^{\alpha'}
\; ,\qquad
K_\gamma{}^{\uM\Lambda}K_{\uM\Lambda}{}^{\gamma'}
=\delta_\gamma{}^{\gamma'}
\; ,
\end{equation}
and analogously in the other sectors. In orthonormal coordinates the dual is the Hermitian adjoint, or the transpose for a real basis. In undressed E$_{6(6)}$ coordinates it includes the vacuum field pairing. All conversions of $K$ and its dual to the coordinates of the parent operators are made together. The invariant labels may be organised into representations of the commutant of $Q_z$ on the seed, acting trivially on the harmonic factor, or resolved into eigenspaces of the total component charges $F$ and $r$ described above.

\subsection{$\mathrm{U}(1)_z$-invariant mass matrices}
\label{sec:InvariantMassContractions}

We use the putative mass matrices collected in appendix~C of \cite{Bhattacharya:2026ypg}, based on \cite{Pico:2026kch}, and retain their notation. Their eigenvalues are squared masses for gravitons, vectors and scalars, and linear masses for two-forms and fermions. We evaluate these matrices on the BBBW vacuum at fixed KK level $k$, suppressing their dependence on $k$ and $z$. The matrices and invariant tensors are expressed in the same field coordinates. In particular, the scalar matrix is converted from the coset coordinates of \cite{Bhattacharya:2026ypg} to the antisymmetric, symplectic-traceless USp$(8)$ tensors $\phi^{ijkl}$ using the vacuum coset representative.

Contraction with the tensors above and their duals gives
\begin{equation}
\label{eq:InvariantMassGravitonNew}
\begin{aligned}
(\cM^2_{\textrm{graviton}})^{\alpha'}{}_{\alpha}
&=K_\alpha{}^\Lambda
(\cM^2_{\textrm{graviton}})^\Sigma{}_{\Lambda}
K_\Sigma{}^{\alpha'}\; ,\\
(\cM_{\textrm{gravitino}})^{\beta'}{}_{\beta}
&=K_\beta{}^{i\Lambda}\Omega_{ji}
(\cM_{\textrm{gravitino}})^{\ell j\,\Sigma}{}_{\Lambda}
K_{\ell\Sigma}{}^{\beta'}\; ,\\
(\cM^2_{\textrm{vector}})^{\gamma'}{}_{\gamma}
&=K_\gamma{}^{\uM\Lambda}
(\cM^2_{\textrm{vector}})^{\uN\Sigma}{}_{\uM\Lambda}
K_{\uN\Sigma}{}^{\gamma'}\; ,\\
(\cM_{\textrm{tensor}})^{\gamma'}{}_{\gamma}
&=\widetilde K_{\gamma\,\uM}{}^\Lambda
(\cM_{\textrm{tensor}})^{\uM\Sigma}{}_{\uN\Lambda}
\widetilde K^{\uN}{}_{\Sigma}{}^{\gamma'}\; ,\\
(\cM_{\textrm{spin-$\frac12$}})^{\delta'}{}_{\delta}
&=K_\delta{}^{ijk\Lambda}
(\cM_{\textrm{spin-$\frac12$}})^{\ell mn\,\Sigma}{}_{ijk\,\Lambda}
K_{\ell mn\Sigma}{}^{\delta'}\; ,\\
(\cM^2_{\textrm{scalar}})^{\zeta'}{}_{\zeta}
&=K_\zeta{}^{ijkl\Lambda}
(\cM^2_{\textrm{scalar}})^{mnpq\Sigma}{}_{ijkl\Lambda}
K_{mnpq\Sigma}{}^{\zeta'}\; .
\end{aligned}
\end{equation}
The invariant indices on the left distinguish the restricted matrices from the parent matrices on the right. These are the U$(1)_z$-invariant mass matrices introduced in section~\ref{sec:U1zMassMatrices} of the main text and diagonalised in section~\ref{sec:UniversalSpectrum}. In the gravitino contraction, $\Omega_{ji}$ implements the identification $\psi_j=\Omega_{ji}\psi^i$. Multiplying the squared-mass matrices by $L_{\rm AdS}^2$ and the linear-mass matrices by $L_{\rm AdS}$ gives their normalisation in AdS units. Restricting the invariant indices to fixed component charges $F$ and $r$ gives the mass blocks used in section~\ref{sec:UniversalSpectrum}. The parent operators must preserve the selected kernels for these contractions to define restrictions. The resulting matrices are covariant on the invariant spaces; no invariance under the full commutant is required. After diagonalising these blocks and removing tensor-hierarchy, Goldstone and Goldstino modes, we organise the physical eigenstates into supermultiplets with common flavour charge $F$ and the required component R-charges. Their SCPs determine the labels $(\eta,\nu)$ of (\ref{eq:ProtectedMasterCharge}).

\section[Component content of $\mathcal N=1$ superconformal multiplets]{Component content of $\mathcal N=1$ superconformal multiplets}
\label{app:NOneMultipletContent}

Here, we list the conformal primaries of $\mathrm{SU}(2,2|1)$ multiplets with
SCP spins $(0,0)$, $(\tfrac12,0)$,
$(0,\tfrac12)$ or $(\tfrac12,\tfrac12)$. We follow the notation of \cite{Cordova:2016emh}, except that our Lorentz labels are half-integral  rather than integral. In $X\bar Y[E;j_1,j_2;r]$, $E$ and $r$ denote the SCP dimension
and R-charge; a common flavour charge $F$ is suppressed.
Each component $(j_1,j_2;r_{\mathrm{comp}})$ is listed at its dimension
$E+n/2$, where $n$ is the supercharge level, with
$Q:(\tfrac12,0;-1)$ and $\bar Q:(0,\tfrac12;+1)$.
Spacetime-derivative descendants are omitted.
The first row is the SCP, and ``short'' includes semishort.
The conditions alongside each multiplet label give its unitary range;
any fixed $E,r$ values are to be substituted in the component rows.
Free-field multiplets and the identity are listed separately for completeness, even if they do not enter the spectrum considered in this paper. A graviton multiplet has a vector SCP, with $(j_1,j_2)=(\frac12,\frac12)$; a gravitino multiplet has a spinor SCP, with $(j_1,j_2)=(0,\frac12)$ or $(\frac12,0)$; and a vector multiplet has a scalar SCP, with $(j_1,j_2)=(0,0)$. The tensor and chiral multiplets likewise have spinor and scalar SCPs, respectively.

\begin{table}[!htbp]
\centering
\caption{Graviton multiplets. The SCP is a vector. The $A_1\bar A_1$ multiplet contains the stress tensor and is the massless bulk graviton multiplet.}
\label{tab:NOneGravitonContent}
\small
\renewcommand{\arraystretch}{1.15}
\setlength{\tabcolsep}{5pt}
\begin{tabular*}{\textwidth}{@{}>{$}l<{$}@{\extracolsep{\fill}}>{$}l<{$}@{}}
\toprule
\multicolumn{1}{@{}l}{Dimension} & \multicolumn{1}{l@{}}{Conformal-primary content $(j_1,j_2;r_{\mathrm{comp}})$} \\
\midrule
\multicolumn{2}{@{}p{\textwidth}@{}}{Long graviton: $L\bar L[E;\tfrac12,\tfrac12;r]$\hfill $E>3+\tfrac32|r|$} \\[4pt]
E & (\tfrac12,\tfrac12;r) \\[2pt]
E+\tfrac12 & (0,\tfrac12;r-1)\oplus (1,\tfrac12;r-1)\oplus (\tfrac12,0;r+1)\oplus (\tfrac12,1;r+1) \\[2pt]
E+1 & \begin{aligned}[t]&(\tfrac12,\tfrac12;r-2)\oplus (0,0;r)\oplus (0,1;r)\\[-1pt]&\mathrel{\oplus} (1,0;r)\oplus (1,1;r)\oplus (\tfrac12,\tfrac12;r+2)\end{aligned} \\[2pt]
E+\tfrac32 & (\tfrac12,0;r-1)\oplus (\tfrac12,1;r-1)\oplus (0,\tfrac12;r+1)\oplus (1,\tfrac12;r+1) \\[2pt]
E+2 & (\tfrac12,\tfrac12;r) \\[2pt]
\addlinespace[3pt]
\midrule
\multicolumn{2}{@{}p{\textwidth}@{}}{Short graviton: $L\bar A_1[E;\tfrac12,\tfrac12;r]$\hfill $E=3+\tfrac32r,\quad r>0$} \\[4pt]
E & (\tfrac12,\tfrac12;r) \\[2pt]
E+\tfrac12 & (0,\tfrac12;r-1)\oplus (1,\tfrac12;r-1)\oplus (\tfrac12,1;r+1) \\[2pt]
E+1 & (\tfrac12,\tfrac12;r-2)\oplus (0,1;r)\oplus (1,1;r) \\[2pt]
E+\tfrac32 & (\tfrac12,1;r-1) \\[2pt]
\addlinespace[3pt]
\midrule
\multicolumn{2}{@{}p{\textwidth}@{}}{Short graviton: $A_1\bar L[E;\tfrac12,\tfrac12;r]$\hfill $E=3-\tfrac32r,\quad r<0$} \\[4pt]
E & (\tfrac12,\tfrac12;r) \\[2pt]
E+\tfrac12 & (1,\tfrac12;r-1)\oplus (\tfrac12,0;r+1)\oplus (\tfrac12,1;r+1) \\[2pt]
E+1 & (1,0;r)\oplus (1,1;r)\oplus (\tfrac12,\tfrac12;r+2) \\[2pt]
E+\tfrac32 & (1,\tfrac12;r+1) \\[2pt]
\addlinespace[3pt]
\midrule
\multicolumn{2}{@{}p{\textwidth}@{}}{Massless graviton: $A_1\bar A_1[E;\tfrac12,\tfrac12;r]$\hfill $E=3,\quad r=0$} \\[4pt]
E & (\tfrac12,\tfrac12;r) \\[2pt]
E+\tfrac12 & (1,\tfrac12;r-1)\oplus (\tfrac12,1;r+1) \\[2pt]
E+1 & (1,1;r) \\[2pt]
\bottomrule
\end{tabular*}
\end{table}

\clearpage

\begin{table}[!htbp]
\centering
\caption{Gravitino multiplets with SCP spins $(\tfrac12,0)$. The top-spin component has Lorentz spins $(1,\tfrac12)$. The doubly shortened multiplet contains a conserved supercurrent.}
\label{tab:NOneGravitinoLeftContent}
\small
\renewcommand{\arraystretch}{1.15}
\setlength{\tabcolsep}{5pt}
\begin{tabular*}{\textwidth}{@{}>{$}l<{$}@{\extracolsep{\fill}}>{$}l<{$}@{}}
\toprule
\multicolumn{1}{@{}l}{Dimension} & \multicolumn{1}{l@{}}{Conformal-primary content $(j_1,j_2;r_{\mathrm{comp}})$} \\
\midrule
\multicolumn{2}{@{}p{\textwidth}@{}}{Long gravitino: $L\bar L[E;\tfrac12,0;r]$\hfill $E>\tfrac52+\tfrac32|r-\tfrac13|$} \\[4pt]
E & (\tfrac12,0;r) \\[2pt]
E+\tfrac12 & (0,0;r-1)\oplus (1,0;r-1)\oplus (\tfrac12,\tfrac12;r+1) \\[2pt]
E+1 & (\tfrac12,0;r-2)\oplus (0,\tfrac12;r)\oplus (1,\tfrac12;r)\oplus (\tfrac12,0;r+2) \\[2pt]
E+\tfrac32 & (\tfrac12,\tfrac12;r-1)\oplus (0,0;r+1)\oplus (1,0;r+1) \\[2pt]
E+2 & (\tfrac12,0;r) \\[2pt]
\addlinespace[3pt]
\midrule
\multicolumn{2}{@{}p{\textwidth}@{}}{Short gravitino: $A_1\bar L[E;\tfrac12,0;r]$\hfill $E=3-\tfrac32r,\quad r<\tfrac13$} \\[4pt]
E & (\tfrac12,0;r) \\[2pt]
E+\tfrac12 & (1,0;r-1)\oplus (\tfrac12,\tfrac12;r+1) \\[2pt]
E+1 & (1,\tfrac12;r)\oplus (\tfrac12,0;r+2) \\[2pt]
E+\tfrac32 & (1,0;r+1) \\[2pt]
\addlinespace[3pt]
\midrule
\multicolumn{2}{@{}p{\textwidth}@{}}{Short gravitino: $L\bar A_2[E;\tfrac12,0;r]$\hfill $E=2+\tfrac32r,\quad r>\tfrac13$} \\[4pt]
E & (\tfrac12,0;r) \\[2pt]
E+\tfrac12 & (0,0;r-1)\oplus (1,0;r-1)\oplus (\tfrac12,\tfrac12;r+1) \\[2pt]
E+1 & (\tfrac12,0;r-2)\oplus (0,\tfrac12;r)\oplus (1,\tfrac12;r) \\[2pt]
E+\tfrac32 & (\tfrac12,\tfrac12;r-1) \\[2pt]
\addlinespace[3pt]
\midrule
\multicolumn{2}{@{}p{\textwidth}@{}}{Massless gravitino: $A_1\bar A_2[E;\tfrac12,0;r]$\hfill $E=\tfrac52,\quad r=\tfrac13$} \\[4pt]
E & (\tfrac12,0;r) \\[2pt]
E+\tfrac12 & (1,0;r-1)\oplus (\tfrac12,\tfrac12;r+1) \\[2pt]
E+1 & (1,\tfrac12;r) \\[2pt]
\bottomrule
\end{tabular*}
\end{table}

\clearpage

\begin{table}[!htbp]
\centering
\caption{Gravitino multiplets with SCP spins $(0,\tfrac12)$. These are the CPT conjugates of table~\ref{tab:NOneGravitinoLeftContent}: interchange $j_1$ and $j_2$, reverse all R-charges and exchange barred and unbarred shortening labels.}
\label{tab:NOneGravitinoRightContent}
\small
\renewcommand{\arraystretch}{1.15}
\setlength{\tabcolsep}{5pt}
\begin{tabular*}{\textwidth}{@{}>{$}l<{$}@{\extracolsep{\fill}}>{$}l<{$}@{}}
\toprule
\multicolumn{1}{@{}l}{Dimension} & \multicolumn{1}{l@{}}{Conformal-primary content $(j_1,j_2;r_{\mathrm{comp}})$} \\
\midrule
\multicolumn{2}{@{}p{\textwidth}@{}}{Long gravitino: $L\bar L[E;0,\tfrac12;r]$\hfill $E>\tfrac52+\tfrac32|r+\tfrac13|$} \\[4pt]
E & (0,\tfrac12;r) \\[2pt]
E+\tfrac12 & (\tfrac12,\tfrac12;r-1)\oplus (0,0;r+1)\oplus (0,1;r+1) \\[2pt]
E+1 & (0,\tfrac12;r-2)\oplus (\tfrac12,0;r)\oplus (\tfrac12,1;r)\oplus (0,\tfrac12;r+2) \\[2pt]
E+\tfrac32 & (0,0;r-1)\oplus (0,1;r-1)\oplus (\tfrac12,\tfrac12;r+1) \\[2pt]
E+2 & (0,\tfrac12;r) \\[2pt]
\addlinespace[3pt]
\midrule
\multicolumn{2}{@{}p{\textwidth}@{}}{Short gravitino: $L\bar A_1[E;0,\tfrac12;r]$\hfill $E=3+\tfrac32r,\quad r>-\tfrac13$} \\[4pt]
E & (0,\tfrac12;r) \\[2pt]
E+\tfrac12 & (\tfrac12,\tfrac12;r-1)\oplus (0,1;r+1) \\[2pt]
E+1 & (0,\tfrac12;r-2)\oplus (\tfrac12,1;r) \\[2pt]
E+\tfrac32 & (0,1;r-1) \\[2pt]
\addlinespace[3pt]
\midrule
\multicolumn{2}{@{}p{\textwidth}@{}}{Short gravitino: $A_2\bar L[E;0,\tfrac12;r]$\hfill $E=2-\tfrac32r,\quad r<-\tfrac13$} \\[4pt]
E & (0,\tfrac12;r) \\[2pt]
E+\tfrac12 & (\tfrac12,\tfrac12;r-1)\oplus (0,0;r+1)\oplus (0,1;r+1) \\[2pt]
E+1 & (\tfrac12,0;r)\oplus (\tfrac12,1;r)\oplus (0,\tfrac12;r+2) \\[2pt]
E+\tfrac32 & (\tfrac12,\tfrac12;r+1) \\[2pt]
\addlinespace[3pt]
\midrule
\multicolumn{2}{@{}p{\textwidth}@{}}{Massless gravitino: $A_2\bar A_1[E;0,\tfrac12;r]$\hfill $E=\tfrac52,\quad r=-\tfrac13$} \\[4pt]
E & (0,\tfrac12;r) \\[2pt]
E+\tfrac12 & (\tfrac12,\tfrac12;r-1)\oplus (0,1;r+1) \\[2pt]
E+1 & (\tfrac12,1;r) \\[2pt]
\bottomrule
\end{tabular*}
\end{table}

\clearpage

\begin{table}[!htbp]
\centering
\caption{Vector multiplets. The SCP is a scalar. The $A_2\bar A_2$ multiplet contains the conserved flavour current and is the massless bulk vector multiplet.}
\label{tab:NOneVectorContent}
\small
\renewcommand{\arraystretch}{1.15}
\setlength{\tabcolsep}{5pt}
\begin{tabular*}{\textwidth}{@{}>{$}l<{$}@{\extracolsep{\fill}}>{$}l<{$}@{}}
\toprule
\multicolumn{1}{@{}l}{Dimension} & \multicolumn{1}{l@{}}{Conformal-primary content $(j_1,j_2;r_{\mathrm{comp}})$} \\
\midrule
\multicolumn{2}{@{}p{\textwidth}@{}}{Long vector: $L\bar L[E;0,0;r]$\hfill $E>2+\tfrac32|r|$} \\[4pt]
E & (0,0;r) \\[2pt]
E+\tfrac12 & (\tfrac12,0;r-1)\oplus (0,\tfrac12;r+1) \\[2pt]
E+1 & (0,0;r-2)\oplus (\tfrac12,\tfrac12;r)\oplus (0,0;r+2) \\[2pt]
E+\tfrac32 & (0,\tfrac12;r-1)\oplus (\tfrac12,0;r+1) \\[2pt]
E+2 & (0,0;r) \\[2pt]
\addlinespace[3pt]
\midrule
\multicolumn{2}{@{}p{\textwidth}@{}}{Short vector: $L\bar A_2[E;0,0;r]$\hfill $E=2+\tfrac32r,\quad r>0$} \\[4pt]
E & (0,0;r) \\[2pt]
E+\tfrac12 & (\tfrac12,0;r-1)\oplus (0,\tfrac12;r+1) \\[2pt]
E+1 & (0,0;r-2)\oplus (\tfrac12,\tfrac12;r) \\[2pt]
E+\tfrac32 & (0,\tfrac12;r-1) \\[2pt]
\addlinespace[3pt]
\midrule
\multicolumn{2}{@{}p{\textwidth}@{}}{Short vector: $A_2\bar L[E;0,0;r]$\hfill $E=2-\tfrac32r,\quad r<0$} \\[4pt]
E & (0,0;r) \\[2pt]
E+\tfrac12 & (\tfrac12,0;r-1)\oplus (0,\tfrac12;r+1) \\[2pt]
E+1 & (\tfrac12,\tfrac12;r)\oplus (0,0;r+2) \\[2pt]
E+\tfrac32 & (\tfrac12,0;r+1) \\[2pt]
\addlinespace[3pt]
\midrule
\multicolumn{2}{@{}p{\textwidth}@{}}{Massless vector: $A_2\bar A_2[E;0,0;r]$\hfill $E=2,\quad r=0$} \\[4pt]
E & (0,0;r) \\[2pt]
E+\tfrac12 & (\tfrac12,0;r-1)\oplus (0,\tfrac12;r+1) \\[2pt]
E+1 & (\tfrac12,\tfrac12;r) \\[2pt]
\bottomrule
\end{tabular*}
\end{table}

\clearpage

\begin{table}[!htbp]
\centering
\caption{Tensor and chiral multiplets. Tensor multiplets contain a two-form component, in $(1,0)$ or $(0,1)$. Chiral multiplets have top spin $\tfrac12$. Each family is displayed together with its CPT conjugate.}
\label{tab:NOneTensorChiralContent}
\small
\renewcommand{\arraystretch}{1.15}
\setlength{\tabcolsep}{5pt}
\begin{tabular*}{\textwidth}{@{}>{$}l<{$}@{\extracolsep{\fill}}>{$}l<{$}@{}}
\toprule
\multicolumn{1}{@{}l}{Dimension} & \multicolumn{1}{l@{}}{Conformal-primary content $(j_1,j_2;r_{\mathrm{comp}})$} \\
\midrule
\multicolumn{2}{@{}p{\textwidth}@{}}{Tensor: $L\bar B_1[E;\tfrac12,0;r]$\hfill $E=\tfrac32r,\quad r>1$} \\[4pt]
E & (\tfrac12,0;r) \\[2pt]
E+\tfrac12 & (0,0;r-1)\oplus (1,0;r-1) \\[2pt]
E+1 & (\tfrac12,0;r-2) \\[2pt]
\addlinespace[3pt]
\midrule
\multicolumn{2}{@{}p{\textwidth}@{}}{Conjugate tensor: $B_1\bar L[E;0,\tfrac12;r]$\hfill $E=-\tfrac32r,\quad r<-1$} \\[4pt]
E & (0,\tfrac12;r) \\[2pt]
E+\tfrac12 & (0,0;r+1)\oplus (0,1;r+1) \\[2pt]
E+1 & (0,\tfrac12;r+2) \\[2pt]
\addlinespace[3pt]
\midrule
\multicolumn{2}{@{}p{\textwidth}@{}}{Chiral: $L\bar B_1[E;0,0;r]$\hfill $E=\tfrac32r,\quad r>\tfrac23$} \\[4pt]
E & (0,0;r) \\[2pt]
E+\tfrac12 & (\tfrac12,0;r-1) \\[2pt]
E+1 & (0,0;r-2) \\[2pt]
\addlinespace[3pt]
\midrule
\multicolumn{2}{@{}p{\textwidth}@{}}{Conjugate chiral: $B_1\bar L[E;0,0;r]$\hfill $E=-\tfrac32r,\quad r<-\tfrac23$} \\[4pt]
E & (0,0;r) \\[2pt]
E+\tfrac12 & (0,\tfrac12;r+1) \\[2pt]
E+1 & (0,0;r+2) \\[2pt]
\bottomrule
\end{tabular*}
\end{table}

\begin{table}[!htbp]
\centering
\caption{Free-field multiplets and identity. The free vector entries describe the two conjugate field-strength multiplets. They are distinct from the massless bulk vector multiplet in table~\ref{tab:NOneVectorContent}.}
\label{tab:NOneFreeContent}
\small
\renewcommand{\arraystretch}{1.15}
\setlength{\tabcolsep}{5pt}
\begin{tabular*}{\textwidth}{@{}>{$}l<{$}@{\extracolsep{\fill}}>{$}l<{$}@{}}
\toprule
\multicolumn{1}{@{}l}{Dimension} & \multicolumn{1}{l@{}}{Conformal-primary content $(j_1,j_2;r_{\mathrm{comp}})$} \\
\midrule
\multicolumn{2}{@{}p{\textwidth}@{}}{Free vector: $A_1\bar B_1[E;\tfrac12,0;r]$\hfill $E=\tfrac32,\quad r=1$} \\[4pt]
E & (\tfrac12,0;r) \\[2pt]
E+\tfrac12 & (1,0;r-1) \\[2pt]
\addlinespace[3pt]
\midrule
\multicolumn{2}{@{}p{\textwidth}@{}}{Conjugate free vector: $B_1\bar A_1[E;0,\tfrac12;r]$\hfill $E=\tfrac32,\quad r=-1$} \\[4pt]
E & (0,\tfrac12;r) \\[2pt]
E+\tfrac12 & (0,1;r+1) \\[2pt]
\addlinespace[3pt]
\midrule
\multicolumn{2}{@{}p{\textwidth}@{}}{Free chiral: $A_2\bar B_1[E;0,0;r]$\hfill $E=1,\quad r=\tfrac23$} \\[4pt]
E & (0,0;r) \\[2pt]
E+\tfrac12 & (\tfrac12,0;r-1) \\[2pt]
\addlinespace[3pt]
\midrule
\multicolumn{2}{@{}p{\textwidth}@{}}{Conjugate free chiral: $B_1\bar A_2[E;0,0;r]$\hfill $E=1,\quad r=-\tfrac23$} \\[4pt]
E & (0,0;r) \\[2pt]
E+\tfrac12 & (0,\tfrac12;r+1) \\[2pt]
\addlinespace[3pt]
\midrule
\multicolumn{2}{@{}p{\textwidth}@{}}{Identity: $B_1\bar B_1[E;0,0;r]$\hfill $E=0,\quad r=0$} \\[4pt]
E & (0,0;r) \\[2pt]
\bottomrule
\end{tabular*}
\end{table}

\newpage


\section{\texorpdfstring{Spectral polynomials}{Spectral polynomials}}
\label{app:Spectral polynomials}


This appendix constructs the polynomial families whose roots determine the SCP dimensions of all long multiplets in the $\mathrm{U}(1)_z$-invariant BBBW spectrum $\textrm{Spec}^0(z)$ in (\ref{eq:SpectrumFourfoldSplit}). The polynomials $G_{\kappa\nu}(x;z)$, $H_{\kappa\nu\eta}(x;z)$, $P_{\kappa\nu}(x;z)$ and $Q_{\kappa\nu\eta}(x;z)$ presented in appendix~\ref{app:ChargedPolynomialsNew} are defined on the label ranges specified below. For $\nu\geq1$, they describe the flavour-charged contribution $\cL_{\textrm{$F$-charged}}(z)$ of section~\ref{sec:FlavourChargedLong}, while $\nu=0$ gives the flavour-neutral contribution $\cL_{\textrm{$F$-neutral}}(z)$ of section~\ref{sec:FlavourNeutralLong}. Some roots also correspond to short multiplets discussed in sections~\ref{sec:FlavourNeutralProtected} and~\ref{sec:FlavourChargedProtected}. These are excluded from the long spectrum through the restrictions on label ranges imposed in the main text. Such restrictions select the long multiplet contributions without modifying the polynomial definitions, which remain valid throughout the ranges stated here. The general polynomials are presented in appendix~\ref{app:ChargedPolynomialsNew}. Their flavour-neutral, $\nu=0$, specialisations are given in appendix~\ref{app:NeutralPolynomials}.

\subsection{\texorpdfstring{General polynomials}{General polynomials}}
\label{app:ChargedPolynomialsNew}

The infinite families of functions $G_{\kappa\nu}(x;z)$, $H_{\kappa\nu\eta}(x;z)$, $P_{\kappa\nu}(x;z)$ and $Q_{\kappa\nu\eta}(x;z)$ are labelled by the non-negative integers $\kappa$ and $\nu$ of (\ref{eq:kappanuRange}). For the ranges specified in (\ref{eq:MatOrdCharged}) below, $\nu_{\textrm{min}}=0$, while $\kappa_{\textrm{min}}$ depends on the polynomial family. The additional charged boundary members $P_{0\nu}$ and $P_{1\nu}$, with $\nu\geq1$, are defined separately in appendix~\ref{app:NewChargedInnerVector}. A further label is the constant offset $\eta$ of (\ref{eq:etaDef}). This offset vanishes for $G_{\kappa\nu}(x;z)$ and $P_{\kappa\nu}(x;z)$, and is therefore omitted from their notation, while $\eta=\pm1$ for $H_{\kappa\nu\eta}(x;z)$ and $\eta=\pm2$ for $Q_{\kappa\nu\eta}(x;z)$. These functions are polynomials in the auxiliary variable $x$. Within these ranges, their degrees 
\begin{equation}
\label{eq:xdegCharged}
\begin{aligned}
\deg_x G_{\kappa\nu} &= \check{\kappa}+1 \quad (\kappa\geq 0) \, ,
&\qquad \deg_x H_{\kappa\nu\eta} &= 2\hat{\kappa}-1 \quad (\kappa\geq1) \, ,\\[5pt]
\deg_x P_{\kappa\nu} &= 2\check{\kappa} \quad (\kappa\geq2) \, ,
&\qquad \deg_x Q_{\kappa\nu\eta} &= \check{\kappa}-1 \quad (\kappa\geq4) \, ,
\end{aligned}
\end{equation}
depend only on $\kappa$ through $\check{\kappa}$ and $\hat{\kappa}$ in (\ref{eq:NeutralParityDefinitions}), and are independent of $\nu$ and $\eta$. Explicitly, these functions are the characteristic polynomials
\begin{equation}
\label{eq:PolyMatCharged}
\begin{aligned}
G_{\kappa\nu}(x;z)
&=\det\big[x\mathbbm{1}_{\check{\kappa}+1}-\cG_{\kappa\nu}(z)\big] \, ,
&\qquad
H_{\kappa\nu\eta}(x;z)
&=\det\big[x\mathbbm{1}_{2\hat{\kappa}-1}-\cH_{\kappa\nu\eta}(z)\big] \, ,\\[5pt]
P_{\kappa\nu}(x;z)
&=\det\big[x\mathbbm{1}_{2\check{\kappa}}-\cP_{\kappa\nu}(z)\big] \, ,
&\qquad
Q_{\kappa\nu\eta}(x;z)
&=\det\big[x\mathbbm{1}_{\check{\kappa}-1}-\cQ_{\kappa\nu\eta}(z)\big] \, ,
\end{aligned}
\end{equation}
of square matrices $\cG_{\kappa\nu}(z)$, $\cH_{\kappa\nu\eta}(z)$, $\cP_{\kappa\nu}(z)$, $\cQ_{\kappa\nu\eta}(z)$ carrying the same labels, of sizes
\begin{equation}
\label{eq:MatOrdCharged}
\begin{aligned}
\textrm{ord}\,\cG_{\kappa\nu}(z)
&=\check{\kappa}+1 \quad (\kappa\geq0) \, ,
&\qquad
\textrm{ord}\,\cH_{\kappa\nu\eta}(z)
&=2\hat{\kappa}-1 \quad (\kappa\geq1) \, ,\\[5pt]
\textrm{ord}\,\cP_{\kappa\nu}(z)
&=2\check{\kappa} \quad (\kappa\geq2) \, ,
&\qquad
\textrm{ord}\,\cQ_{\kappa\nu\eta}(z)
&=\check{\kappa}-1 \quad (\kappa\geq4) \, ,
\end{aligned}
\end{equation}
matching (\ref{eq:xdegCharged}).

The matrix elements of $\cG_{\kappa\nu}(z)$, $\cH_{\kappa\nu\eta}(z)$, $\cP_{\kappa\nu}(z)$, $\cQ_{\kappa\nu\eta}(z)$ are typically discontinuous and non-monotonic functions of the BBBW twist parameter $z$. Their dependence on $z$ is through the R-symmetry mixing parameter $\epsilon(z)$ of (\ref{eq:SpectrumExactRCharge}) and the quantities $F_{\nu,z}$ and $\rho_{\nu,z}$ defined in (\ref{eq:ProtectedMasterCharge}) through the twisting data (\ref{eq:SpectrumTwistData2}). At fixed integer labels, the matrix entries are polynomial in these parameters on each branch obtained by resolving the absolute values. The labels $\kappa$ and $\eta$ generally appear explicitly, whereas $\nu$ enters only through $F_{\nu,z}$ and $\rho_{\nu,z}$. Consequently, the coefficients of the characteristic polynomials (\ref{eq:PolyMatCharged}) are polynomial in $\epsilon(z)$, $F_{\nu,z}$ and $\rho_{\nu,z}$ on each such branch. The spectral functions remain polynomials in the auxiliary variable $x$ throughout. The matrices are constructed explicitly in appendices~\ref{app:ChargedMatrixGraviton}--\ref{app:NewChargedOuterVector}.

\subsubsection{\texorpdfstring{General graviton
polynomials $G_{\kappa\nu}(x;z)$}
{General graviton polynomials}}
\label{app:ChargedMatrixGraviton}

The $(\check{\kappa}+1)$-dimensional matrix
$\cG_{\kappa\nu}(z)$ that determines the polynomial
$G_{\kappa\nu}(x;z)$ through (\ref{eq:PolyMatCharged})
can be written, for all $\nu\geq0$, $\kappa \geq 0$ as follows.
For $\kappa=0,1$, it is the one-dimensional matrix
\begin{equation}
\label{eq:NewChargedGMatrixn01}
\kappa=0,1:\qquad
\cG_{\kappa\nu}(z)=\bigl(a_{\kappa\nu,0}(z)\bigr).
\end{equation}
For $\kappa\geq2$, it has the tridiagonal form
{\setlength{\arraycolsep}{2pt}
\begin{equation}
\label{eq:NewChargedGMatrix}
\cG_{\kappa\nu}(z)=
\begin{pmatrix}
a_{\kappa\nu,0}(z)&1&0&\cdots&0\\
b_{\kappa\nu,0}(z)&a_{\kappa\nu,1}(z)&1&\ddots&\vdots\\
0&\ddots&\ddots&\ddots&0\\
\vdots&\ddots&
b_{\kappa\nu,\check{\kappa}-2}(z)&
a_{\kappa\nu,\check{\kappa}-1}(z)&1\\
0&\cdots&0&
b_{\kappa\nu,\check{\kappa}-1}(z)&
a_{\kappa\nu,\check{\kappa}}(z)
\end{pmatrix}.
\end{equation}
}The coefficients are given, for all $z$, by
{\setlength{\arraycolsep}{2pt}
\begin{eqnarray}
\label{eq:NewChargedGCoefficients}
a_{\kappa\nu,i}(z)
&=&
3\kappa(\kappa+3)
-3\bigl[\check{\kappa}
       +2i(\check{\kappa}-i)\bigr](1+3\epsilon^2)
\nonumber\\
&&{}
+6(2\kappa-2\check{\kappa}+1)
  (2i-\check{\kappa})\epsilon
\nonumber\\
&&{}
+\frac{9}{16}
\Bigl[
(1-\epsilon)
\left|\rho_{\nu,z}-(1+\epsilon)F_{\nu,z}\right|
+
(1+\epsilon)
\left|\rho_{\nu,z}+(1-\epsilon)F_{\nu,z}\right|
\Bigr]^2
\nonumber\\
&&{}
+\frac34(1-\epsilon)
 \bigl[4\kappa+5-2i+3(2i+1)\epsilon\bigr]
 \left|\rho_{\nu,z}-(1+\epsilon)F_{\nu,z}\right|
\nonumber\\
&&{}
+\frac34(1+\epsilon)
 \bigl[4\kappa+5-2\check{\kappa}+2i
       -3(2\check{\kappa}-2i+1)\epsilon\bigr]
 \left|\rho_{\nu,z}+(1-\epsilon)F_{\nu,z}\right|
\, ,
\\[4pt]
b_{\kappa\nu,i}(z)
&=&
9(i+1)(\check{\kappa}-i)
(1-\epsilon^2)(1-9\epsilon^2)
\left[
i+1+\frac12
\left|\rho_{\nu,z}+(1-\epsilon)F_{\nu,z}\right|
\right]
\nonumber\\
&&{}\times
\left[
\check{\kappa}-i+\frac12
\left|\rho_{\nu,z}-(1+\epsilon)F_{\nu,z}\right|
\right]
\; .
\nonumber
\end{eqnarray}
}Here $i=0,\ldots,\check{\kappa}$ for
$a_{\kappa\nu,i}(z)$ and
$i=0,\ldots,\check{\kappa}-1$ for
$b_{\kappa\nu,i}(z)$.
%
With these definitions, the polynomials
$G_{\kappa\nu}(x;z)$ in (\ref{eq:PolyMatCharged}) are,
for all $\nu\geq0$ and the first few
$\kappa=0,\ldots,5$,
\begin{equation}
\label{eq:ChargedGFirstMembers}
\begin{aligned}
G_{\kappa\nu}(x;z)
&=x-a_{\kappa\nu,0}(z),
&&\kappa=0,1,
\\
G_{\kappa\nu}(x;z)
&=\bigl(x-a_{\kappa\nu,0}(z)\bigr)
  \bigl(x-a_{\kappa\nu,1}(z)\bigr)
  -b_{\kappa\nu,0}(z),
&&\kappa=2,3,
\\
G_{\kappa\nu}(x;z)
&=\bigl(x-a_{\kappa\nu,0}(z)\bigr)
  \bigl(x-a_{\kappa\nu,1}(z)\bigr)
  \bigl(x-a_{\kappa\nu,2}(z)\bigr)
\\
&\quad
-b_{\kappa\nu,0}(z)
 \bigl(x-a_{\kappa\nu,2}(z)\bigr)
-b_{\kappa\nu,1}(z)
 \bigl(x-a_{\kappa\nu,0}(z)\bigr),
&&\kappa=4,5.
\end{aligned}
\end{equation}

\subsubsection{General gravitino polynomials
$H_{\kappa\nu\eta}(x;z)$}
\label{app:NewChargedGravitino}

The $(2\hat{\kappa}-1)$-dimensional matrix
$\cH_{\kappa\nu\eta}(z)$ that determines the gravitino
multiplet polynomial $H_{\kappa\nu\eta}(x;z)$ through
(\ref{eq:PolyMatCharged}) can be written for all $\nu\geq0$, 
$\kappa\geq1$, $\eta=\pm1$ as follows.
For $\kappa=1,2$ ($\hat{\kappa}=1$) the matrix is
one-dimensional,
\begin{equation}
\label{eq:NewChargedHMatrixn12}
\kappa=1,2:\qquad
\cH_{\kappa\nu\eta}(z)
=\bigl(d_{\kappa\nu\eta}(z)\bigr).
\end{equation}
For $\kappa=3,4$ ($\hat{\kappa}=2$) it is the
$3\times3$ matrix
\begin{equation}
\label{eq:NewChargedHMatrixn34}
\kappa=3,4:\qquad
\cH_{\kappa\nu\eta}(z)
=\begin{pmatrix}
d_{\kappa\nu\eta}(z)&E_{\kappa\nu\eta}(z)\\[3pt]
F_{\kappa\nu\eta}(z)&D_{\kappa\nu\eta,1}(z)
\end{pmatrix},
\end{equation}
and for $\kappa\geq5$ ($\hat{\kappa}\geq3$),
{\setlength{\arraycolsep}{2pt}
\begin{equation}
\label{eq:NewChargedHMatrixngeq5}
\cH_{\kappa\nu\eta}(z)
=\begin{pmatrix}
d_{\kappa\nu\eta}(z)&E_{\kappa\nu\eta}(z)
&0_{1\times2}&\cdots&0_{1\times2}\\
F_{\kappa\nu\eta}(z)&D_{\kappa\nu\eta,1}(z)
&M_{\kappa\nu\eta,1}(z)&\ddots&\vdots\\
0_{2\times1}&N_{\kappa\nu\eta,1}(z)
&D_{\kappa\nu\eta,2}(z)&\ddots&0_{2\times2}\\
\vdots&\ddots&\ddots&\ddots
&M_{\kappa\nu\eta,\hat{\kappa}-2}(z)\\
0_{2\times1}&\cdots&0_{2\times2}
&N_{\kappa\nu\eta,\hat{\kappa}-2}(z)
&D_{\kappa\nu\eta,\hat{\kappa}-1}(z)
\end{pmatrix}.
\end{equation}
}


In (\ref{eq:NewChargedHMatrixn12})--%
(\ref{eq:NewChargedHMatrixngeq5}),
$d_{\kappa\nu\eta}(z)$ is the scalar
\begin{equation}
\label{eq:NewChargedHBoundaryDiagonalCoefficient}
\begin{aligned}
d_{\kappa\nu\eta}(z)
={}&\Biggl[
2\kappa-\hat{\kappa}-\frac12
-3(\hat{\kappa}-1)\epsilon
+\frac34(1-\epsilon)
\left|\rho_{\nu,z}+2\eta-(1+\epsilon)F_{\nu,z}\right|
\\
&\qquad
+\frac34(1+\epsilon)
\left|\rho_{\nu,z}+2\eta+(1-\epsilon)F_{\nu,z}\right|
\Biggr]^2-\frac14
\\
&+3(\hat{\kappa}-1)(1-\epsilon)(1+3\epsilon)
\left[
\hat{\kappa}-1+\frac12
\left|\rho_{\nu,z}+2\eta-(1+\epsilon)F_{\nu,z}\right|
\right].
\end{aligned}
\end{equation}
For $\hat{\kappa}\geq2$, $E_{\kappa\nu\eta}(z)$ and
$F_{\kappa\nu\eta}(z)$ are the $1\times2$ and
$2\times1$ matrices
\begin{equation}
\label{eq:NewChargedHBoundaryBlocks}
\begin{aligned}
E_{\kappa\nu\eta}(z)
={}&
3(\epsilon-1)
\left[
\hat{\kappa}-1+\frac12
\left|\rho_{\nu,z}+2\eta-(1+\epsilon)F_{\nu,z}\right|
\right]
\\
&\quad\times
\begin{pmatrix}
6(\hat{\kappa}-1)\epsilon
&
\displaystyle
\left[
\frac{3\epsilon-1}{2}
\left|\rho_{\nu,z}+2\eta+(1-\epsilon)F_{\nu,z}\right|
-1
\right]
\sqrt{\frac{\hat{\kappa}-1}{2\hat{\kappa}}}
\end{pmatrix},
\\[8pt]
F_{\kappa\nu\eta}(z)
={}&
\begin{pmatrix}
\frac32(1+3\epsilon)\\[4pt]
3(1+\epsilon)(1+3\epsilon)
\sqrt{2\hat{\kappa}(\hat{\kappa}-1)}
\end{pmatrix}.
\end{aligned}
\end{equation}
The latter depends on $z$ only through $\epsilon(z)$,
and is also independent of the labels $\nu$ and $\eta$.
Finally, $D_{\kappa\nu\eta,i}(z)$,
$M_{\kappa\nu\eta,i}(z)$,
$N_{\kappa\nu\eta,i}(z)$ are the $2\times2$ matrices
\begin{equation}
\label{eq:NewChargedHBlockMatrices}
\begin{aligned}
D_{\kappa\nu\eta,i}(z)
&=
\begin{pmatrix}
D^{11}_{\kappa\nu\eta,i}(z)&D^{12}_{\kappa\nu\eta,i}(z)\\[3pt]
D^{21}_{\kappa\nu\eta,i}(z)&D^{22}_{\kappa\nu\eta,i}(z)
\end{pmatrix},
\\[6pt]
M_{\kappa\nu\eta,i}(z)
&=
\begin{pmatrix}
M^{11}_{\kappa\nu\eta,i}(z)&M^{12}_{\kappa\nu\eta,i}(z)\\[3pt]
M^{21}_{\kappa\nu\eta,i}(z)&M^{22}_{\kappa\nu\eta,i}(z)
\end{pmatrix},
\\[6pt]
N_{\kappa\nu\eta,i}(z)
&=
\begin{pmatrix}
N^{11}_{\kappa\nu\eta,i}(z)&N^{12}_{\kappa\nu\eta,i}(z)\\[3pt]
N^{21}_{\kappa\nu\eta,i}(z)&N^{22}_{\kappa\nu\eta,i}(z)
\end{pmatrix}.
\end{aligned}
\end{equation}
The entries of $D_{\kappa\nu\eta,i}(z)$ are, for
$i=1,\ldots,\hat{\kappa}-1$,
\begin{equation}
\label{eq:NewChargedHDiagonalCoefficients}
\begin{aligned}
&D^{11}_{\kappa\nu\eta,i}(z)
=\Biggl[
2\kappa-\hat{\kappa}-3i-\frac12
-3(\hat{\kappa}-i)\epsilon
+\frac34(1-\epsilon)
\left|\rho_{\nu,z}+2\eta-(1+\epsilon)F_{\nu,z}\right|
\\
&\;+\frac34(1+\epsilon)
\left|\rho_{\nu,z}+2\eta+(1-\epsilon)F_{\nu,z}\right|
\Biggr]^2-\frac14
+6i(1+\epsilon)
\Biggl[2\kappa-\hat{\kappa}-i-\frac12
\\
&\quad
-3\left(\hat{\kappa}-i-\frac12\right)\epsilon
+\frac34(1-\epsilon)
\left|\rho_{\nu,z}+2\eta-(1+\epsilon)F_{\nu,z}\right|
+\left|\rho_{\nu,z}+2\eta+(1-\epsilon)F_{\nu,z}\right|
\Biggr]
\\
&\quad
+3(\hat{\kappa}-i)(1-\epsilon)(1+3\epsilon)
\left[
\hat{\kappa}-i+\frac12
\left|\rho_{\nu,z}+2\eta-(1+\epsilon)F_{\nu,z}\right|
\right],
\\[6pt]
&D^{12}_{\kappa\nu\eta,i}(z)
=
3\sqrt{\frac{i}{
(i+1)(\hat{\kappa}-i)(\hat{\kappa}-i+1)}}
\\
&\quad\times
\Biggl[
(2\kappa-3\hat{\kappa}+i+1)\epsilon
-\hat{\kappa}+i
+\epsilon
\left|\rho_{\nu,z}+2\eta+(1-\epsilon)F_{\nu,z}\right|
\\
&\qquad
+\frac{\epsilon-1}{4}
\left(
\left|\rho_{\nu,z}+2\eta-(1+\epsilon)F_{\nu,z}\right|-2
\right)
\Biggr],
\\[6pt]
&D^{21}_{\kappa\nu\eta,i}(z)
=
18\epsilon(1+\epsilon)
\sqrt{i(i+1)(\hat{\kappa}-i)(\hat{\kappa}-i+1)},
\\[6pt]
&D^{22}_{\kappa\nu\eta,i}(z)
=\Biggl[
2\kappa-\hat{\kappa}-3i-\frac12
-3(\hat{\kappa}-i-1)\epsilon
+\frac34(1-\epsilon)
\left|\rho_{\nu,z}+2\eta-(1+\epsilon)F_{\nu,z}\right|
\\
&\;+\frac34(1+\epsilon)
\left|\rho_{\nu,z}+2\eta+(1-\epsilon)F_{\nu,z}\right|
\Biggr]^2-\frac14
+6i(1+\epsilon)
\Biggl[2\kappa-\hat{\kappa}-i-\frac12
\\
&\quad
-3\left(\hat{\kappa}-i-\frac12\right)\epsilon
+\frac34(1-\epsilon)
\left|\rho_{\nu,z}+2\eta-(1+\epsilon)F_{\nu,z}\right|
+\left|\rho_{\nu,z}+2\eta+(1-\epsilon)F_{\nu,z}\right|
\Biggr]
\\
&\quad
+3(\hat{\kappa}-i-1)(1-\epsilon)(1+3\epsilon)
\left[
\hat{\kappa}-i-1+\frac12
\left|\rho_{\nu,z}+2\eta-(1+\epsilon)F_{\nu,z}\right|
\right].
\end{aligned}
\end{equation}
The coefficient $D^{21}_{\kappa\nu\eta,i}(z)$ depends
on $z$ only through $\epsilon(z)$, and is also
independent of the labels $\nu$ and $\eta$.
For $i=1,\ldots,\hat{\kappa}-2$, the components of
$M_{\kappa\nu\eta,i}(z)$ are
\begin{equation}
\label{eq:NewChargedHUpperCoefficients}
\begin{aligned}
M^{11}_{\kappa\nu\eta,i}(z)
={}&
\frac{3(\epsilon-1)(\hat{\kappa}-i-1)}{\hat{\kappa}-i}
\sqrt{\frac{i(\hat{\kappa}-i)}
{(i+1)(\hat{\kappa}-i+1)}}
\\
&\quad\times
\left[
\hat{\kappa}-i-1+\frac12
\left|\rho_{\nu,z}+2\eta-(1+\epsilon)F_{\nu,z}\right|
\right]
\\
&\quad\times
\left[
3(i+1)\epsilon-i
+\frac{3\epsilon-1}{2}
\left|\rho_{\nu,z}+2\eta+(1-\epsilon)F_{\nu,z}\right|
\right],
\\[6pt]
M^{12}_{\kappa\nu\eta,i}(z)
={}&
\frac{3(1-\epsilon)}{2(\hat{\kappa}-i)}
\sqrt{\frac{i(\hat{\kappa}-i-1)}
{(i+2)(\hat{\kappa}-i+1)}}
\!\left[
\hat{\kappa}-i-1+\frac12
\left|\rho_{\nu,z}+2\eta-(1+\epsilon)F_{\nu,z}\right|
\right],
\\[6pt]
M^{21}_{\kappa\nu\eta,i}(z)
={}&
18\epsilon(\epsilon-1)(\hat{\kappa}-i-1)
\left[
\hat{\kappa}-i-1+\frac12
\left|\rho_{\nu,z}+2\eta-(1+\epsilon)F_{\nu,z}\right|
\right],
\\[6pt]
M^{22}_{\kappa\nu\eta,i}(z)
={}&
\frac{3(\epsilon-1)}{\hat{\kappa}-i}
\sqrt{\frac{(i+1)(\hat{\kappa}-i)(\hat{\kappa}-i-1)}
{i+2}}
\\
&\quad\times
\left[
\hat{\kappa}-i-1+\frac12
\left|\rho_{\nu,z}+2\eta-(1+\epsilon)F_{\nu,z}\right|
\right]
\\
&\quad\times
\left[
3i\epsilon-i-1
+\frac{3\epsilon-1}{2}
\left|\rho_{\nu,z}+2\eta+(1-\epsilon)F_{\nu,z}\right|
\right].
\end{aligned}
\end{equation}
Finally, the matrix $N_{\kappa\nu\eta,i}(z)$,
$i=1,\ldots,\hat{\kappa}-2$, depends on $z$ only
through $\epsilon(z)$, and is also independent of
the labels $\nu$ and $\eta$. Its components read
\begin{equation}
\label{eq:NewChargedHLowerCoefficients}
\begin{aligned}
N^{11}_{\kappa\nu\eta,i}(z)
&=3(1+\epsilon)(1+3\epsilon)
\sqrt{i(i+1)(\hat{\kappa}-i)(\hat{\kappa}-i+1)},
\\[6pt]
N^{12}_{\kappa\nu\eta,i}(z)
&=\frac32(1+3\epsilon),
\\[6pt]
N^{21}_{\kappa\nu\eta,i}(z)
&=0,
\\[6pt]
N^{22}_{\kappa\nu\eta,i}(z)
&=3(1+\epsilon)(1+3\epsilon)
\sqrt{(i+1)(i+2)(\hat{\kappa}-i-1)(\hat{\kappa}-i)}.
\end{aligned}
\end{equation}
With these definitions, the first few polynomials
$H_{\kappa\nu\eta}(x;z)$, $\kappa=1,\ldots,4$, in
(\ref{eq:PolyMatCharged}) are, for all $\nu\geq0$
and $\eta=\pm1$,
\begin{equation}
\label{eq:ChargedHFirstMembers}
\begin{aligned}
H_{\kappa\nu\eta}(x;z)
={}&x+\frac14-
\Biggl[
2\kappa-\frac32
+\frac34(1-\epsilon)
\left|\rho_{\nu,z}+2\eta-(1+\epsilon)F_{\nu,z}\right|
\\
&\qquad
+\frac34(1+\epsilon)
\left|\rho_{\nu,z}+2\eta+(1-\epsilon)F_{\nu,z}\right|
\Biggr]^2,
\qquad \kappa=1,2,
\\[8pt]
H_{\kappa\nu\eta}(x;z)
={}&\bigl(x-d_{\kappa\nu\eta}(z)\bigr)
\Bigl[
\bigl(x-D^{11}_{\kappa\nu\eta,1}(z)\bigr)
\bigl(x-D^{22}_{\kappa\nu\eta,1}(z)\bigr)
-D^{12}_{\kappa\nu\eta,1}(z)D^{21}_{\kappa\nu\eta,1}(z)
\Bigr]
\\
&-E_{\kappa\nu\eta}(z)
\begin{pmatrix}
x-D^{22}_{\kappa\nu\eta,1}(z)
&
D^{12}_{\kappa\nu\eta,1}(z)
\\[3pt]
D^{21}_{\kappa\nu\eta,1}(z)
&
x-D^{11}_{\kappa\nu\eta,1}(z)
\end{pmatrix}
F_{\kappa\nu\eta}(z),
\qquad \kappa=3,4.
\end{aligned}
\end{equation}

The signs $s_{\kappa\nu\eta a} = \pm1 $ in
(\ref{eq:GinoRCharge}) are given, provided the root is not
a common zero of the auxiliary polynomials $H_{(1)\kappa\nu\eta}(x;z)$, $H_{(2)\kappa\nu\eta}(x;z)$ defined below, by
\begin{equation}
\label{eq:NewChargedHSignRule}
s_{\kappa\nu\eta a}(z)
=\operatorname{sgn}\!\left[
H_{(1)\kappa\nu\eta}(h_{\kappa\nu\eta a, z};z)\,
H_{(2)\kappa\nu\eta}(h_{\kappa\nu\eta a, z};z)
\right].
\end{equation}
Here $H_{(1)\kappa\nu\eta}(x;z)$ and
$H_{(2)\kappa\nu\eta}(x;z)$ are auxiliary polynomials,
\begin{equation}
\label{eq:NewChargedHAuxiliaryLeading}
\begin{aligned}
H_{(1)\kappa\nu\eta}(x;z)
&=x^{\hat{\kappa}-1}
  +\textrm{terms of lower degree in $x$}\, ,\\[4pt]
H_{(2)\kappa\nu\eta}(x;z)
&=B_{\kappa\nu\eta}(z)\,x^{\hat{\kappa}-1}
  +\textrm{terms of lower degree in $x$}\, ,
\end{aligned}
\end{equation}
with
\begin{equation}
\label{eq:NewChargedHAuxiliaryCoefficient}
\begin{aligned}
B_{\kappa\nu\eta}(z)=\eta\Biggl[
&2\kappa-\hat{\kappa}-\frac12\\
&+\frac34\bigl(1-\epsilon(z)\bigr)
 \left|\rho_{\nu,z}+2\eta
       -\bigl(1+\epsilon(z)\bigr)F_{\nu,z}\right|\\
&+\frac34\bigl(1+\epsilon(z)\bigr)
 \left|\rho_{\nu,z}+2\eta
       +\bigl(1-\epsilon(z)\bigr)F_{\nu,z}\right|
\Biggr]\, ,
\end{aligned}
\end{equation}
that decompose $H_{\kappa\nu\eta}(x;z)$ as
\begin{equation}
\label{eq:NewChargedHAuxiliaryFactorization}
H_{\kappa\nu\eta}(x;z)
=\bigl(x+\tfrac14\bigr)H_{(1)\kappa\nu\eta}(x;z)^2
-H_{(2)\kappa\nu\eta}(x;z)^2\, .
\end{equation}
This decomposition is imposed as a polynomial identity before
specialisation to a particular background. For
$\hat{\kappa}=1$, the terms of lower degree in $x$ are absent.
The SCP with Lorentz spins $(0,\frac12)$ then has R-charge $\sigma_{\kappa \nu \eta a}$ in (\ref{eq:GinoRCharge}) and flavour charge $s_{\kappa\nu\eta a}(z)F_{\nu,z}$, while its CPT partner has opposite charges.

To include common zeros at a fixed background, define
\begin{equation}
\label{eq:NewChargedHCommonFactor}
\begin{aligned}
D_{\kappa\nu\eta}(x;z)
&=\gcd_x\!\left(
H_{(1)\kappa\nu\eta}(x;z),
H_{(2)\kappa\nu\eta}(x;z)
\right),\\[4pt]
\overline H_{(i)\kappa\nu\eta}(x;z)
&=\frac{H_{(i)\kappa\nu\eta}(x;z)}
        {D_{\kappa\nu\eta}(x;z)}\, ,
\qquad i=1,2\, ,
\end{aligned}
\end{equation}
where the greatest common divisor is monic and is taken after
specialising $z$. A root of $D_{\kappa\nu\eta}(x;z)$ of
multiplicity $q$ gives $2q$ roots of
$H_{\kappa\nu\eta}(x;z)$, counted with algebraic multiplicity:
$q$ are assigned $s_{\kappa\nu\eta a}(z)=+1$ and $q$ are
assigned $s_{\kappa\nu\eta a}(z)=-1$.
For the remaining roots of
$H_{\kappa\nu\eta}(x;z)/D_{\kappa\nu\eta}(x;z)^2$, use
(\ref{eq:NewChargedHSignRule}) with
$H_{(i)\kappa\nu\eta}(x;z)$ replaced by
$\overline H_{(i)\kappa\nu\eta}(x;z)$.
Coincident roots are assigned separate indices $a$, and
protected multiplets are excluded individually as in
section~\ref{sec:FlavourChargedLong}. The assignment of signs to roots ordered by dimension need not remain fixed. In appendix~\ref{app:NeutralMatrixGravitino} we will discuss examples in the simpler $\nu=0$ case.

\subsubsection{General vector multiplet polynomials
$P_{\kappa\nu}(x;z)$}
\label{app:NewChargedInnerVector}

For $\nu\geq1$, the charged boundary members are
\begin{equation*}
P_{\kappa\nu}(x;z)
=x-\frac94 A_{\nu,z}^{\,2}
  +\frac92\nu f_z(1-\epsilon^2)
  -6\kappa A_{\nu,z},
\qquad \kappa=0,1,
\end{equation*}
where $A_{\nu,z}=\nu(f_z+|\epsilon|\lambda_z)$ and
$\epsilon=\epsilon(z)$. These polynomials have degree one
and correspond to $1\times1$ matrices. Both boundary
members occur physically for $|z|<1$; only $\kappa=1$
occurs at $|z|=1$, and neither occurs for $|z|>1$.
They do not introduce any flavour-neutral members.
Protected occurrences are excluded from the long spectrum
according to the label restrictions in
section~\ref{sec:FlavourChargedLong}.

The $2\check{\kappa}$-dimensional matrix
$\cP_{\kappa\nu}(z)$ that determines the polynomial
$P_{\kappa\nu}(x;z)$ through (\ref{eq:PolyMatCharged})
can be written, for all $\nu\geq0$, $\kappa \geq 2$ as follows. For $\kappa=2,3$, the matrix is
\begin{equation}
\label{eq:ChargedPUniformSmallMatrix}
\cP_{\kappa\nu}(z)=
\begin{pmatrix}
P^{11}_{\kappa\nu}(z)&P^{12}_{\kappa\nu}(z)\\[3pt]
P^{21}_{\kappa\nu}(z)&P^{22}_{\kappa\nu}(z)
\end{pmatrix},
\qquad \kappa=2,3,
\end{equation}
with components
\begin{equation}
\label{eq:ChargedPUniformSmallEntries}
\begin{aligned}
&P^{11}_{\kappa\nu}(z)
=\frac9{16}
\Bigl[
(1-\epsilon)
\left|\rho_{\nu,z}-(1+\epsilon)F_{\nu,z}\right|
+(1+\epsilon)
\left|\rho_{\nu,z}+(1-\epsilon)F_{\nu,z}\right|
\Bigr]^2
\\
&\quad
+\frac34(1-\epsilon)(4\kappa-11-3\epsilon)
\left|\rho_{\nu,z}-(1+\epsilon)F_{\nu,z}\right|
\\
&\quad
+\frac34(1+\epsilon)(4\kappa-11+3\epsilon)
\left|\rho_{\nu,z}+(1-\epsilon)F_{\nu,z}\right|,
\\[5pt]
&P^{12}_{\kappa\nu}(z)=1,
\\[5pt]
&P^{21}_{\kappa\nu}(z)
=27\epsilon
\Bigl[
(1+\epsilon)(1-3\epsilon)
\left|\rho_{\nu,z}+(1-\epsilon)F_{\nu,z}\right|
\\
&\qquad
-(1-\epsilon)(1+3\epsilon)
\left|\rho_{\nu,z}-(1+\epsilon)F_{\nu,z}\right|
\Bigr]
\\
&\quad
-\frac92(1-\epsilon^2)
\Bigl[
8(\kappa-2)
+3(1-\epsilon)
\left|\rho_{\nu,z}-(1+\epsilon)F_{\nu,z}\right|
+3(1+\epsilon)
\left|\rho_{\nu,z}+(1-\epsilon)F_{\nu,z}\right|
\Bigr]
\\
&\quad\times
\Bigl[
\kappa(\kappa-1)
+2\left|\rho_{\nu,z}-(1+\epsilon)F_{\nu,z}\right|
+2\left|\rho_{\nu,z}+(1-\epsilon)F_{\nu,z}\right|
\Bigr],
\\[5pt]
&P^{22}_{\kappa\nu}(z)
=6(4\kappa-7+3\epsilon^2)
\\
&\quad
+\frac9{16}
\Bigl[
(1-\epsilon)
\left|\rho_{\nu,z}-(1+\epsilon)F_{\nu,z}\right|
+(1+\epsilon)
\left|\rho_{\nu,z}+(1-\epsilon)F_{\nu,z}\right|
\Bigr]^2
\\
&\quad
+\frac34(1-\epsilon)(4\kappa+3+3\epsilon)
\left|\rho_{\nu,z}-(1+\epsilon)F_{\nu,z}\right|
\\
&\quad
+\frac34(1+\epsilon)(4\kappa+3-3\epsilon)
\left|\rho_{\nu,z}+(1-\epsilon)F_{\nu,z}\right|.
\end{aligned}
\end{equation}
For $\kappa=4, 5$, it is given by the 
$4\times4$ matrix
\begin{equation}
\label{eq:ChargedPUniformFirstMatriceskappa45}
\cP_{\kappa\nu}(z)
=Y_{\kappa\nu}(z),
 \qquad \kappa=4,5,
\end{equation}
for $\kappa=6,7$, it is given by the 
$6\times6$ matrix
\begin{equation}
\label{eq:ChargedPUniformFirstMatriceskappa67}
\cP_{\kappa\nu}(z)
=
\begin{pmatrix}
Y_{\kappa\nu}(z)&E_{\kappa\nu}(z)\\[3pt]
F_{\kappa\nu}(z)&X_{\kappa\nu,2}(z)
\end{pmatrix}, \qquad \kappa=6,7,
\end{equation}
for $\kappa=8,9$ it is the 
$8\times8$ matrix
\begin{equation}
\label{eq:ChargedPUniformMatrix89}
\begin{aligned}
\cP_{\kappa\nu}(z)
&=
\begin{pmatrix}
Y_{\kappa\nu}(z)
&E_{\kappa\nu}(z)
&Z_{\kappa\nu}(z)
\\[3pt]
F_{\kappa\nu}(z)
&X_{\kappa\nu,2}(z)
&U_{\kappa\nu,3}(z)
\\[3pt]
0_{2\times4}
&\ell_{\kappa\nu,2}(z)\mathbbm{1}_2
&X_{\kappa\nu,3}(z)
\end{pmatrix},
&&\kappa=8,9.
\end{aligned}
\end{equation}
and, finally, for $\kappa\geq10$, it takes on the general form
{\setlength{\arraycolsep}{2pt}
\begin{equation}
\label{eq:ChargedPUniformMatrix}
\cP_{\kappa\nu}(z)=
\begin{pmatrix}
Y_{\kappa\nu}(z)
&E_{\kappa\nu}(z)
&Z_{\kappa\nu}(z)
&\cdots
&0_{4\times2}
\\[3pt]
F_{\kappa\nu}(z)
&X_{\kappa\nu,2}(z)
&U_{\kappa\nu,3}(z)
&\ddots
&\vdots
\\[3pt]
0_{2\times4}
&\ddots
&\ddots
&\ddots
&V_{\kappa\nu,\check{\kappa}-1}(z)
\\[3pt]
0_{2\times4}
&\cdots
&\ell_{\kappa\nu,\check{\kappa}-3}(z)\mathbbm{1}_2
&X_{\kappa\nu,\check{\kappa}-2}(z)
&U_{\kappa\nu,\check{\kappa}-1}(z)
\\[3pt]
0_{2\times4}
&\cdots
&0_{2\times2}
&\ell_{\kappa\nu,\check{\kappa}-2}(z)\mathbbm{1}_2
&X_{\kappa\nu,\check{\kappa}-1}(z)
\end{pmatrix}.
\end{equation}
}All blocks outside the displayed bands vanish.
The matrices $E_{\kappa\nu}(z)$ and $Z_{\kappa\nu}(z)$
have size $4\times2$, while $F_{\kappa\nu}(z)$ has
size $2\times4$. The blocks $X_{\kappa\nu,i}(z)$, $U_{\kappa\nu,i}(z)$ and  $V_{\kappa\nu,i}(z)$ are $2 \times 2$, and $\ell_{\kappa\nu,i}(z)$ is a scalar.

The $4 \times 4$ block $Y_{\kappa\nu} (z)$ has components
\begin{equation}
\label{eq:ChargedPUniformBoundaryMatrix}
Y_{\kappa\nu}(z)=
\begin{pmatrix}
Y^{11}_{\kappa\nu}(z)
&Y^{12}_{\kappa\nu}(z)
&Y^{13}_{\kappa\nu}(z)
&Y^{14}_{\kappa\nu}(z)
\\[3pt]
Y^{21}_{\kappa\nu}(z)
&Y^{22}_{\kappa\nu}(z)
&Y^{23}_{\kappa\nu}(z)
&Y^{24}_{\kappa\nu}(z)
\\[3pt]
Y^{31}_{\kappa\nu}(z)
&Y^{32}_{\kappa\nu}(z)
&Y^{33}_{\kappa\nu}(z)
&Y^{34}_{\kappa\nu}(z)
\\[3pt]
Y^{41}_{\kappa\nu}(z)
&Y^{42}_{\kappa\nu}(z)
&Y^{43}_{\kappa\nu}(z)
&Y^{44}_{\kappa\nu}(z)
\end{pmatrix} ,
\end{equation}
that can be written as follows. The first-row entries are
\begin{equation}
\label{eq:ChargedPUniformBoundaryEntries}
\begin{aligned}
&Y^{11}_{\kappa\nu}(z)
=24(1+\kappa-2\check{\kappa})
\\
&\quad
+\frac9{16}
\Bigl[
(1-\epsilon)
\left|\rho_{\nu,z}-(1+\epsilon)F_{\nu,z}\right|
+(1+\epsilon)
\left|\rho_{\nu,z}+(1-\epsilon)F_{\nu,z}\right|
\Bigr]^2
\\
&\quad
+\frac34(1-\epsilon)
(4\kappa-8\check{\kappa}+13+9\epsilon)
\left|\rho_{\nu,z}-(1+\epsilon)F_{\nu,z}\right|
\\
&\quad
+\frac34(1+\epsilon)
(4\kappa-8\check{\kappa}+13-9\epsilon)
\left|\rho_{\nu,z}+(1-\epsilon)F_{\nu,z}\right|,
\\[5pt]
&Y^{12}_{\kappa\nu}(z)
=12(1-\kappa+2\check{\kappa})
\\
&\quad
-\frac92
\Bigl[
(1-\epsilon)
\left|\rho_{\nu,z}-(1+\epsilon)F_{\nu,z}\right|
+(1+\epsilon)
\left|\rho_{\nu,z}+(1-\epsilon)F_{\nu,z}\right|
\Bigr],
\\[5pt]
&Y^{13}_{\kappa\nu}(z)
=-12(7+2\kappa-4\check{\kappa})
\\
&\quad
-9
\Bigl[
(1-\epsilon)
\left|\rho_{\nu,z}-(1+\epsilon)F_{\nu,z}\right|
+(1+\epsilon)
\left|\rho_{\nu,z}+(1-\epsilon)F_{\nu,z}\right|
\Bigr],
\\[5pt]
&Y^{14}_{\kappa\nu}(z)
=6(1-\epsilon^2)\kappa(\kappa-1)
\\
&\quad
+\frac32(1-\epsilon)
\bigl[8\check{\kappa}+(8\check{\kappa}-1)\epsilon
      -3\epsilon^2\bigr]
\left|\rho_{\nu,z}-(1+\epsilon)F_{\nu,z}\right|
\\
&\quad
+\frac32(1+\epsilon)
\bigl[8\check{\kappa}-(8\check{\kappa}-1)\epsilon
      -3\epsilon^2\bigr]
\left|\rho_{\nu,z}+(1-\epsilon)F_{\nu,z}\right|.
\end{aligned}
\end{equation}
The second-row entries are
\begin{equation}
\label{eq:ChargedPUniformBoundaryRelations}
\begin{aligned}
&Y^{21}_{\kappa\nu}(z)
=-3(1-\epsilon^2)\bigl[\kappa(\kappa-1)+4\bigr]
\\
&\quad
-\frac32(1-\epsilon)
\bigl[4\check{\kappa}+(4\check{\kappa}-1)\epsilon
      -3\epsilon^2\bigr]
\left|\rho_{\nu,z}-(1+\epsilon)F_{\nu,z}\right|
\\
&\quad
-\frac32(1+\epsilon)
\bigl[4\check{\kappa}-(4\check{\kappa}-1)\epsilon
      -3\epsilon^2\bigr]
\left|\rho_{\nu,z}+(1-\epsilon)F_{\nu,z}\right|,
\\[5pt]
&Y^{22}_{\kappa\nu}(z)
=24(\kappa-2\check{\kappa})-6+18\epsilon^2
\\
&\quad
+\frac9{16}
\Bigl[
(1-\epsilon)
\left|\rho_{\nu,z}-(1+\epsilon)F_{\nu,z}\right|
+(1+\epsilon)
\left|\rho_{\nu,z}+(1-\epsilon)F_{\nu,z}\right|
\Bigr]^2
\\
&\quad
+\frac34(1-\epsilon)
(4\kappa-8\check{\kappa}+11+3\epsilon)
\left|\rho_{\nu,z}-(1+\epsilon)F_{\nu,z}\right|
\\
&\quad
+\frac34(1+\epsilon)
(4\kappa-8\check{\kappa}+11-3\epsilon)
\left|\rho_{\nu,z}+(1-\epsilon)F_{\nu,z}\right|,
\\[5pt]
&Y^{23}_{\kappa\nu}(z)
=24
-3(1-\epsilon)(1+3\epsilon)
\left|\rho_{\nu,z}-(1+\epsilon)F_{\nu,z}\right|
\\
&\quad
-3(1+\epsilon)(1-3\epsilon)
\left|\rho_{\nu,z}+(1-\epsilon)F_{\nu,z}\right|,
\\[5pt]
&Y^{24}_{\kappa\nu}(z)
=-6(1-\epsilon^2)\kappa(\kappa-1)
\\
&\quad
-3(1-\epsilon)
\bigl[4\check{\kappa}+(4\check{\kappa}-1)\epsilon
      -3\epsilon^2\bigr]
\left|\rho_{\nu,z}-(1+\epsilon)F_{\nu,z}\right|
\\
&\quad
-3(1+\epsilon)
\bigl[4\check{\kappa}-(4\check{\kappa}-1)\epsilon
      -3\epsilon^2\bigr]
\left|\rho_{\nu,z}+(1-\epsilon)F_{\nu,z}\right|.
\end{aligned}
\end{equation}
The third-row entries are
\begin{equation}
\label{eq:ChargedPUniformBoundaryThirdRow}
\begin{aligned}
&Y^{31}_{\kappa\nu}(z)
=6(1-\epsilon^2)(\check{\kappa}-1)
(2\kappa-2\check{\kappa}+1)
\\
&\quad
+\frac32(1-\epsilon)
\bigl[4\check{\kappa}-5+(4\check{\kappa}-7)\epsilon\bigr]
\left|\rho_{\nu,z}-(1+\epsilon)F_{\nu,z}\right|
\\
&\quad
+\frac32(1+\epsilon)
\bigl[4\check{\kappa}-5-(4\check{\kappa}-7)\epsilon\bigr]
\left|\rho_{\nu,z}+(1-\epsilon)F_{\nu,z}\right|,
\\[5pt]
&Y^{32}_{\kappa\nu}(z)
=6+12\kappa-24\check{\kappa}
\\
&\quad
+\frac92
\Bigl[
(1-\epsilon)
\left|\rho_{\nu,z}-(1+\epsilon)F_{\nu,z}\right|
+(1+\epsilon)
\left|\rho_{\nu,z}+(1-\epsilon)F_{\nu,z}\right|
\Bigr],
\\[5pt]
&Y^{33}_{\kappa\nu}(z)
=102+48\kappa-96\check{\kappa}-18\epsilon^2
\\
&\quad
+\frac9{16}
\Bigl[
(1-\epsilon)
\left|\rho_{\nu,z}-(1+\epsilon)F_{\nu,z}\right|
+(1+\epsilon)
\left|\rho_{\nu,z}+(1-\epsilon)F_{\nu,z}\right|
\Bigr]^2
\\
&\quad
+\frac34(1-\epsilon)
(4\kappa-8\check{\kappa}+25+9\epsilon)
\left|\rho_{\nu,z}-(1+\epsilon)F_{\nu,z}\right|
\\
&\quad
+\frac34(1+\epsilon)
(4\kappa-8\check{\kappa}+25-9\epsilon)
\left|\rho_{\nu,z}+(1-\epsilon)F_{\nu,z}\right|,
\\[5pt]
&Y^{34}_{\kappa\nu}(z)
=-3(1-\epsilon^2)\kappa(\kappa-1)
\\
&\quad
-\frac32(1-\epsilon)
\bigl[4\check{\kappa}+(4\check{\kappa}-1)\epsilon
      -3\epsilon^2\bigr]
\left|\rho_{\nu,z}-(1+\epsilon)F_{\nu,z}\right|
\\
&\quad
-\frac32(1+\epsilon)
\bigl[4\check{\kappa}-(4\check{\kappa}-1)\epsilon
      -3\epsilon^2\bigr]
\left|\rho_{\nu,z}+(1-\epsilon)F_{\nu,z}\right|.
\end{aligned}
\end{equation}
Finally, the fourth-row entries are
\begin{equation}
\label{eq:ChargedPUniformBoundaryFourthRow}
\begin{aligned}
&Y^{41}_{\kappa\nu}(z)
=-6(3+2\kappa-4\check{\kappa})
\\
&\quad
-\frac92
\Bigl[
(1-\epsilon)
\left|\rho_{\nu,z}-(1+\epsilon)F_{\nu,z}\right|
+(1+\epsilon)
\left|\rho_{\nu,z}+(1-\epsilon)F_{\nu,z}\right|
\Bigr],
\\[5pt]
&Y^{42}_{\kappa\nu}(z)=-18,
\qquad
Y^{43}_{\kappa\nu}(z)=18,
\\[5pt]
&Y^{44}_{\kappa\nu}(z)
=\frac9{16}
\Bigl[
(1-\epsilon)
\left|\rho_{\nu,z}-(1+\epsilon)F_{\nu,z}\right|
+(1+\epsilon)
\left|\rho_{\nu,z}+(1-\epsilon)F_{\nu,z}\right|
\Bigr]^2
\\
&\quad
+\frac34(1-\epsilon)
(4\kappa-8\check{\kappa}-3-3\epsilon)
\left|\rho_{\nu,z}-(1+\epsilon)F_{\nu,z}\right|
\\
&\quad
+\frac34(1+\epsilon)
(4\kappa-8\check{\kappa}-3+3\epsilon)
\left|\rho_{\nu,z}+(1-\epsilon)F_{\nu,z}\right|.
\end{aligned}
\end{equation}

The blocks $E_{\kappa\nu}(z)$ and $Z_{\kappa\nu}(z)$ read
\begin{equation}
\label{eq:ChargedPUniformUpperBoundary}
E_{\kappa\nu}(z)=
\begin{pmatrix}
54&0\\
-36&-18\\
E^{31}_{\kappa\nu}(z)&-18\\
0&0
\end{pmatrix},
\qquad
Z_{\kappa\nu}(z)=
\begin{pmatrix}
0&0\\
0&0\\
108&0\\
0&0
\end{pmatrix},
\end{equation}
where
\begin{equation}
\label{eq:ChargedPUniformUpperBoundaryEntry}
\begin{aligned}
E^{31}_{\kappa\nu}(z)
={}&-18(13+2\kappa-4\check{\kappa})
\\
&-\frac{27}{2}
\Bigl[
(1-\epsilon)
\left|\rho_{\nu,z}-(1+\epsilon)F_{\nu,z}\right|
+(1+\epsilon)
\left|\rho_{\nu,z}+(1-\epsilon)F_{\nu,z}\right|
\Bigr] ,
\end{aligned}
\end{equation}
while $F_{\kappa\nu}(z)$ is 
\begin{equation}
\label{eq:ChargedPUniformLowerBoundary}
F_{\kappa\nu}(z)=
\begin{pmatrix}
0&0&F^{13}_{\kappa\nu}(z)&0\\[3pt]
F^{21}_{\kappa\nu}(z)
&F^{22}_{\kappa\nu}(z)
&F^{23}_{\kappa\nu}(z)
&F^{24}_{\kappa\nu}(z)
\end{pmatrix},
\end{equation}
with nonzero components
\begin{equation}
\label{eq:ChargedPUniformLowerBoundaryEntries}
\begin{aligned}
&F^{13}_{\kappa\nu}(z)
=6(1-\epsilon^2)(\check{\kappa}-2)
\\
&\quad\times
\Bigl[
2\kappa-2\check{\kappa}+3
+\left|\rho_{\nu,z}-(1+\epsilon)F_{\nu,z}\right|
+\left|\rho_{\nu,z}+(1-\epsilon)F_{\nu,z}\right|
\Bigr],
\\[5pt]
&F^{21}_{\kappa\nu}(z)
=-\frac12(1-\epsilon^2)(\check{\kappa}-2)
\\
&\quad\times
\Bigl[
2\kappa-2\check{\kappa}+3
+\left|\rho_{\nu,z}-(1+\epsilon)F_{\nu,z}\right|
+\left|\rho_{\nu,z}+(1-\epsilon)F_{\nu,z}\right|
\Bigr]
\\
&\quad\times
\Bigl[
(1-\epsilon)(1+3\epsilon)
\left|\rho_{\nu,z}-(1+\epsilon)F_{\nu,z}\right|
+(1+\epsilon)(1-3\epsilon)
\left|\rho_{\nu,z}+(1-\epsilon)F_{\nu,z}\right|
\Bigr],
\\[5pt]
&F^{22}_{\kappa\nu}(z)
=(1-\epsilon^2)(\check{\kappa}-2)
\\
&\quad\times
\Bigl[
2\kappa-2\check{\kappa}+3
+\left|\rho_{\nu,z}-(1+\epsilon)F_{\nu,z}\right|
+\left|\rho_{\nu,z}+(1-\epsilon)F_{\nu,z}\right|
\Bigr]
\\
&\quad\times
\Bigl[
2+4\kappa-8\check{\kappa}
+\frac32(1-\epsilon)
\left|\rho_{\nu,z}-(1+\epsilon)F_{\nu,z}\right|
+\frac32(1+\epsilon)
\left|\rho_{\nu,z}+(1-\epsilon)F_{\nu,z}\right|
\Bigr],
\\[5pt]
&F^{23}_{\kappa\nu}(z)
=-4(1-\epsilon^2)(\check{\kappa}-2)
\\
&\quad\times
\Bigl[
2\kappa-2\check{\kappa}+3
+\left|\rho_{\nu,z}-(1+\epsilon)F_{\nu,z}\right|
+\left|\rho_{\nu,z}+(1-\epsilon)F_{\nu,z}\right|
\Bigr],
\\[5pt]
&F^{24}_{\kappa\nu}(z)
=-(1-\epsilon^2)(\check{\kappa}-2)
\\
&\quad\times
\Bigl[
2\kappa-2\check{\kappa}+3
+\left|\rho_{\nu,z}-(1+\epsilon)F_{\nu,z}\right|
+\left|\rho_{\nu,z}+(1-\epsilon)F_{\nu,z}\right|
\Bigr]
\\
&\quad\times
\Biggl[
(1-\epsilon^2)\kappa(\kappa-1)
\\
&\qquad
+\frac12(1-\epsilon)
\bigl[4\check{\kappa}+(4\check{\kappa}-1)\epsilon
      -3\epsilon^2\bigr]
\left|\rho_{\nu,z}-(1+\epsilon)F_{\nu,z}\right|
\\
&\qquad
+\frac12(1+\epsilon)
\bigl[4\check{\kappa}-(4\check{\kappa}-1)\epsilon
      -3\epsilon^2\bigr]
\left|\rho_{\nu,z}+(1-\epsilon)F_{\nu,z}\right|
\Biggr].
\end{aligned}
\end{equation}
The blocks $E_{\kappa\nu}(z)$ and $F_{\kappa\nu}(z)$
are required only for $\kappa\geq6$, and
$Z_{\kappa\nu}(z)$ only for $\kappa\geq8$.

The scalar coefficients $\ell_{\kappa\nu,i}(z)$ are, for $i=0,\ldots,\check{\kappa}-2$,
\begin{equation}
\label{eq:ChargedPCompactLowerBand}
\begin{aligned}
\ell_{\kappa\nu,i}(z)
={}&6(1-\epsilon^2)(\check{\kappa}-i-1)
\\
&\times
\Bigl[
2\kappa-2\check{\kappa}+2i+1
+\left|\rho_{\nu,z}-(1+\epsilon)F_{\nu,z}\right|
+\left|\rho_{\nu,z}+(1-\epsilon)F_{\nu,z}\right|
\Bigr] .
\end{aligned}
\end{equation}
The $2\times2$ matrices $X_{\kappa\nu,i}(z)$ and $U_{\kappa\nu,i}(z)$ are
\begin{equation}
\label{eq:ChargedPBlockMatrices}
X_{\kappa\nu,i}(z)
=
\begin{pmatrix}
X^{11}_{\kappa\nu,i}(z)&X^{12}_{\kappa\nu,i}(z)\\[3pt]
X^{21}_{\kappa\nu,i}(z)&X^{22}_{\kappa\nu,i}(z)
\end{pmatrix},
\qquad
U_{\kappa\nu,i}(z)
=
\begin{pmatrix}
U^{11}_{\kappa\nu,i}(z)&U^{12}_{\kappa\nu,i}(z)\\[3pt]
U^{21}_{\kappa\nu,i}(z)&U^{22}_{\kappa\nu,i}(z)
\end{pmatrix}.
\end{equation}
For $i=0,\ldots,\check{\kappa}-1$, the components of
$X_{\kappa\nu,i}(z)$ are
\begin{equation}
\label{eq:ChargedPCompactDiagonalBlocks}
\begin{aligned}
&X^{11}_{\kappa\nu,i}(z)
=3(i+1)
\bigl[(11-3\epsilon^2)i+8+8\kappa-16\check{\kappa}\bigr]
\\
&\quad
+\frac9{16}
\Bigl[
(1-\epsilon)
\left|\rho_{\nu,z}-(1+\epsilon)F_{\nu,z}\right|
+(1+\epsilon)
\left|\rho_{\nu,z}+(1-\epsilon)F_{\nu,z}\right|
\Bigr]^2
\\
&\quad
+3(2i+2+\kappa-2\check{\kappa}+\delta_{i0})
\Bigl[
(1-\epsilon)
\left|\rho_{\nu,z}-(1+\epsilon)F_{\nu,z}\right|
+(1+\epsilon)
\left|\rho_{\nu,z}+(1-\epsilon)F_{\nu,z}\right|
\Bigr]
\\
&\quad
+\frac94(1-\epsilon^2)(2i+1-2\delta_{i0})
\Bigl[
\left|\rho_{\nu,z}-(1+\epsilon)F_{\nu,z}\right|
+\left|\rho_{\nu,z}+(1-\epsilon)F_{\nu,z}\right|
\Bigr],
\\[5pt]
&X^{12}_{\kappa\nu,i}(z)
=18+
\Biggl[
6+12\kappa-24\check{\kappa}
\\
&\qquad
+\frac92
\Bigl[
(1-\epsilon)
\left|\rho_{\nu,z}-(1+\epsilon)F_{\nu,z}\right|
+(1+\epsilon)
\left|\rho_{\nu,z}+(1-\epsilon)F_{\nu,z}\right|
\Bigr]
\Biggr]\delta_{i0},
\\[5pt]
&X^{21}_{\kappa\nu,i}(z)
=-24i(i+1)
-3(1-\epsilon^2)
\bigl[(\kappa-2)(\kappa+1)-2i(i-1)+6\bigr]
\\
&\quad
+\frac12(9\epsilon^2-12i-15)
\Bigl[
(1-\epsilon)
\left|\rho_{\nu,z}-(1+\epsilon)F_{\nu,z}\right|
+(1+\epsilon)
\left|\rho_{\nu,z}+(1-\epsilon)F_{\nu,z}\right|
\Bigr]
\\
&\quad
+\frac32(1-\epsilon^2)(6i+7-4\check{\kappa})
\Bigl[
\left|\rho_{\nu,z}-(1+\epsilon)F_{\nu,z}\right|
+\left|\rho_{\nu,z}+(1-\epsilon)F_{\nu,z}\right|
\Bigr],
\\[5pt]
&X^{22}_{\kappa\nu,i}(z)
=24i(i-1+\kappa-2\check{\kappa})
+9(1-\epsilon^2)(i+2)(i-1)
\\
&\quad
+\frac9{16}
\Bigl[
(1-\epsilon)
\left|\rho_{\nu,z}-(1+\epsilon)F_{\nu,z}\right|
+(1+\epsilon)
\left|\rho_{\nu,z}+(1-\epsilon)F_{\nu,z}\right|
\Bigr]^2
\\
&\quad
+3(2i-1+\kappa-2\check{\kappa})
\Bigl[
(1-\epsilon)
\left|\rho_{\nu,z}-(1+\epsilon)F_{\nu,z}\right|
+(1+\epsilon)
\left|\rho_{\nu,z}+(1-\epsilon)F_{\nu,z}\right|
\Bigr]
\\
&\quad
+\frac94(1-\epsilon^2)(2i+1)
\Bigl[
\left|\rho_{\nu,z}-(1+\epsilon)F_{\nu,z}\right|
+\left|\rho_{\nu,z}+(1-\epsilon)F_{\nu,z}\right|
\Bigr].
\end{aligned}
\end{equation}
For $i=1,\ldots,\check{\kappa}-1$, the components of
$U_{\kappa\nu,i}(z)$ are
\begin{equation}
\label{eq:ChargedPCompactFirstUpperBand}
\begin{aligned}
&U^{11}_{\kappa\nu,i}(z)
=-6(i+1)(5i+3+2\kappa-4\check{\kappa})
-12\delta_{i1}
\\
&\quad
-\frac92(i+1)
\Bigl[
(1-\epsilon)
\left|\rho_{\nu,z}-(1+\epsilon)F_{\nu,z}\right|
+(1+\epsilon)
\left|\rho_{\nu,z}+(1-\epsilon)F_{\nu,z}\right|
\Bigr],
\\[4pt]
&U^{12}_{\kappa\nu,i}(z)
=-18(1+\delta_{i1}),
\\[4pt]
&U^{21}_{\kappa\nu,i}(z)
=24i(i+1)
\\
&\quad
-\frac32(i+1)
\Bigl[
(1-\epsilon)(1+3\epsilon)
\left|\rho_{\nu,z}-(1+\epsilon)F_{\nu,z}\right|
+(1+\epsilon)(1-3\epsilon)
\left|\rho_{\nu,z}+(1-\epsilon)F_{\nu,z}\right|
\Bigr],
\\[4pt]
&U^{22}_{\kappa\nu,i}(z)
=-6\bigl[
5i^2-10i-1+2(\kappa-2\check{\kappa})(i-1)
\bigr]
\\
&\quad
-\frac92(i-1)
\Bigl[
(1-\epsilon)
\left|\rho_{\nu,z}-(1+\epsilon)F_{\nu,z}\right|
+(1+\epsilon)
\left|\rho_{\nu,z}+(1-\epsilon)F_{\nu,z}\right|
\Bigr].
\end{aligned}
\end{equation}
Finally, $V_{\kappa\nu,i}(z)$ is the $z$-independent matrix 
\begin{equation}
\label{eq:ChargedPCompactSecondUpperBand}
V_{\kappa\nu,i}(z)=
\begin{pmatrix}
9i(i+1)&0\\
-6i(i+1)&9i(i-3)
\end{pmatrix},
\qquad i=2,\ldots,\check{\kappa}-1.
\end{equation}
%

\subsubsection{General vector multiplet polynomials
$Q_{\kappa\nu\eta}(x;z)$}
\label{app:NewChargedOuterVector}

The $(\check{\kappa}-1)$-dimensional matrix
$\cQ_{\kappa\nu\eta}(z)$ that determines the polynomial
$Q_{\kappa\nu\eta}(x;z)$ through (\ref{eq:PolyMatCharged})
can be written, for all $\nu\geq0$, $\kappa \geq 4$ and $\eta = \pm2$ as follows. For $\kappa=4,5$,
\begin{equation}
\label{eq:NewChargedQMatrixn45}
\cQ_{\kappa\nu\eta}(z)
=\bigl(r_{\kappa\nu\eta,2}(z)\bigr) \; , \qquad \kappa=4,5 \; ,
\end{equation}
and for $\kappa\geq6$ takes the tridiagonal form
{\setlength{\arraycolsep}{3pt}
\begin{equation}
\label{eq:NewChargedQMatrix}
\cQ_{\kappa\nu\eta}(z)=
\begin{pmatrix}
r_{\kappa\nu\eta,2}(z)
&1&0&\cdots&0\\
s_{\kappa\nu\eta,2}(z)
&r_{\kappa\nu\eta,3}(z)
&1&\ddots&\vdots\\
0&s_{\kappa\nu\eta,3}(z)
&r_{\kappa\nu\eta,4}(z)
&\ddots&0\\
\vdots&\ddots&\ddots&\ddots&1\\
0&\cdots&0
&s_{\kappa\nu\eta,\check{\kappa}-1}(z)
&r_{\kappa\nu\eta,\check{\kappa}}(z)
\end{pmatrix}.
\end{equation}
}
The diagonal coefficients, for all $z$ and
$i=2,\ldots,\check{\kappa}$, are
\begin{equation}
\label{eq:NewChargedQDiagonalCoefficients}
\begin{aligned}
r_{\kappa\nu\eta,i}(z)
={}&3+3(1-\epsilon)(\kappa-3)(\kappa-2)
-3(1-\epsilon)(1-3\epsilon)(i-2)(i-1)
\\
&+6\epsilon
\bigl[
4(i-2)^2+(1+2\kappa-4\check{\kappa})(2i-3)
\bigr]-18\epsilon^2
\\
&+\frac{9}{16}
\Bigl[
(1-\epsilon)
\left|\rho_{\nu,z}+2\eta-(1+\epsilon)F_{\nu,z}\right|
+(1+\epsilon)
\left|\rho_{\nu,z}+2\eta+(1-\epsilon)F_{\nu,z}\right|
\Bigr]^2
\\
&+\frac34(1-\epsilon)
\bigl[4\kappa-2i-7+3\epsilon(2i-3)\bigr]
\left|\rho_{\nu,z}+2\eta-(1+\epsilon)F_{\nu,z}\right|
\\
&+\frac34
\Bigl[
4\kappa-2i-7
+(4\kappa+24i-16\check{\kappa}-30)\epsilon
-3(2i-3)\epsilon^2
\Bigr]
\\
&\quad\times
\left|\rho_{\nu,z}+2\eta+(1-\epsilon)F_{\nu,z}\right| ,
\end{aligned}
\end{equation}
while the lower-band coefficients, for all $z$ and
$i=2,\ldots,\check{\kappa}-1$, are
\begin{equation}
\label{eq:NewChargedQLowerCoefficients}
\begin{aligned}
s_{\kappa\nu\eta,i}(z)
={}&
18\epsilon(1-\epsilon)(1-3\epsilon)
(i-1)(\check{\kappa}-i)
\\
&\times
\Bigl[
2(i-1)
+\left|\rho_{\nu,z}+2\eta+(1-\epsilon)F_{\nu,z}\right|
\Bigr]
\\
&\times
\Bigl[
2\kappa-2\check{\kappa}+2i-5
+\left|\rho_{\nu,z}+2\eta-(1+\epsilon)F_{\nu,z}\right|
+\left|\rho_{\nu,z}+2\eta+(1-\epsilon)F_{\nu,z}\right|
\Bigr].
\end{aligned}
\end{equation}

With these definitions, the first few polynomials are
\begin{equation}
\label{eq:ChargedQFirstMembers}
\begin{aligned}
Q_{\kappa\nu\eta}(x;z)
&=x-r_{\kappa\nu\eta,2}(z),
&&\kappa=4,5,
\\[4pt]
Q_{\kappa\nu\eta}(x;z)
&=\bigl(x-r_{\kappa\nu\eta,2}(z)\bigr)
  \bigl(x-r_{\kappa\nu\eta,3}(z)\bigr)
  -s_{\kappa\nu\eta,2}(z),
&&\kappa=6,7,
\\[4pt]
Q_{\kappa\nu\eta}(x;z)
&=\bigl(x-r_{\kappa\nu\eta,2}(z)\bigr)
  \bigl(x-r_{\kappa\nu\eta,3}(z)\bigr)
  \bigl(x-r_{\kappa\nu\eta,4}(z)\bigr)
\\
&\quad
-s_{\kappa\nu\eta,2}(z)
 \bigl(x-r_{\kappa\nu\eta,4}(z)\bigr)
-s_{\kappa\nu\eta,3}(z)
 \bigl(x-r_{\kappa\nu\eta,2}(z)\bigr),
&&\kappa=8,9.
\end{aligned}
\end{equation}

The polynomial definitions extend beyond the physical label ranges.
In the exterior, $|z|>1$, and for $\nu\geq1$, the formal polynomial
$Q_{4\nu,-2}$ does not correspond to a physical vector pair, except
when $\nu=1$ at a distinguished value of $z$ in
(\ref{eq:Distinguishedz}), where it gives a long pair. Outside this
exception, the $\kappa=5$ member gives a protected vector pair and
the long family starts at $\kappa=6$. Similarly, the formal
polynomial $H_{1\nu,-1}$ does not correspond to a physical
gravitino pair in the exterior. The label restrictions in
table~\ref{tab:NewChargedExteriorLong} therefore enforce physical
existence as well as excluding protected occurrences.

%

\newpage 

\subsection{\texorpdfstring{Flavour-neutral polynomials}{Flavour-neutral polynomials}}
\label{app:NeutralPolynomials}

The polynomials $G_\kappa(x;z)$, $H_\kappa(x;z)$, $P_\kappa(x;z)$ and $Q_\kappa(x;z)$, of degree
\begin{equation}
\label{eq:xdeg}
\begin{aligned}
\deg_x G_\kappa &= \check{\kappa} +1 \quad (\kappa \geq 0) \, ,
&\qquad \deg_x H_\kappa &= 2 \hat{\kappa} -1 \quad ( \kappa  \geq1 ) \, ,\\[5pt]
\deg_x P_\kappa &= 2 \check{\kappa} \quad ( \kappa \geq2) \, ,
&\qquad \deg_x Q_\kappa &= \check{\kappa} -1 \quad ( \kappa  \geq4) \, ,
\end{aligned}
\end{equation}
governing the long flavour-neutral spectrum $\cL_{\textrm{$F$-neutral}}(z)$ of section~\ref{sec:FlavourNeutralLong} are obtained by setting $\nu=0$ in the general families introduced in appendix~\ref{app:ChargedPolynomialsNew}:
\begin{equation}
\label{eq:NeutralChargedCorr}
\begin{aligned}
G_\kappa(x;z) &= G_{\kappa0}(x;z) \, ,
&\qquad H_\kappa(x;z) &= H_{\kappa0\eta}(x;z) \, ,\\[5pt]
P_\kappa(x;z) &= P_{\kappa0}(x;z) \, ,
&\qquad Q_\kappa(x;z) &= Q_{\kappa0\eta}(x;z) \, .
\end{aligned}
\end{equation}
Their matrix construction therefore follows directly by evaluating (\ref{eq:PolyMatCharged}) on the $\nu=0$ matrices of appendix~\ref{app:ChargedPolynomialsNew}. Since $F_{\nu,z}=\rho_{\nu,z}=0$ at $\nu=0$, the dependence on $z$ reduces to dependence on $\epsilon(z)$ alone. Although the matrix entries contain terms linear in $\epsilon$, their characteristic polynomials are polynomials in $\epsilon^2$. At $\nu=0$, the dependence on the label $\eta$ is exclusively through $|\eta|$ or $\eta^2$, so the two choices $\eta=\pm1$ give the same $H_\kappa(x;z)$, while $\eta=\pm2$ give the same $Q_\kappa(x;z)$. The label $\eta$ can therefore be omitted from all flavour-neutral polynomials. The following subsections present these polynomials explicitly.

\subsubsection{\texorpdfstring{Flavour-neutral graviton
polynomials $G_{\kappa}(x;z)$}
{Flavour-neutral graviton polynomials}}
\label{app:NeutralMatrixGraviton}

Setting $\nu=0$ in (\ref{eq:NewChargedGCoefficients}), the coefficients of the flavour-neutral graviton matrix $\cG_{\kappa0}(z)$ in (\ref{eq:NewChargedGMatrixn01}), (\ref{eq:NewChargedGMatrix}) reduce, for all $\kappa\geq0$, to
\begin{equation}
\label{eq:NewChargedGCoefficientsNeutral}
\begin{aligned}
a_{\kappa0,i}(z)
&=3\kappa(\kappa+3)-3\bigl[\check{\kappa}+2i(\check{\kappa}-i)\bigr](1+3\epsilon^2)
+6(2\kappa-2\check{\kappa}+1)(2i-\check{\kappa})\epsilon,
\\[4pt]
b_{\kappa0,i}(z)
&=9(i+1)^2(\check{\kappa}-i)^2(1-\epsilon^2)(1-9\epsilon^2).
\end{aligned}
\end{equation}
The polynomials $G_\kappa(x;z)$ are then obtained by substituting this matrix into the determinant formula (\ref{eq:PolyMatCharged}). For $\kappa=0,\ldots,5$, the expressions (\ref{eq:ChargedGFirstMembers}), evaluated at $\nu=0$ with the coefficients (\ref{eq:NewChargedGCoefficientsNeutral}), give
\begin{equation}
\label{eq:NeutralGFirstMembers}
\begin{aligned}
G_0(x;z)
&=x,
\\[5pt]
G_1(x;z)
&=x-12,
\\[5pt]
G_2(x;z)
&=x^2+(18\epsilon^2-54)x-720\epsilon^2+720,
\\[5pt]
G_3(x;z)
&=x^2+(18\epsilon^2-102)x-1728\epsilon^2+2592,
\\[5pt]
G_4(x;z)
&=x^3+(72\epsilon^2-228)x^2
 +(972\epsilon^4-13896\epsilon^2+17244)x
\\
&\quad-190512\epsilon^4+622944\epsilon^2-432432,
\\[5pt]
G_5(x;z)
&=x^3+(72\epsilon^2-336)x^2
 +(972\epsilon^4-22536\epsilon^2+37548)x
\\
&\quad-349920\epsilon^4+1589760\epsilon^2-1395360.
\end{aligned}
\end{equation}
%

\subsubsection{\texorpdfstring{Flavour-neutral gravitino
polynomials $H_{\kappa}(x;z)$}
{Flavour-neutral gravitino polynomials}}
\label{app:NeutralMatrixGravitino}

The flavour-neutral gravitino polynomials $H_\kappa(x;z)$ are similarly obtained by substituting the matrix $\cH_{\kappa\nu\eta}(z)$ of appendix~\ref{app:NewChargedGravitino}, evaluated at $\nu=0$, into the determinant formula (\ref{eq:PolyMatCharged}). As discussed at the beginning of appendix~\ref{app:NeutralPolynomials}, this specialisation leaves the matrix dependent on $z$ only through $\epsilon(z)$ and removes its dependence on $\eta$. The corresponding matrix entries follow straightforwardly from the expressions in appendix~\ref{app:NewChargedGravitino}, but remain lengthy and will not be reproduced here. For $\kappa=1,\ldots,4$, the polynomials (\ref{eq:ChargedHFirstMembers}) reduce at $\nu=0$ to
\begin{equation}
\label{eq:NeutralHFirstMembers}
\begin{aligned}
H_1(x;z)
&=x-12,
\\[4pt]
H_2(x;z)
&=x-30,
\\[4pt]
H_3(x;z)
&=x^3+(54\epsilon^2-150)x^2
 +(729\epsilon^4-5769\epsilon^2+7452)x
\\
&\quad-48438\epsilon^4+154062\epsilon^2-122472,
\\[4pt]
H_4(x;z)
&=x^3+(54\epsilon^2-240)x^2
 +(729\epsilon^4-9801\epsilon^2+19152)x
\\
&\quad-96228\epsilon^4+442260\epsilon^2-508032.
\end{aligned}
\end{equation}

The SCP R-charges $\sigma_{\kappa a}=\pm1$ associated with the
roots $h_{\kappa a, z}$ follow from the general sign prescription
of appendix~\ref{app:NewChargedGravitino}. Define its flavour-neutral, $\nu=0$,
specialisation by
\begin{equation} \label{eq:AuxGinoPolNeutral}
 H_{(i)\kappa}(x;z)
 :=H_{(i)\kappa0,+1}(x;z),
 \qquad i=1,2.
\end{equation}
These auxiliary polynomials have coefficients that are polynomial in
$\epsilon^2$ and satisfy
\begin{equation}
 H_\kappa(x;z)
 =\bigl(x+\tfrac14\bigr)H_{(1)\kappa}(x;z)^2
  -H_{(2)\kappa}(x;z)^2.
\end{equation}
Here $H_{(1)\kappa}$ is monic of degree $\hat{\kappa}-1$,
while $H_{(2)\kappa}$ has the same degree and leading
coefficient $\kappa+\check{\kappa}+\tfrac52$.
The identity is imposed before specialising to a particular
background. For a root $h_{\kappa a, z}$ that is not a common zero of the auxiliary
polynomials (\ref{eq:AuxGinoPolNeutral}), the SCP with Lorentz spins $(0,\frac12)$ has
R-charge
\begin{equation}
\label{eq:GenericFormulaRcharges}
 \sigma_{\kappa a}
 =\operatorname{sgn}\!\left[
 H_{(1)\kappa}(h_{\kappa a, z};z)\,
 H_{(2)\kappa}(h_{\kappa a, z};z)
 \right],
 \qquad a=1,\ldots,2\hat{\kappa}-1.
\end{equation}
Its CPT partner has spins $(\frac12,0)$ and charge
$-\sigma_{\kappa a}$. In particular,
$\sigma_{11}=\sigma_{21}=+1$; the $\kappa=1$ pair is
protected, whereas the pairs with $\kappa\geq2$ are long. At common zeros, the prescription
(\ref{eq:NewChargedHCommonFactor}) applies. Namely, at fixed
$z=z_0$, let
\begin{equation}
\label{eq:NeutralGravitinoCommonFactor}
 D_\kappa(x;z_0)
 =\gcd_x\!\left(
 H_{(1)\kappa}(x;z_0),
 H_{(2)\kappa}(x;z_0)
 \right),
\end{equation}
with monic greatest common divisor. A zero of $D_\kappa$
of multiplicity $q$ contributes $2q$ root occurrences of
$H_\kappa$, with $q$ of each sign. For the remaining roots
of $H_\kappa/D_\kappa^2$, use
(\ref{eq:GenericFormulaRcharges}) with each auxiliary
polynomial divided by $D_\kappa$. Contributions at the
same numerical root are added. The sign
counts are
\begin{equation}
\label{eq:NeutralGravitinoSignCounts}
 n_{\kappa,+}
 :=\#\{a:\sigma_{\kappa a}=+1\}=\hat{\kappa},
 \qquad
 n_{\kappa,-}
 :=\#\{a:\sigma_{\kappa a}=-1\}=\hat{\kappa}-1,
\end{equation}
throughout $|\epsilon|<1/\sqrt3$, including degenerate
backgrounds with the multiplicities prescribed above.

A branch followed within a fixed charge--chirality sector
retains its sign, since $F=0$ implies $r=R_0$.
Nevertheless, opposite-sign branches can cross at common
zeros of the auxiliary polynomials, changing the sign
sequence when roots are ordered by increasing dimension.
A degeneracy need not entail such an exchange. For example, at $\kappa=3,4$ the ordered sequence is
$(+,-,+)$ throughout $0<\epsilon^2<1/3$, and the degeneracy
at $\epsilon=0$ does not reverse this ordering. At
$\kappa=5$, however, the second and third roots cross at
\begin{equation}
\label{eq:NeutralGravitinoK5Crossing}
 \epsilon(z_*)^2=\frac{19}{184},
 \qquad
 h_{52, z_*}=h_{53, z_*}=\frac{2256}{23}.
\end{equation}
With $h_{51, z}\leq\cdots\leq h_{55, z}$, the ordered signs are
\begin{equation}
\label{eq:NeutralGravitinoK5OrderedSigns}
 (\sigma_{51},\ldots,\sigma_{55})
 =
 \begin{cases}
 (+,+,-,-,+),
 &0<\epsilon^2<\dfrac{19}{184},\\[5pt]
 (+,-,+,-,+),
 &\dfrac{19}{184}<\epsilon^2<\dfrac13.
 \end{cases}
\end{equation}
At the crossing, the coincident occurrences carry one
sign each, and both CPT pairs remain long. Finally, the
$\epsilon^2$ dependence of the auxiliary polynomials
makes the ordered sign assignments invariant under
$\epsilon\mapsto-\epsilon$, equivalently $z\mapsto-z$,
up to the ordering of coincident occurrences.

\subsubsection{\texorpdfstring{Flavour-neutral vector
polynomials $P_{\kappa}(x;z)$}
{Flavour-neutral vector polynomials}}
\label{app:NeutralMatrixVectorP}

For the flavour-neutral vector sector, we set $\nu=0$, $F_{\nu,z}=\rho_{\nu,z}=0$, in the construction of appendix~\ref{app:NewChargedInnerVector}. The resulting matrices depend on $z$ only through $\epsilon(z)$, but their explicit expressions again remain lengthy and will be omitted. Substitution into the determinant formula (\ref{eq:PolyMatCharged}) gives the neutral polynomials $P_\kappa(x;z)$. The matrix entries, and hence the coefficients of $P_\kappa(x;z)$, are polynomials in $\epsilon^2$. For illustration, we list below the first few examples, with $\kappa=2,\ldots,5$:
\begin{equation}
\label{eq:NeutralPFirstMembers}
\begin{aligned}
P_2(x;z)
&=x\bigl(x-18\epsilon^2-6\bigr),
\\[5pt]
P_3(x;z)
&=x^2-(18\epsilon^2+30)x-216\epsilon^2+216,
\\[5pt]
P_4(x;z)
&=x^4-120x^3-(324\epsilon^4+1800\epsilon^2-5292)x^2
\\
&\quad-(23328\epsilon^4-103680\epsilon^2+101088)x
+699840\epsilon^4-1399680\epsilon^2+699840,
\\[5pt]
P_5(x;z)
&=x^4-216x^3-(324\epsilon^4+4680\epsilon^2-17388)x^2
\\
&\quad-(90720\epsilon^4-501120\epsilon^2+617760)x
+5054400\epsilon^4-13219200\epsilon^2+8164800.
\\[5pt]
\end{aligned}
\end{equation}

\subsubsection{\texorpdfstring{Flavour-neutral vector
polynomials $Q_{\kappa}(x;z)$}
{Flavour-neutral vector polynomials}}
\label{app:NeutralMatrixVectorQ}

For the vector family $Q_\kappa(x;z)$, the neutral limit
$\nu=0$ gives simple expressions
for the entries of $\cQ_{\kappa0\eta}(z)$. The latter matrix retains the
forms (\ref{eq:NewChargedQMatrixn45}), (\ref{eq:NewChargedQMatrix}),
while its coefficients (\ref{eq:NewChargedQDiagonalCoefficients}), (\ref{eq:NewChargedQLowerCoefficients}) simplify, for $\kappa\geq4$, to
\begin{equation}
\label{eq:NewChargedQCoefficientsNeutral}
\begin{aligned}
r_{\kappa0\eta,i}(z)
&=9+3(1-\epsilon)\kappa(\kappa+3)-3i(i+1)
 -9\bigl[i(i+1)-2\bigr]\epsilon^2
\\
&\quad+12\epsilon\bigl[(i-1)(3i+1)
 +(2i-1)(\kappa-2\check{\kappa})\bigr],
\\[4pt]
s_{\kappa0\eta,i}(z)
&=36\epsilon(1-\epsilon)(1-3\epsilon)(i^2-1)
 (\check{\kappa}-i)(2\kappa-2\check{\kappa}+2i+3).
\end{aligned}
\end{equation}
As usual, they become functions of $z$ through $\epsilon(z)$ only. 
Both coefficients (\ref{eq:NewChargedQCoefficientsNeutral}) are also independent of $\eta=\pm2$. With these entries, the determinant (\ref{eq:PolyMatCharged})
defines $Q_\kappa(x;z)$. The first few members,
$\kappa=4,\ldots,9$, follow by bringing (\ref{eq:NewChargedQCoefficientsNeutral}) to 
(\ref{eq:ChargedQFirstMembers}), namely,
\begin{equation}
\label{eq:NeutralQFirstMembers}
\begin{aligned}
Q_4(x;z)
&=x+36\epsilon^2-75,
\\[5pt]
Q_5(x;z)
&=x+36\epsilon^2-111,
\\[5pt]
Q_6(x;z)
&=x^2+(126\epsilon^2-288)x
 +3240\epsilon^4-19098\epsilon^2+20655,
\\[5pt]
Q_7(x;z)
&=x^2+(126\epsilon^2-384)x
 +3240\epsilon^4-26298\epsilon^2+36783,
\\[5pt]
Q_8(x;z)
&=x^3+(288\epsilon^2-705)x^2
 +(23652\epsilon^4-142776\epsilon^2+165231)x
\\
&\quad+524880\epsilon^6-6427836\epsilon^4
 +17552376\epsilon^2-12872655,
\\[5pt]
Q_9(x;z)
&=x^3+(288\epsilon^2-885)x^2
 +(23652\epsilon^4-183096\epsilon^2+260631)x
\\
&\quad+524880\epsilon^6-8469036\epsilon^4
 +28831896\epsilon^2-25540515.
\end{aligned}
\end{equation}

\section{$\cN=2$ enhancement at MN2}
\label{app:N=2Enhancement}

Let us exhibit the reassembly of the $\cN=1$ $\mathrm{SU}(2,2|1) \times \mathrm{U}(1)_+$ multiplets listed for the MN2 spectrum in tables~\ref{tab:FirstZ-neutral-short}, \ref{tab:FirstZ-neutral-long}, \ref{tab:MN2ProtectedMultiplets} and~\ref{tab:NewChargedMN2Long} of section~\ref{sec:UniversalSpectrum} into the $\cN=2$ $\mathrm{SU}(2,2|2)$ multiplet content (\ref{eq:MN2ReassembledSpectrum}). To see this, it is convenient to proceed in reverse, by branching $\mathrm{SU}(2,2|2)$ into $\mathrm{SU}(2,2|1) \times \mathrm{U}(1)_+$ multiplets, and tracking the $\cN=2$ content of (\ref{eq:MN2ReassembledSpectrum}) through the $\cN=1$ tables of section~\ref{sec:UniversalSpectrum}. 

For the $A_2\bar A_2$ tower in (\ref{eq:MN2ReassembledSpectrum}) we find, at $k=0$, the branchings
\begin{equation}
\label{eq:MN2StressTensorBranching}
\begin{split}
 A_2\bar A_2[2;0,0;0,0]\longrightarrow{}&
 A_1\bar A_1[3;\half,\half;0]_0
 \oplus A_2\bar A_2[2;0,0;0]_0
\\[5pt]
&\oplus A_1\bar A_2[\tfrac52;\half,0;\tfrac13]_{-1}
 \oplus A_2\bar A_1[\tfrac52;0,\half;-\tfrac13]_{1}\; ,
\end{split}
\end{equation}
while for all other $k \geq 1$,
\begin{equation}
\label{eq:MN2ShortGravitonBranching}
\begin{aligned}
&A_2\bar A_2[2k+2;0,0;k,0]
\\[5pt]
&\quad\longrightarrow
 L\bar A_2[2k+2;0,0;\tfrac{4k}{3}]_{-k}
 \oplus A_2\bar L[2k+2;0,0;-\tfrac{4k}{3}]_{k}
\\[5pt]
&\qquad\oplus L\bar A_1[2k+3;\half,\half;\tfrac{4k}{3}]_{-k}
 \oplus A_1\bar L[2k+3;\half,\half;-\tfrac{4k}{3}]_{k}
\\[5pt]
&\qquad\oplus L\bar A_2[2k+\tfrac52;\half,0;\tfrac{4k+1}{3}]_{-k-1}
 \oplus A_2\bar L[2k+\tfrac52;0,\half;-\tfrac{4k+1}{3}]_{k+1}
\\[5pt]
&\qquad\oplus L\bar A_1[2k+\tfrac52;0,\half;\tfrac{4k-1}{3}]_{1-k}
 \oplus A_1\bar L[2k+\tfrac52;\half,0;-\tfrac{4k-1}{3}]_{k-1}
\\[5pt]
&\qquad\oplus\bigoplus_{t=-k+1}^{k-1}\Bigl(
 L\bar L[2k+2;0,0;\tfrac{4t}{3}]_{-t}
 \oplus L\bar L[2k+3;\half,\half;\tfrac{4t}{3}]_{-t}
\\[5pt]
&\qquad\qquad\qquad\qquad
 \oplus L\bar L[2k+\tfrac52;\half,0;\tfrac{4t+1}{3}]_{-t-1}
 \oplus L\bar L[2k+\tfrac52;0,\half;\tfrac{4t-1}{3}]_{1-t}
 \Bigr)\; .
\end{aligned}
\end{equation}
The $B_1\bar B_1$ tower branches, for all $k\geq0$, as
\begin{equation}
\label{eq:MN2BPSBranching}
\begin{aligned}
&B_1\bar B_1[2k+4;0,0;k+2,0]
\\[4pt]
&\quad\longrightarrow
 L\bar B_1[2k+4;0,0;\tfrac{4(k+2)}{3}]_{-k-2}
 \oplus B_1\bar L[2k+4;0,0;-\tfrac{4(k+2)}{3}]_{k+2}
\\[4pt]
&\qquad\oplus L\bar A_2[2k+4;0,0;\tfrac{4(k+1)}{3}]_{-k-1}
 \oplus A_2\bar L[2k+4;0,0;-\tfrac{4(k+1)}{3}]_{k+1}
\\
&\qquad\oplus\bigoplus_{t=-k}^{k}
 L\bar L[2k+4;0,0;\tfrac{4t}{3}]_{-t}\; .
\end{aligned}
\end{equation}
Finally, a long $\cN=2$ multiplet with a scalar SCP of dimension $E$, U$(1)$ R-symmetry $n=0$, and SU$(2)$ R-symmetry spin $\ell$ splits as
\begin{equation}
\label{eq:MN2LongBranching}
\begin{aligned}
&L\bar L[E;0,0;\ell,0]\longrightarrow
 \bigoplus_{t=-\ell}^{\ell}\Bigl(
 L\bar L[E;0,0;\tfrac{4t}{3}]_{-t}
\\
&\qquad\oplus L\bar L[E+\tfrac12;\half,0;\tfrac{4t+1}{3}]_{-t-1}
 \oplus L\bar L[E+\tfrac12;0,\half;\tfrac{4t-1}{3}]_{1-t}
\\
&\qquad\oplus L\bar L[E+1;\half,\half;\tfrac{4t}{3}]_{-t}
\\
&\qquad\oplus L\bar L[E+1;0,0;\tfrac{4t+2}{3}]_{-t-2}
 \oplus L\bar L[E+1;0,0;\tfrac{4t-2}{3}]_{2-t}
\\
&\qquad\oplus L\bar L[E+\tfrac32;0,\half;\tfrac{4t+1}{3}]_{-t-1}
 \oplus L\bar L[E+\tfrac32;\half,0;\tfrac{4t-1}{3}]_{1-t}
\\
&\qquad\oplus L\bar L[E+2;0,0;\tfrac{4t}{3}]_{-t}
 \Bigr)\; .
\end{aligned}
\end{equation}

The comparison with the tables of section~\ref{sec:UniversalSpectrum} is now explicit. The neutral graviton and vector in (\ref{eq:MN2StressTensorBranching}) are the $k=0$ entries of table~\ref{tab:FirstZ-neutral-short}, while its charged gravitino pair is the isolated $k=0$ entry of table~\ref{tab:MN2ProtectedMultiplets}. In (\ref{eq:MN2ShortGravitonBranching}), the short vector and graviton pairs reproduce the $\nu=k$ families of table~\ref{tab:MN2ProtectedMultiplets}, and the two short gravitino pairs have $\nu=k+1$ and $\nu=k-1$, respectively. At $k=1$, the latter gives instead the neutral short gravitino pair of table~\ref{tab:FirstZ-neutral-short}. The chiral and short-vector pairs in (\ref{eq:MN2BPSBranching}) reproduce the remaining protected families, with $\nu=k+2$ and $\nu=k+1$. The long sums in these two branchings also contribute to the spectrum. In (\ref{eq:MN2ShortGravitonBranching}), $t=0$ gives the neutral $\eta=0$ vector and graviton of dimensions $2k+2$ and $2k+3$, while the gravitino terms at $t=\pm1$ give the neutral CPT pair of dimension $2k+\frac52$ for $k\geq2$. Equation~(\ref{eq:MN2BPSBranching}) supplies the neutral $\eta=0$ vector of dimension $2k+4$ at $t=0$, including the $\kappa=2$, $k=0$ boundary member of the $P$ family. These are entries of table~\ref{tab:FirstZ-neutral-long}. The charged terms reproduce the $2k+3$ graviton row, the two $2k+\frac52$ gravitino rows and the $2k+2$ vector row of table~\ref{tab:NewChargedMN2Long} from (\ref{eq:MN2ShortGravitonBranching}), and the $2k+4$ vector row from (\ref{eq:MN2BPSBranching}); both vector rows have $\eta=0$.

The remaining long multiplets follow from (\ref{eq:MN2LongBranching}) with $E=D_{kb}$ and $\ell=k-2b$, for $b=1,\ldots,\lfloor k/2\rfloor$. The neutral terms at $t=0$ give the graviton of dimension $D_{kb}+1$ and the $\eta=0$ vectors of dimensions $D_{kb}$ and $D_{kb}+2$. The gravitino terms at $t=\pm1$, present when $\ell\geq1$, give neutral CPT pairs of dimensions $D_{kb}+\frac12$ and $D_{kb}+\frac32$, with $\sigma=+1$ and $\sigma=-1$, respectively. The $\eta=2$ and $\eta=-2$ vector terms become neutral at $t=2$ and $t=-2$, respectively, giving the $r=\pm2$ pair of dimension $D_{kb}+1$ when $\ell\geq2$. These complete all four families in table~\ref{tab:FirstZ-neutral-long}. The charged terms supply every remaining row of table~\ref{tab:NewChargedMN2Long}: the $D_{kb}+1$ gravitons, the four gravitino rows of dimensions $D_{kb}+\frac12$ and $D_{kb}+\frac32$, the $\eta=0$ vectors of dimensions $D_{kb}$ and $D_{kb}+2$, and the three $\eta=\pm2$ vector rows of dimension $D_{kb}+1$. Throughout the long sums, the representative with $F=-\nu<0$ has $t=\nu$ for gravitons and $\eta=0$ vectors, $t=\nu\pm1$ for gravitini, $t=\nu+2$ for $\eta=2$ vectors and $t=\nu-2$ for $\eta=-2$ vectors. In particular, $t=\nu-2=-1$ gives the separately listed $\nu=1$, $\eta=-2$ vector row. These identifications, the summation ranges and the level relations in the tables reproduce all restrictions on $\nu$, $\kappa$ and $b$. The opposite-flavour terms provide the CPT partners. After this discussion, every multiplet in the branchings (\ref{eq:MN2StressTensorBranching})--(\ref{eq:MN2LongBranching}) and every entry in the MN2 tables, \ref{tab:FirstZ-neutral-short}, \ref{tab:FirstZ-neutral-long}, \ref{tab:MN2ProtectedMultiplets} and~\ref{tab:NewChargedMN2Long} of section~\ref{sec:UniversalSpectrum}, is accounted for.

At $z=1$, the $\mathrm{U}(1)_+\times\mathrm{U}(1)_r$ flavour and
superconformal R-symmetries of the $\cN=1$ description are embedded
in the $\mathrm{SU}(2)\times\mathrm{U}(1)$ R-symmetry of the
$\cN=2$ MN2 solution. An $\mathrm{SU}(2)\times\mathrm{U}(1)$ representation
with spin $\ell$ and charge $n$ splits under $\mathrm{U}(1)_+\times\mathrm{U}(1)_r$ as
\begin{equation}
\label{eq:MN2ChargeDictionary}
(\mathbf{2\ell+1})_n
\longrightarrow
\bigoplus_{m=-\ell}^{\ell}
\mathbf{1}_{\left(n-m,\;\frac{2n+4m}{3}\right)}\; ,
\end{equation}
where the subscripts on the right-hand side denote the
$\mathrm{U}(1)_+$ flavour charge $F$ and the
$\mathrm{U}(1)_r$ superconformal R-charge $r$, respectively.
The sum runs over the $\mathrm{SU}(2)$ Cartan weights in unit steps.
The reference R-charge is $R_0=n+m$, so the embedding gives
$F=n-m$ and $r=R_0-\frac13F$, consistently with (\ref{eq:R0FfromJpm}), (\ref{eq:SpectrumExactRCharge}) and $\epsilon=-\frac13$ at MN2. Conversely, a state with charges $(R_0,F)$ has $n=\frac12(R_0+F)$ and
$m=\frac12(R_0-F)$. The $\mathrm{SU}(2,2|2) \supset \mathrm{SU}(2,2|1) \times \mathrm{U}(1)_+$ branchings (\ref{eq:MN2StressTensorBranching})--(\ref{eq:MN2LongBranching}) utilise the $\mathrm{SU}(2) \times \mathrm{U}(1) \supset \mathrm{U}(1)_+\times\mathrm{U}(1)_r$ splitting (\ref{eq:MN2ChargeDictionary}).

\begingroup
\DeclareRobustCommand{\PutativeSkull}{\raisebox{-0.1ex}{\includegraphics[height=.65em]{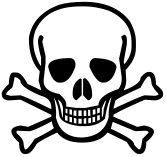}}}
\section{Putative spectrum}
\label{app:PutativeSpectrum}
\label{app:PutativeSpectralFlows}

Let us comment on the putative spectrum discussed in the introduction.
This is naturally obtained by diagonalising the trombone KK mass
matrices of~\cite{Pico:2026kch}, without imposing
$\mathrm{U}(1)_z$ invariance. The spectrum is naturally organised by the KK level
$k$ defined in appendix~\ref{app:InvariantMassMatrices}, namely, the
degree of the $S^4$ harmonic in
(\ref{eq:SeedTimesHarmonicRepresentations}). At each level we retain
the full field representations, remove the Higgsed and spurious states,
and allocate the remaining mass eigenstates into
$\mathrm{SU}(2,2|1)\times\mathrm{U}(1)_+$ supermultiplets.
The resulting spectrum is putative because the underlying maximal
truncation is only locally defined: its modes need not extend to
globally defined fluctuations on the internal space. In particular,
some multiplets contain unphysical states, as their dimensions lie
below the superconformal unitarity bounds. The physical sector studied
in the main text is obtained by restricting the mass matrices to
$\mathrm{U}(1)_z$-invariant fields before diagonalising them.


Figure~\ref{fig:BBBWPutativeUnitarityFlows} shows the SCP dimensions
$E$ as functions of $z$ at $k=0,\ldots,3$. Gravitons, gravitini and vectors are shown
in blue, green and red, respectively. The red/pink distinction between
the invariant vector families governed by $P_\kappa$ and $Q_\kappa$
is not made for the full putative spectrum. The $k=0$ chiral pair has
$E=3$ and coincides with the dotted stress-tensor curve.

For a long multiplet $\LL[E;j_1,j_2;r]_F$, the unitarity conditions
of appendix~\ref{app:NOneMultipletContent} read
\begin{equation}
\label{eq:PutativeLongUnitarityBound}
 E>E_{\mathrm{LL}}\, ,\qquad
 E_{\mathrm{LL}}=
 \max\!\left\{2+2j_1-\frac32r\,,\;
                 2+2j_2+\frac32r\right\}\; .
\end{equation}
Since $r=R_0+\epsilon(z)F$, the bound depends on the branch as well
as on $z$. The lower panels of the figure plot $E-E_{\mathrm{LL}}$
close to zero: curves below the dashed line violate the bound.
Short multiplets obey their own shortening conditions and are omitted
from these lower panels. At a shortening point, a long multiplet
splits into short multiplets, so the continuous curves alone need
not give the complete multiplet content at that point. Tensor multiplets provide an example. At the levels shown, they occur
at isolated shortening points at fixed $k$, and hence do not appear as
continuous tensor curves. At $z=3$, $k=0$, the putative spectrum contains
\begin{equation}
\label{eq:PutativeTensorExample}
 L\bar B_1\left[\frac{17+2\sqrt7}{6};\frac12,0;
                    \frac{17+2\sqrt7}{9}\right]_{-1}
 \oplus\mathrm{CPT}\; ,
\end{equation}
which is precisely the tensor pair retained in
table~\ref{tab:RepresentativeThree}. It arises when a long gravitino
multiplet splits into a short gravitino and a tensor multiplet.
There are no tensor multiplets through $k=2$ at the interior value
$z=1/3$ considered below. Two-form fields are nevertheless present as
components of other multiplets.

\begin{figure}[p]
\centering
\captionsetup[subfigure]{position=top}
\begin{subfigure}[t]{.47\textwidth}
\centering
\caption{$k=0$}\label{fig:BBBWPutativeUnitarityKK0}
\includegraphics[width=\linewidth]{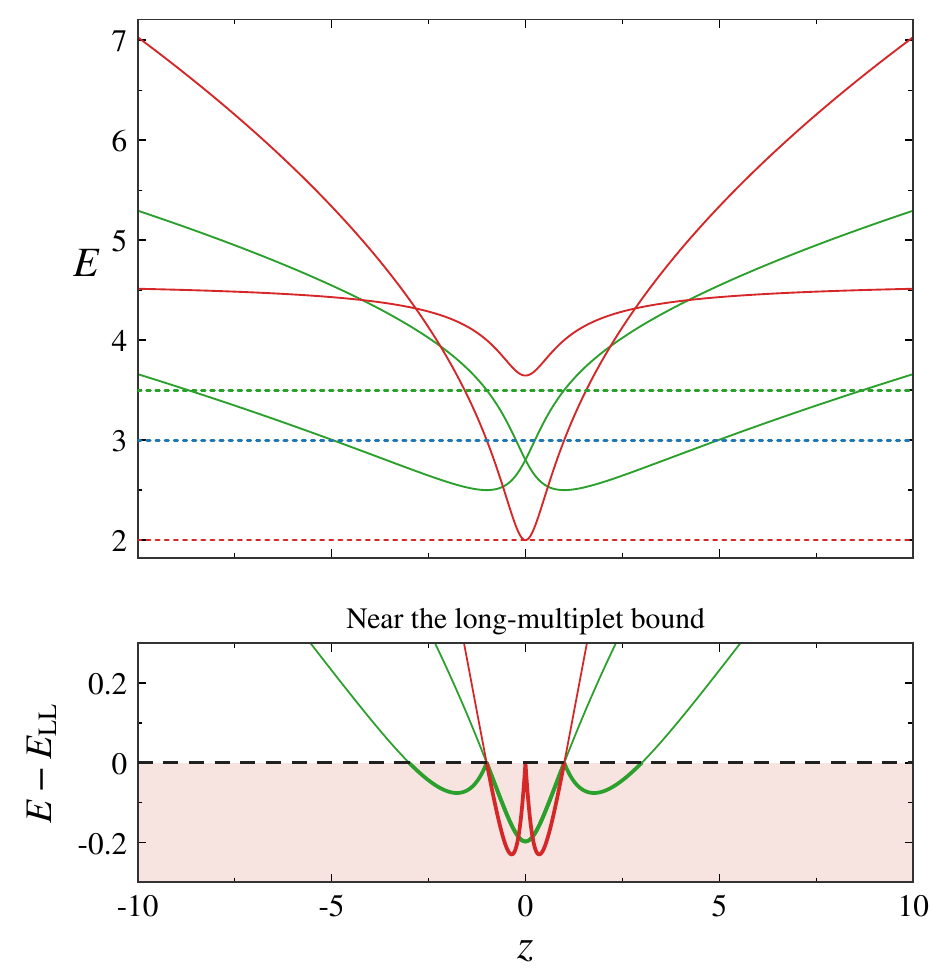}
\end{subfigure}\hfill
\begin{subfigure}[t]{.47\textwidth}
\centering
\caption{$k=1$}\label{fig:BBBWPutativeUnitarityKK1}
\includegraphics[width=\linewidth]{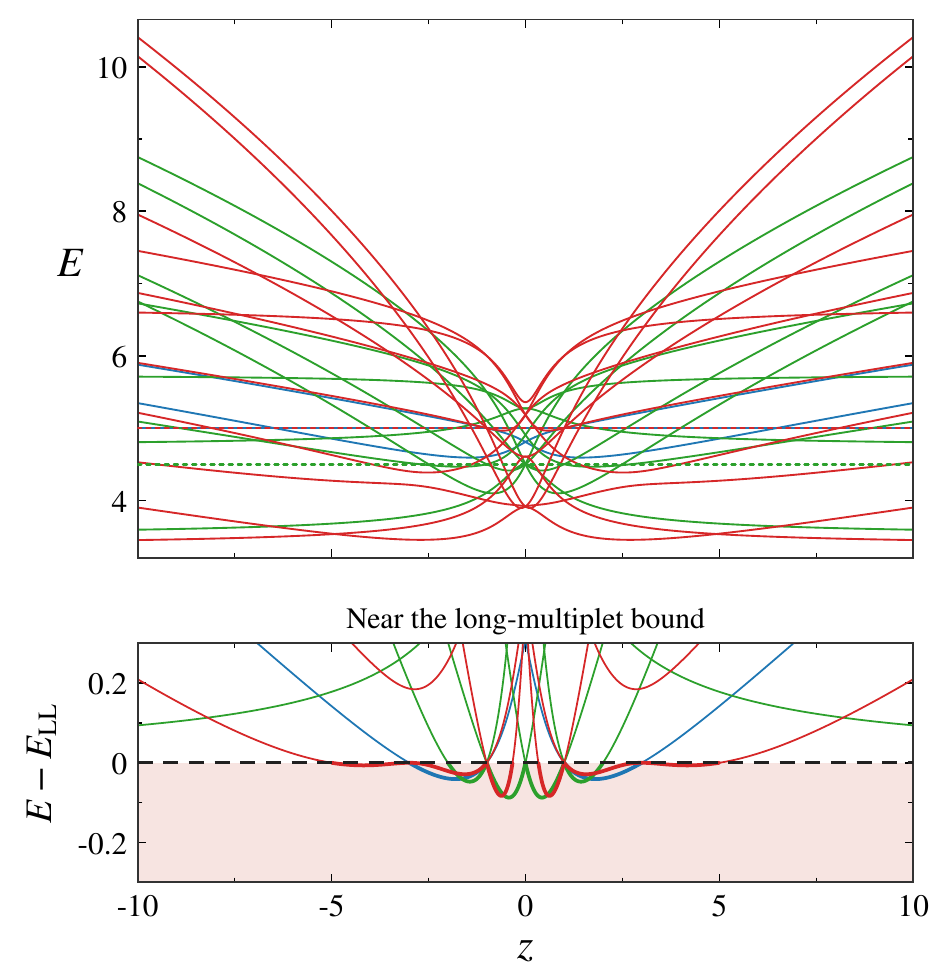}
\end{subfigure}\\[8pt]
\begin{subfigure}[t]{.47\textwidth}
\centering
\caption{$k=2$}\label{fig:BBBWPutativeUnitarityKK2}
\includegraphics[width=\linewidth]{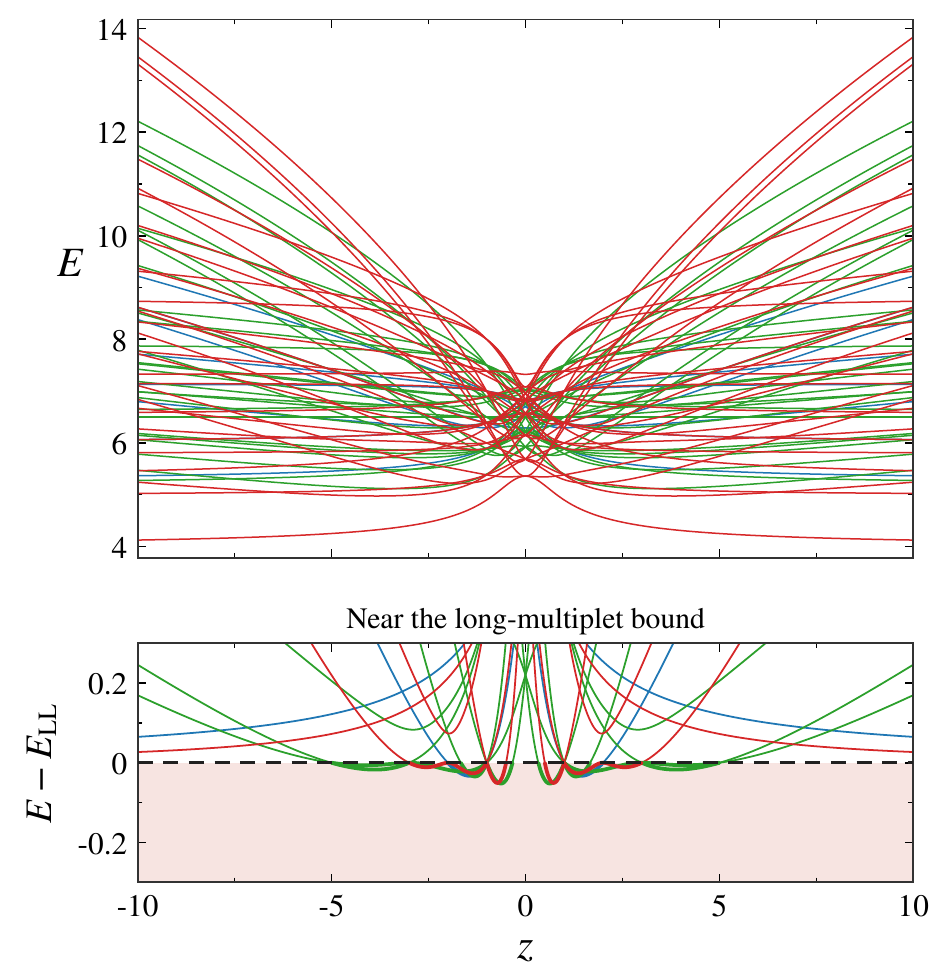}
\end{subfigure}\hfill
\begin{subfigure}[t]{.47\textwidth}
\centering
\caption{$k=3$}\label{fig:BBBWPutativeUnitarityKK3}
\includegraphics[width=\linewidth]{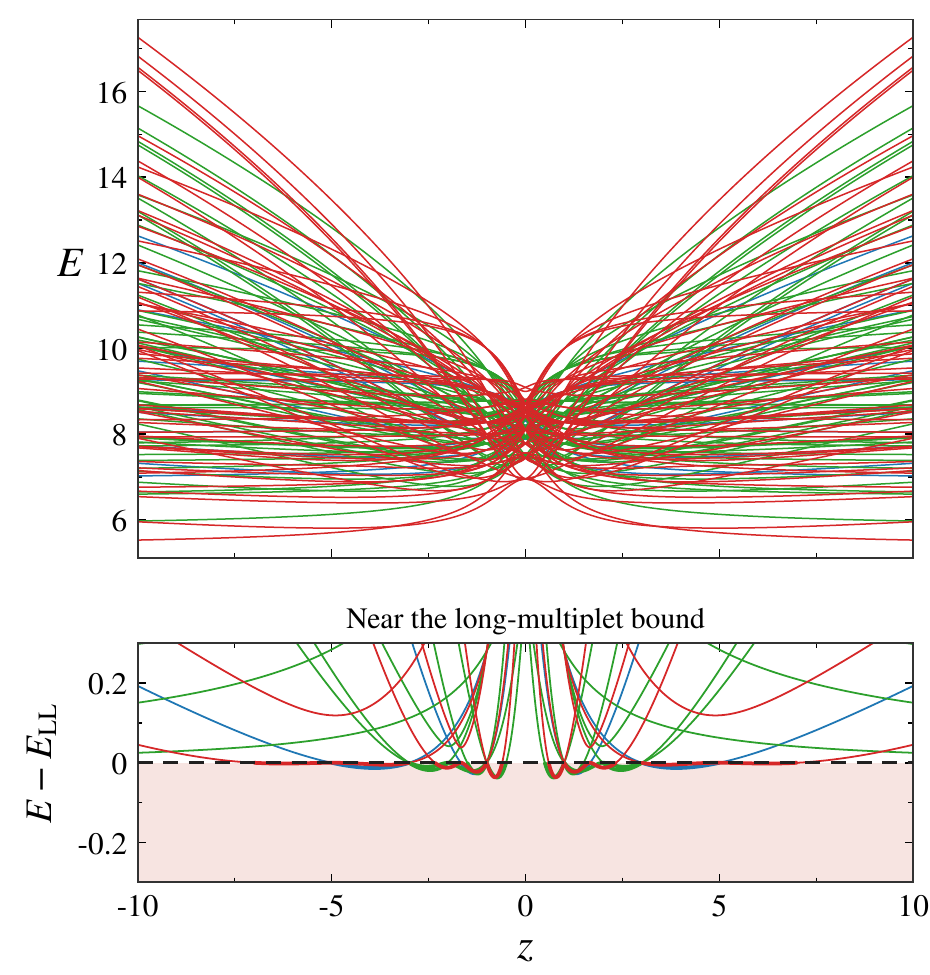}
\end{subfigure}
\caption{The putative BBBW spectrum at $k=0,\ldots,3$, with SCP
dimensions $E$ plotted against $z$ for $-10\leq z\leq10$.
Graviton, gravitino and vector multiplets are blue, green and red,
respectively. Solid curves denote generically long multiplets and
dotted curves denote short families present for all $z$. The $k=0$
chiral pair coincides with the stress-tensor curve at $E=3$.
The lower panels show $E-E_{\mathrm{LL}}$ for the long branches in
the range $[-0.3,0.3]$; the shaded region lies below the unitarity
bound. Additional short multiplets present only at isolated $z$,
including tensor multiplets, are not drawn as continuous curves.
Coincident curves conceal multiplicities. The plots interpolate the
allowed quantised twisting data.}
\label{fig:BBBWPutativeUnitarityFlows}
\end{figure}


Let us conclude by looking at a specific example. For comparison with section~\ref{sec:RepresentativeOneThird}, we
choose the distinguished interior value $z=1/3$. The primitive
twisting data are $(\widehat p,\widehat q)=(2,1)$ and
$\epsilon=1-2\sqrt3/3$. Table~\ref{tab:PutativeRepresentativeOneThird}
gives the putative spectrum through $k=2$, in the same notation as
table~\ref{tab:RepresentativeOneThird} of the main text. A superscript $*$ marks each
multiplet that is also retained in table~\ref{tab:RepresentativeOneThird}. A superscript
${}^{\PutativeSkull}$ marks a multiplet that violates a superconformal
unitarity bound. These two sets are disjoint. In particular, the
absence of a star does not necessarily imply a unitarity violation.

There are $13$, $49$ and $128$ putative multiplets at
$k=0,1,2$, respectively, counting CPT partners separately. Of these,
$3$, $7$ and $14$ belong to the invariant spectrum, while
$6$, $2$ and $0$ violate unitarity. The additional short
multiplets at this distinguished value of $z$ are included
separately. Dimensions and R-charges are rounded to three decimal
places, as in section~\ref{sec:RepresentativeSpectra}. For example, the $k=0$ vector with $(F,R_0)=(-2,0)$ has
\begin{equation}
\label{eq:PutativeUnitarityExample}
 r=\frac{4\sqrt3}{3}-2\simeq0.30940\, ,\qquad
 E\simeq2.23482<2+\frac32|r|=2\sqrt3-1\simeq2.46410\; .
\end{equation}
\begingroup
\fontsize{8}{10}\selectfont
\thinmuskip=2mu\medmuskip=3mu\thickmuskip=4mu
\renewcommand{\arraystretch}{1.0}
\setlength{\tabcolsep}{4pt}
\setlength{\LTpre}{6pt}
\setlength{\LTcapwidth}{\textwidth}
\setlength{\LTpost}{10pt}
\begin{longtable}{@{}p{0.105\textwidth}|p{\dimexpr0.895\textwidth-8.4pt\relax}@{}}
\caption[The full putative $z=1/3$ spectrum at $k=0,\ldots,2$.]{The full putative $z=\frac13$ spectrum at $k=0,\ldots,2$.
Dimensions and noninteger R-charges are rounded to three decimal places.
A superscript $*$ marks the multiplets retained in the invariant spectrum
of table~\ref{tab:RepresentativeOneThird}; a superscript ${}^{\PutativeSkull}$ marks
a unitarity violation. A mark on a correlated-sign or parenthesised CPT pair applies to
both members. Explicit prefactors give the multiplicity of each representation.}
\label{tab:PutativeRepresentativeOneThird}\\
\hline\hline
\textbf{Multiplet} & $z=\frac13$ \\
\hline
\endfirsthead
\multicolumn{2}{c}{\footnotesize\tablename\ \thetable\ (continued): $z=\frac13$}\\
\hline\hline
\textbf{Multiplet} & Representations \\
\hline
\endhead
\hline\multicolumn{2}{r}{\textit{Continued on the next page}}\\
\endfoot
\hline\hline
\endlastfoot
\multicolumn{2}{l}{\rule{0pt}{13pt}$k=0$}\\*
Graviton & \raggedright \mbox{\(A_1\bar A_1[3.000;\tfrac{1}{2},\tfrac{1}{2};0]_{0}^{*}\)} \arraybackslash \\[1pt]
Gravitino & \raggedright \mbox{\((L\bar A_2[3.500;\tfrac{1}{2},0;1]_{0}\oplus\mathrm{CPT})\)} \allowbreak
\mbox{\(\oplus\,(L\bar L[2.607;\tfrac{1}{2},0;0.155]_{-1}\oplus\mathrm{CPT})^{\PutativeSkull}\)} \allowbreak
\mbox{\(\oplus\,(L\bar L[3.071;0,\tfrac{1}{2};0.155]_{-1}\oplus\mathrm{CPT})^{\PutativeSkull}\)} \arraybackslash \\[1pt]
Vector & \raggedright \mbox{\(A_2\bar A_2[2.000;0,0;0]_{0}^{*}\)} \allowbreak
\mbox{\(\oplus\,L\bar L[3.726;0,0;0]_{0}^{*}\)} \allowbreak
\mbox{\(\oplus\,L\bar L[2.235;0,0;\pm 0.309]_{\mp 2}^{\PutativeSkull}\)} \arraybackslash \\[1pt]
Chiral & \raggedright \mbox{\((L\bar B_1[3.000;0,0;2]_{0}\oplus\mathrm{CPT})\)} \arraybackslash \\[1pt]
\hline\hline
\multicolumn{2}{l}{\rule{0pt}{13pt}$k=1$}\\*
Graviton & \raggedright \mbox{\(L\bar L[5.000;\tfrac{1}{2},\tfrac{1}{2};0]_{0}^{*}\)} \allowbreak
\mbox{\(\oplus\,L\bar L[4.711;\tfrac{1}{2},\tfrac{1}{2};\pm 0.845]_{\pm 1}\)} \allowbreak
\mbox{\(\oplus\,L\bar L[4.894;\tfrac{1}{2},\tfrac{1}{2};\pm 1.155]_{\mp 1}\)} \arraybackslash \\[1pt]
Gravitino & \raggedright \mbox{\((L\bar L[4.400;\tfrac{1}{2},0;1]_{0}\oplus\mathrm{CPT})\)} \allowbreak
\mbox{\(\oplus\,(L\bar A_1[4.500;0,\tfrac{1}{2};1]_{0}\oplus\mathrm{CPT})^{*}\)} \allowbreak
\mbox{\(\oplus\,(L\bar L[5.237;0,\tfrac{1}{2};1]_{0}\oplus\mathrm{CPT})\)} \allowbreak
\mbox{\(\oplus\,(L\bar L[5.337;\tfrac{1}{2},0;1]_{0}\oplus\mathrm{CPT})\)} \allowbreak
\mbox{\(\oplus\,(L\bar L[4.690;\tfrac{1}{2},0;0.155]_{-1}\oplus\mathrm{CPT})\)} \allowbreak
\mbox{\(\oplus\,(L\bar L[5.154;0,\tfrac{1}{2};0.155]_{-1}\oplus\mathrm{CPT})\)} \arraybackslash \\[1pt]
Gravitino\newline\textit{(cont.)} & \raggedright \mbox{\(\oplus\,(L\bar L[4.220;\tfrac{1}{2},0;-0.691]_{-2}\oplus\mathrm{CPT})\)} \allowbreak
\mbox{\(\oplus\,(L\bar L[4.416;\tfrac{1}{2},0;1.309]_{-2}\oplus\mathrm{CPT})\)} \allowbreak
\mbox{\(\oplus\,(L\bar L[4.684;0,\tfrac{1}{2};-0.691]_{-2}\oplus\mathrm{CPT})\)} \allowbreak
\mbox{\(\oplus\,(L\bar L[4.880;0,\tfrac{1}{2};1.309]_{-2}\oplus\mathrm{CPT})^{\PutativeSkull}\)} \arraybackslash \\[1pt]
Vector & \raggedright \mbox{\(L\bar L[4.416;0,0;0]_{0}^{*}\)} \allowbreak
\mbox{\(\oplus\,L\bar L[5.557;0,0;0]_{0}^{*}\)} \allowbreak
\mbox{\(\oplus\,(L\bar A_2[5.000;0,0;2]_{0}\oplus\mathrm{CPT})\)} \allowbreak
\mbox{\(\oplus\,L\bar L[3.832;0,0;\pm 0.845]_{\pm 1}\)} \allowbreak
\mbox{\(\oplus\,L\bar L[3.938;0,0;\pm 1.155]_{\mp 1}\)} \allowbreak
\mbox{\(\oplus\,L\bar L[4.946;0,0;\pm 1.155]_{\mp 1}\)} \arraybackslash \\[1pt]
Vector\newline\textit{(cont.)} & \raggedright \mbox{\(\oplus\,L\bar L[4.992;0,0;\pm 0.845]_{\pm 1}\)} \allowbreak
\mbox{\(\oplus\,L\bar L[5.380;0,0;\pm 0.845]_{\pm 1}\)} \allowbreak
\mbox{\(\oplus\,L\bar L[5.509;0,0;\pm 1.155]_{\mp 1}\)} \allowbreak
\mbox{\(\oplus\,L\bar L[4.678;0,0;\pm 0.309]_{\mp 2}\)} \allowbreak
\mbox{\(\oplus\,L\bar L[3.970;0,0;\pm 0.536]_{\pm 3}\)} \allowbreak
\mbox{\(\oplus\,(L\bar A_2[4.196;0,0;1.464]_{-3}\oplus\mathrm{CPT})^{*}\)} \arraybackslash \\[1pt]
Chiral & \raggedright \mbox{\((L\bar B_1[5.196;0,0;3.464]_{-3}\oplus\mathrm{CPT})\)} \arraybackslash \\[1pt]
\hline\hline
\multicolumn{2}{l}{\rule{0pt}{13pt}$k=2$}\\*
Graviton & \raggedright \mbox{\(L\bar L[6.192;\tfrac{1}{2},\tfrac{1}{2};0]_{0}^{*}\)} \allowbreak
\mbox{\(\oplus\,L\bar L[6.883;\tfrac{1}{2},\tfrac{1}{2};0]_{0}^{*}\)} \allowbreak
\mbox{\(\oplus\,L\bar L[6.529;\tfrac{1}{2},\tfrac{1}{2};\pm 2]_{0}\)} \allowbreak
\mbox{\(\oplus\,L\bar L[6.554;\tfrac{1}{2},\tfrac{1}{2};\pm 0.845]_{\pm 1}\)} \allowbreak
\mbox{\(\oplus\,L\bar L[6.839;\tfrac{1}{2},\tfrac{1}{2};\pm 1.155]_{\mp 1}\)} \allowbreak
\mbox{\(\oplus\,L\bar L[6.300;\tfrac{1}{2},\tfrac{1}{2};\pm 0.309]_{\mp 2}\)} \arraybackslash \\[1pt]
Graviton\newline\textit{(cont.)} & \raggedright \mbox{\(\oplus\,L\bar L[6.407;\tfrac{1}{2},\tfrac{1}{2};\pm 1.691]_{\pm 2}\)} \allowbreak
\mbox{\(\oplus\,L\bar L[6.740;\tfrac{1}{2},\tfrac{1}{2};\pm 2.309]_{\mp 2}\)} \arraybackslash \\[1pt]
Gravitino & \raggedright \mbox{\((L\bar L[6.308;\tfrac{1}{2},0;1]_{0}\oplus\mathrm{CPT})\)} \allowbreak
\mbox{\(\oplus\,(L\bar L[6.500;0,\tfrac{1}{2};1]_{0}\oplus\mathrm{CPT})^{*}\)} \allowbreak
\mbox{\(\oplus\,(L\bar L[6.992;0,\tfrac{1}{2};1]_{0}\oplus\mathrm{CPT})\)} \allowbreak
\mbox{\(\oplus\,(L\bar L[7.184;\tfrac{1}{2},0;1]_{0}\oplus\mathrm{CPT})\)} \allowbreak
\mbox{\(\oplus\,(L\bar L[5.765;\tfrac{1}{2},0;0.155]_{-1}\oplus\mathrm{CPT})\)} \allowbreak
\mbox{\(\oplus\,(L\bar L[5.944;0,\tfrac{1}{2};0.155]_{-1}\oplus\mathrm{CPT})\)} \arraybackslash \\[1pt]
Gravitino\newline\textit{(cont.)} & \raggedright \mbox{\(\oplus\,(L\bar L[6.569;\tfrac{1}{2},0;0.155]_{-1}\oplus\mathrm{CPT})\)} \allowbreak
\mbox{\(\oplus\,(L\bar L[6.626;0,\tfrac{1}{2};0.155]_{-1}\oplus\mathrm{CPT})\)} \allowbreak
\mbox{\(\oplus\,(L\bar L[6.846;\tfrac{1}{2},0;0.155]_{-1}\oplus\mathrm{CPT})\)} \allowbreak
\mbox{\(\oplus\,(L\bar L[7.074;0,\tfrac{1}{2};0.155]_{-1}\oplus\mathrm{CPT})\)} \allowbreak
\mbox{\(\oplus\,(L\bar L[6.074;\tfrac{1}{2},0;-1.845]_{-1}\oplus\mathrm{CPT})\)} \allowbreak
\mbox{\(\oplus\,(L\bar L[6.124;\tfrac{1}{2},0;2.155]_{-1}\oplus\mathrm{CPT})\)} \arraybackslash \\[1pt]
Gravitino\newline\textit{(cont.)} & \raggedright \mbox{\(\oplus\,(L\bar L[6.132;0,\tfrac{1}{2};-1.845]_{-1}\oplus\mathrm{CPT})\)} \allowbreak
\mbox{\(\oplus\,(L\bar L[6.349;0,\tfrac{1}{2};2.155]_{-1}\oplus\mathrm{CPT})\)} \allowbreak
\mbox{\(\oplus\,(L\bar L[6.896;0,\tfrac{1}{2};-1.845]_{-1}\oplus\mathrm{CPT})\)} \allowbreak
\mbox{\(\oplus\,(L\bar L[6.955;\tfrac{1}{2},0;-1.845]_{-1}\oplus\mathrm{CPT})\)} \allowbreak
\mbox{\(\oplus\,(L\bar L[7.005;0,\tfrac{1}{2};2.155]_{-1}\oplus\mathrm{CPT})\)} \allowbreak
\mbox{\(\oplus\,(L\bar L[7.230;\tfrac{1}{2},0;2.155]_{-1}\oplus\mathrm{CPT})\)} \arraybackslash \\[1pt]
Gravitino\newline\textit{(cont.)} & \raggedright \mbox{\(\oplus\,(L\bar L[6.152;\tfrac{1}{2},0;-0.691]_{-2}\oplus\mathrm{CPT})\)} \allowbreak
\mbox{\(\oplus\,(L\bar L[6.446;\tfrac{1}{2},0;1.309]_{-2}\oplus\mathrm{CPT})\)} \allowbreak
\mbox{\(\oplus\,(L\bar L[6.616;0,\tfrac{1}{2};-0.691]_{-2}\oplus\mathrm{CPT})\)} \allowbreak
\mbox{\(\oplus\,(L\bar L[6.910;0,\tfrac{1}{2};1.309]_{-2}\oplus\mathrm{CPT})\)} \allowbreak
\mbox{\(\oplus\,(L\bar L[5.768;\tfrac{1}{2},0;0.464]_{-3}\oplus\mathrm{CPT})^{*}\)} \allowbreak
\mbox{\(\oplus\,(L\bar L[6.232;0,\tfrac{1}{2};0.464]_{-3}\oplus\mathrm{CPT})\)} \arraybackslash \\[1pt]
Gravitino\newline\textit{(cont.)} & \raggedright \mbox{\(\oplus\,(L\bar L[5.881;\tfrac{1}{2},0;-1.536]_{-3}\oplus\mathrm{CPT})\)} \allowbreak
\mbox{\(\oplus\,(L\bar L[6.232;\tfrac{1}{2},0;2.464]_{-3}\oplus\mathrm{CPT})\)} \allowbreak
\mbox{\(\oplus\,(L\bar L[6.345;0,\tfrac{1}{2};-1.536]_{-3}\oplus\mathrm{CPT})\)} \allowbreak
\mbox{\(\oplus\,(L\bar A_1[6.696;0,\tfrac{1}{2};2.464]_{-3}\oplus\mathrm{CPT})^{*}\)} \arraybackslash \\[1pt]
Vector & \raggedright \mbox{\(L\bar L[5.271;0,0;0]_{0}^{*}\)} \allowbreak
\mbox{\(\oplus\,L\bar L[6.308;0,0;0]_{0}^{*}\)} \allowbreak
\mbox{\(\oplus\,2\, L\bar L[6.529;0,0;0]_{0}\)} \allowbreak
\mbox{\(\oplus\,L\bar L[7.043;0,0;0]_{0}^{*}\)} \allowbreak
\mbox{\(\oplus\,L\bar L[7.408;0,0;0]_{0}^{*}\)} \allowbreak
\mbox{\(\oplus\,L\bar L[5.617;0,0;\pm 2]_{0}\)} \arraybackslash \\[1pt]
Vector\newline\textit{(cont.)} & \raggedright \mbox{\(\oplus\,L\bar L[6.698;0,0;\pm 2]_{0}\)} \allowbreak
\mbox{\(\oplus\,L\bar L[6.875;0,0;\pm 2]_{0}\)} \allowbreak
\mbox{\(\oplus\,L\bar L[7.357;0,0;\pm 2]_{0}\)} \allowbreak
\mbox{\(\oplus\,L\bar L[5.921;0,0;\pm 0.845]_{\pm 1}\)} \allowbreak
\mbox{\(\oplus\,L\bar L[6.105;0,0;\pm 1.155]_{\mp 1}\)} \allowbreak
\mbox{\(\oplus\,L\bar L[6.715;0,0;\pm 1.155]_{\mp 1}\)} \arraybackslash \\[1pt]
Vector\newline\textit{(cont.)} & \raggedright \mbox{\(\oplus\,L\bar L[6.831;0,0;\pm 0.845]_{\pm 1}\)} \allowbreak
\mbox{\(\oplus\,L\bar L[7.164;0,0;\pm 0.845]_{\pm 1}\)} \allowbreak
\mbox{\(\oplus\,L\bar L[7.383;0,0;\pm 1.155]_{\mp 1}\)} \allowbreak
\mbox{\(\oplus\,L\bar L[5.347;0,0;\pm 0.309]_{\mp 2}\)} \allowbreak
\mbox{\(\oplus\,L\bar L[6.235;0,0;\pm 0.309]_{\mp 2}\)} \allowbreak
\mbox{\(\oplus\,L\bar L[6.660;0,0;\pm 0.309]_{\mp 2}\)} \arraybackslash \\[1pt]
Vector\newline\textit{(cont.)} & \raggedright \mbox{\(\oplus\,L\bar L[6.907;0,0;\pm 0.309]_{\mp 2}\)} \allowbreak
\mbox{\(\oplus\,L\bar L[5.558;0,0;\pm 1.691]_{\pm 2}\)} \allowbreak
\mbox{\(\oplus\,L\bar L[5.788;0,0;\pm 2.309]_{\mp 2}\)} \allowbreak
\mbox{\(\oplus\,L\bar L[6.479;0,0;\pm 1.691]_{\pm 2}\)} \allowbreak
\mbox{\(\oplus\,L\bar L[6.627;0,0;\pm 2.309]_{\mp 2}\)} \allowbreak
\mbox{\(\oplus\,L\bar L[7.017;0,0;\pm 1.691]_{\pm 2}\)} \arraybackslash \\[1pt]
Vector\newline\textit{(cont.)} & \raggedright \mbox{\(\oplus\,L\bar L[7.311;0,0;\pm 2.309]_{\mp 2}\)} \allowbreak
\mbox{\(\oplus\,L\bar L[6.089;0,0;\pm 0.536]_{\pm 3}\)} \allowbreak
\mbox{\(\oplus\,L\bar L[6.398;0,0;\pm 1.464]_{\mp 3}^{*}\)} \allowbreak
\mbox{\(\oplus\,(L\bar A_2[7.196;0,0;3.464]_{-3}\oplus\mathrm{CPT})\)} \allowbreak
\mbox{\(\oplus\,L\bar L[5.546;0,0;\pm 0.619]_{\mp 4}\)} \allowbreak
\mbox{\(\oplus\,L\bar L[5.670;0,0;\pm 1.381]_{\pm 4}\)} \arraybackslash \\[1pt]
Vector\newline\textit{(cont.)} & \raggedright \mbox{\(\oplus\,L\bar L[6.052;0,0;\pm 2.619]_{\mp 4}\)} \arraybackslash \\[1pt]
\end{longtable}
\endgroup

It and its CPT conjugate carry the skull symbol in the table.
Throughout the table, $L\bar L$ denotes the long multiplet
assignment even when its dimension is below the bound.

\endgroup

\bibliography{references}

\providecommand{\href}[2]{#2}\begingroup\raggedright\begin{thebibliography}{10}

\bibitem{Benini:2009mz}
F.~Benini, Y.~Tachikawa, and B.~Wecht, {\it {Sicilian gauge theories and N=1
  dualities}},  {\em JHEP} {\bf 01} (2010) 088,
  [\href{http://arxiv.org/abs/0909.1327}{{\tt arXiv:0909.1327}}].

\bibitem{Bah:2011vv}
I.~Bah, C.~Beem, N.~Bobev, and B.~Wecht, {\it {AdS/CFT Dual Pairs from
  M5-Branes on Riemann Surfaces}},  {\em Phys. Rev. D} {\bf 85} (2012) 121901,
  [\href{http://arxiv.org/abs/1112.5487}{{\tt arXiv:1112.5487}}].

\bibitem{Bah:2012dg}
I.~Bah, C.~Beem, N.~Bobev, and B.~Wecht, {\it {Four-Dimensional SCFTs from
  M5-Branes}},  {\em JHEP} {\bf 06} (2012) 005,
  [\href{http://arxiv.org/abs/1203.0303}{{\tt arXiv:1203.0303}}].

\bibitem{Gaiotto:2009we}
D.~Gaiotto, {\it {N=2 dualities}},  {\em JHEP} {\bf 08} (2012) 034,
  [\href{http://arxiv.org/abs/0904.2715}{{\tt arXiv:0904.2715}}].

\bibitem{Gaiotto:2009gz}
D.~Gaiotto and J.~Maldacena, {\it {The Gravity duals of N=2 superconformal
  field theories}},  {\em JHEP} {\bf 10} (2012) 189,
  [\href{http://arxiv.org/abs/0904.4466}{{\tt arXiv:0904.4466}}].

\bibitem{Intriligator:2003jj}
K.~A. Intriligator and B.~Wecht, {\it {The Exact superconformal R symmetry
  maximizes a}},  {\em Nucl. Phys. B} {\bf 667} (2003) 183--200,
  [\href{http://arxiv.org/abs/hep-th/0304128}{{\tt hep-th/0304128}}].

\bibitem{Maldacena:2000mw}
J.~M. Maldacena and C.~Nunez, {\it {Supergravity description of field theories
  on curved manifolds and a no go theorem}},  {\em Int. J. Mod. Phys. A} {\bf
  16} (2001) 822--855, [\href{http://arxiv.org/abs/hep-th/0007018}{{\tt
  hep-th/0007018}}].

\bibitem{Baggio:2014hua}
M.~Baggio, N.~Halmagyi, D.~R. Mayerson, D.~Robbins, and B.~Wecht, {\it {Higher
  Derivative Corrections and Central Charges from Wrapped M5-branes}},  {\em
  JHEP} {\bf 12} (2014) 042, [\href{http://arxiv.org/abs/1408.2538}{{\tt
  arXiv:1408.2538}}].

\bibitem{Bah:2020uev}
I.~Bah, F.~Bonetti, and R.~Minasian, {\it {Discrete and higher-form symmetries
  in SCFTs from wrapped M5-branes}},  {\em JHEP} {\bf 03} (2021) 196,
  [\href{http://arxiv.org/abs/2007.15003}{{\tt arXiv:2007.15003}}].

\bibitem{Beem:2012yn}
C.~Beem and A.~Gadde, {\it {The $N=1$ superconformal index for class $S$ fixed
  points}},  {\em JHEP} {\bf 04} (2014) 036,
  [\href{http://arxiv.org/abs/1212.1467}{{\tt arXiv:1212.1467}}].

\bibitem{Maldacena:1997re}
J.~M. Maldacena, {\it {The Large N limit of superconformal field theories and
  supergravity}},  {\em Int. J. Theor. Phys.} {\bf 38} (1999) 1113--1133,
  [\href{http://arxiv.org/abs/hep-th/9711200}{{\tt hep-th/9711200}}]. [Adv.
  Theor. Math. Phys.2,231(1998)].

\bibitem{Gubser:1998bc}
S.~Gubser, I.~R. Klebanov, and A.~M. Polyakov, {\it {Gauge theory correlators
  from noncritical string theory}},  {\em Phys. Lett. B} {\bf 428} (1998)
  105--114, [\href{http://arxiv.org/abs/hep-th/9802109}{{\tt hep-th/9802109}}].

\bibitem{Witten:1998qj}
E.~Witten, {\it {Anti-de Sitter space and holography}},  {\em Adv. Theor. Math.
  Phys.} {\bf 2} (1998) 253--291,
  [\href{http://arxiv.org/abs/hep-th/9802150}{{\tt hep-th/9802150}}].

\bibitem{Bhattacharya:2024tjw}
R.~Bhattacharya, A.~Katyal, and O.~Varela, {\it {Class S Superconformal Indices
  from Maximal Supergravity}},  {\em Phys. Rev. Lett.} {\bf 134} (2025), no.~18
  181601, [\href{http://arxiv.org/abs/2411.16837}{{\tt arXiv:2411.16837}}].

\bibitem{Bhattacharya:2026ypg}
R.~Bhattacharya, A.~Katyal, and O.~Varela, {\it {Maximal $D=5$ trombone
  supergravity from M5-branes and $\mathrm{SU}(2)$-flavoured $\mathcal{N}=1$
  class $\mathcal{S}$ operator spectra}},
  \href{http://arxiv.org/abs/2607.25952}{{\tt arXiv:2607.25952}}.

\bibitem{Malek:2019eaz}
E.~Malek and H.~Samtleben, {\it {Kaluza-Klein Spectrometry for Supergravity}},
  {\em Phys. Rev. Lett.} {\bf 124} (2020), no.~10 101601,
  [\href{http://arxiv.org/abs/1911.12640}{{\tt arXiv:1911.12640}}].

\bibitem{Malek:2020yue}
E.~Malek and H.~Samtleben, {\it {Kaluza-Klein Spectrometry from Exceptional
  Field Theory}},  {\em Phys. Rev. D} {\bf 102} (2020), no.~10 106016,
  [\href{http://arxiv.org/abs/2009.03347}{{\tt arXiv:2009.03347}}].

\bibitem{Varela:2020wty}
O.~Varela, {\it {Super-Chern-Simons spectra from Exceptional Field Theory}},
  {\em JHEP} {\bf 04} (2021) 283, [\href{http://arxiv.org/abs/2010.09743}{{\tt
  arXiv:2010.09743}}].

\bibitem{Cesaro:2020soq}
M.~Ces{\`a}ro and O.~Varela, {\it {Kaluza-Klein fermion mass matrices from
  exceptional field theory and $ \mathcal{N} $ = 1 spectra}},  {\em JHEP} {\bf
  03} (2021) 138, [\href{http://arxiv.org/abs/2012.05249}{{\tt
  arXiv:2012.05249}}].

\bibitem{Pico:2026kch}
M.~Pico and O.~Varela, {\it {Kaluza-Klein trombone mass matrices and universal
  class $\mathcal{R}$ operator spectra}},  {\em JHEP} {\bf 07} (2026) 19,
  [\href{http://arxiv.org/abs/2603.28908}{{\tt arXiv:2603.28908}}].

\bibitem{Coimbra:2011ky}
A.~Coimbra, C.~Strickland-Constable, and D.~Waldram, {\it {$E_{d(d)} \times
  \mathbb{R}^+$ generalised geometry, connections and M theory}},  {\em JHEP}
  {\bf 02} (2014) 054, [\href{http://arxiv.org/abs/1112.3989}{{\tt
  arXiv:1112.3989}}].

\bibitem{Coimbra:2012af}
A.~Coimbra, C.~Strickland-Constable, and D.~Waldram, {\it {Supergravity as
  Generalised Geometry II: $E_{d(d)} \times \mathbb{R}^+$ and M theory}},  {\em
  JHEP} {\bf 03} (2014) 019, [\href{http://arxiv.org/abs/1212.1586}{{\tt
  arXiv:1212.1586}}].

\bibitem{Hohm:2013pua}
O.~Hohm and H.~Samtleben, {\it {Exceptional Form of D=11 Supergravity}},  {\em
  Phys.Rev.Lett.} {\bf 111} (2013) 231601,
  [\href{http://arxiv.org/abs/1308.1673}{{\tt arXiv:1308.1673}}].

\bibitem{Hohm:2013vpa}
O.~Hohm and H.~Samtleben, {\it {Exceptional Field Theory I: $E_{6(6)}$
  covariant Form of M-Theory and Type IIB}},  {\em Phys. Rev.} {\bf D89}
  (2014), no.~6 066016, [\href{http://arxiv.org/abs/1312.0614}{{\tt
  arXiv:1312.0614}}].

\bibitem{LeDiffon:2008sh}
A.~Le~Diffon and H.~Samtleben, {\it {Supergravities without an Action: Gauging
  the Trombone}},  {\em Nucl. Phys. B} {\bf 811} (2009) 1--35,
  [\href{http://arxiv.org/abs/0809.5180}{{\tt arXiv:0809.5180}}].

\bibitem{Varela:2025xeb}
O.~Varela, {\it {Trombone gaugings of five-dimensional maximal supergravity}},
  {\em JHEP} {\bf 02} (2026) 163, [\href{http://arxiv.org/abs/2509.12391}{{\tt
  arXiv:2509.12391}}].

\bibitem{Pico:2025cmc}
M.~Pico and O.~Varela, {\it {Consistent subsectors of maximal supergravity and
  wrapped M5-branes}},  {\em JHEP} {\bf 05} (2026) 003,
  [\href{http://arxiv.org/abs/2511.15892}{{\tt arXiv:2511.15892}}].

\bibitem{Pico:2026rji}
M.~Pico and O.~Varela, {\it {Maximal trombone supergravity from wrapped
  M5-branes}},  {\em JHEP} {\bf 05} (2026) 076,
  [\href{http://arxiv.org/abs/2601.07960}{{\tt arXiv:2601.07960}}].

\bibitem{Cassani:2019vcl}
D.~Cassani, G.~Josse, M.~Petrini, and D.~Waldram, {\it {Systematics of
  consistent truncations from generalised geometry}},  {\em JHEP} {\bf 11}
  (2019) 017, [\href{http://arxiv.org/abs/1907.06730}{{\tt arXiv:1907.06730}}].

\bibitem{Cassani:2020cod}
D.~Cassani, G.~Josse, M.~Petrini, and D.~Waldram, {\it {$\mathcal{N} $ = 2
  consistent truncations from wrapped M5-branes}},  {\em JHEP} {\bf 02} (2021)
  232, [\href{http://arxiv.org/abs/2011.04775}{{\tt arXiv:2011.04775}}].

\bibitem{Josse:2025uro}
G.~Josse, M.~Petrini, and M.~Pico, {\it {Consistent Truncations and Generalised
  Geometry: Scanning through Dimensions and Supersymmetry}},
  \href{http://arxiv.org/abs/2512.03027}{{\tt arXiv:2512.03027}}.

\bibitem{Guarino:2020flh}
A.~Guarino, E.~Malek, and H.~Samtleben, {\it {Stable Nonsupersymmetric
  Anti{\textendash}de Sitter Vacua of Massive IIA Supergravity}},  {\em Phys.
  Rev. Lett.} {\bf 126} (2021), no.~6 061601,
  [\href{http://arxiv.org/abs/2011.06600}{{\tt arXiv:2011.06600}}].

\bibitem{Cesaro:2021haf}
M.~Cesaro, G.~Larios, and O.~Varela, {\it {Supersymmetric spectroscopy on
  AdS$_{4} \times S^{7}$ and AdS$_{4} \times S^{6}$}},  {\em JHEP} {\bf 07}
  (2021) 094, [\href{http://arxiv.org/abs/2103.13408}{{\tt arXiv:2103.13408}}].
  [Erratum: JHEP 05, 084 (2024)].

\bibitem{Giambrone:2021zvp}
A.~Giambrone, E.~Malek, H.~Samtleben, and M.~Trigiante, {\it {Global properties
  of the conformal manifold for S-fold backgrounds}},  {\em JHEP} {\bf 06}
  (2021), no.~111 111, [\href{http://arxiv.org/abs/2103.10797}{{\tt
  arXiv:2103.10797}}].

\bibitem{Cesaro:2021tna}
M.~Ces{\`a}ro, G.~Larios, and O.~Varela, {\it {The spectrum of
  marginally-deformed $ \mathcal{N} $ = 2 CFTs with AdS$_{4}$ S-fold duals of
  type IIB}},  {\em JHEP} {\bf 12} (2021) 214,
  [\href{http://arxiv.org/abs/2109.11608}{{\tt arXiv:2109.11608}}].

\bibitem{Cesaro:2022mbu}
M.~Ces{\`a}ro, G.~Larios, and O.~Varela, {\it {$ \mathcal{N} $ = 1 S-fold
  spectroscopy}},  {\em JHEP} {\bf 08} (2022) 242,
  [\href{http://arxiv.org/abs/2206.04064}{{\tt arXiv:2206.04064}}].

\bibitem{deWit:1986iy}
B.~de~Wit and H.~Nicolai, {\it {The Consistency of the $S^7$ Truncation in
  $D=11$ Supergravity}},  {\em Nucl.Phys.} {\bf B281} (1987) 211.

\bibitem{Lee:2014mla}
K.~Lee, C.~Strickland-Constable, and D.~Waldram, {\it {Spheres, generalised
  parallelisability and consistent truncations}},  {\em Fortsch. Phys.} {\bf
  65} (2017), no.~10-11 1700048, [\href{http://arxiv.org/abs/1401.3360}{{\tt
  arXiv:1401.3360}}].

\bibitem{Guarino:2015jca}
A.~Guarino, D.~L. Jafferis, and O.~Varela, {\it {The string origin of dyonic
  N=8 supergravity and its simple Chern-Simons duals}},  {\em Phys. Rev. Lett.}
  {\bf 115} (2015), no.~9 091601, [\href{http://arxiv.org/abs/1504.08009}{{\tt
  arXiv:1504.08009}}].

\bibitem{Guarino:2015vca}
A.~Guarino and O.~Varela, {\it {Consistent $ \mathcal{N}=8 $ truncation of
  massive IIA on S$^{6}$}},  {\em JHEP} {\bf 12} (2015) 020,
  [\href{http://arxiv.org/abs/1509.02526}{{\tt arXiv:1509.02526}}].

\bibitem{Inverso:2016eet}
G.~Inverso, H.~Samtleben, and M.~Trigiante, {\it {Type II supergravity origin
  of dyonic gaugings}},  {\em Phys. Rev.} {\bf D95} (2017), no.~6 066020,
  [\href{http://arxiv.org/abs/1612.05123}{{\tt arXiv:1612.05123}}].

\bibitem{deWit:1982ig}
B.~de~Wit and H.~Nicolai, {\it {N=8 Supergravity}},  {\em Nucl.Phys.} {\bf
  B208} (1982) 323.

\bibitem{DallAgata:2011aa}
G.~Dall'Agata and G.~Inverso, {\it {On the Vacua of N = 8 Gauged Supergravity
  in 4 Dimensions}},  {\em Nucl.Phys.} {\bf B859} (2012) 70--95,
  [\href{http://arxiv.org/abs/1112.3345}{{\tt arXiv:1112.3345}}].

\bibitem{Guarino:2015qaa}
A.~Guarino and O.~Varela, {\it {Dyonic ISO(7) supergravity and the duality
  hierarchy}},  {\em JHEP} {\bf 02} (2016) 079,
  [\href{http://arxiv.org/abs/1508.04432}{{\tt arXiv:1508.04432}}].

\bibitem{Cordova:2016emh}
C.~Cordova, T.~T. Dumitrescu, and K.~Intriligator, {\it {Multiplets of
  Superconformal Symmetry in Diverse Dimensions}},  {\em JHEP} {\bf 03} (2019)
  163, [\href{http://arxiv.org/abs/1612.00809}{{\tt arXiv:1612.00809}}].

\bibitem{Gauntlett:2007ma}
J.~P. Gauntlett and O.~Varela, {\it {Consistent Kaluza-Klein reductions for
  general supersymmetric AdS solutions}},  {\em Phys.Rev.} {\bf D76} (2007)
  126007, [\href{http://arxiv.org/abs/0707.2315}{{\tt arXiv:0707.2315}}].

\bibitem{Szepietowski:2012tb}
P.~Szepietowski, {\it {Comments on a-maximization from gauged supergravity}},
  {\em JHEP} {\bf 12} (2012) 018, [\href{http://arxiv.org/abs/1209.3025}{{\tt
  arXiv:1209.3025}}].

\bibitem{Faedo:2019cvr}
A.~F. Faedo, C.~Nunez, and C.~Rosen, {\it {Consistent truncations of
  supergravity and $\frac{1}{2}$-BPS RG flows in $4d$ SCFTs}},  {\em JHEP} {\bf
  03} (2020) 080, [\href{http://arxiv.org/abs/1912.13516}{{\tt
  arXiv:1912.13516}}].

\end{thebibliography}\endgroup


\end{document}